\documentclass[aps,preprint,floats,epsf,epsfig,nofootinbib,letter]{revtex4}
\usepackage{epsfig}
\usepackage{graphicx}
\usepackage{dcolumn}
\usepackage{bm}
\usepackage{subfigure}

\def\DESepsf(#1 width #2){\epsfxsize=#2 \epsfbox{#1}}

\begin{document}
\def\be{\begin{eqnarray}}
\def\en{\end{eqnarray}}
\def\non{\nonumber}
\def\la{\langle}
\def\ra{\rangle}
\def\Br{{\mathcal B}}
\def\A{{\mathcal A}}
\def\BB{{{\cal B} \overline {\cal B}}}
\def\BD{{{\cal B} \overline {\cal D}}}
\def\DB{{{\cal D} \overline {\cal B}}}
\def\DD{{{\cal D} \overline {\cal D}}}
\def\sq{\sqrt}


\title{Rates and $CP$ asymmetries of Charmless Two-body Baryonic $B_{u,d,s}$ Decays}

\author{Chun-Khiang Chua}
\affiliation{Department of Physics and Center for High Energy Physics,
Chung Yuan Christian University,
Chung-Li, Taiwan 320, Republic of China}

\date{\today}

\begin{abstract}
With the experimental evidences of $\overline B {}^0\to p \overline{p}$ and $B^-\to\Lambda\overline p$ decays, 
it is now possible to extract both tree and penguin amplitudes of the charmless two-body baryonic $B$ decays
for the first time.
The extracted penguin-tree ratio agrees with the expectation.
Using the topological amplitude approach with the experimental results on $\overline B {}^0\to p \overline{p}$ and $B^-\to\Lambda\bar p$ decay rates as input,
predictions on all other $\overline B_q\to {\cal B} \overline {\cal B}$, 
${\cal B} \overline {\cal D}$, 
${\cal D} \overline {\cal B}$ and 
${\cal D} \overline {\cal D}$ decay rates, where ${\mathcal B}$ and ${\cal D}$ are the low lying octet and decuplet baryons, respectively, are given.
It is non-trivial that the results do not violate any existing experimental upper limit.
From the analysis it is understandable that why $\overline B {}^0\to p \overline{p}$ and $B^-\to\Lambda\bar p$ modes are the first two modes with experimental evidences.
Relations on rates are verified using the numerical results.
We note that the predicted $B^-\to p\overline{\Delta^{++}}$ rate is close to the experimental bound,
which has not been updated in the last ten years.
Direct $CP$ asymmetries of all $\overline B_q\to {\cal B} \overline {\cal B}$, 
${\cal B} \overline {\cal D}$, 
${\cal D} \overline {\cal B}$ and 
${\cal D} \overline {\cal D}$ modes are explored.
Relations on $CP$ asymmetries are examined using the numerical results.
The direct $CP$ asymmetry of $\overline B {}^0\to p \overline{p}$ decay can be as large as $\pm 50\%$.
Some of the $CP$ asymmetries can serve as tests of the Standard Model.
Most of them are pure penguin modes, which are expected to be sensitive to New Physics contributions.
In particular, 
$\overline B{}^0_s\to \Xi^{-}\overline{\Xi^{-}}$, 
$\overline B{}^0\to \Xi^{-}\overline{\Sigma^{*-}}$, 
$\overline B{}^0\to \Omega^-\overline{\Xi^{-}}$, 
$\overline B{}^0_s\to \Sigma^{*-} \overline{\Sigma^{*-}}$, 
$\overline B{}^0_s\to \Omega^{-} \overline{\Omega^{-}}$, 
$\overline B{}^0_s\to \Xi^{-}\overline{\Xi^{*-}}$, 
$\overline B{}^0_s\to \Xi^{*-}\overline{\Xi^{-}}$,  
$\overline B{}^0\to \Xi^{*-} \overline{\Sigma^{*-}}$, 
$\overline B{}^0\to \Omega^{-} \overline{\Xi^{*-}}$ 
and
$\overline B{}^0_s\to \Xi^{*-} \overline{\Xi^{*-}}$ 
decays 
are $\Delta S=-1$ pure penguin modes with unsuppressed rates, which can be searched in near future. 
Their $CP$ asymmetries are constrained to be of few $\%$ and are good candidates to be added to the list of the tests of the Standard Model.

\end{abstract}


\maketitle

\section{Introduction}

Recently, 
LHCb collaboration reported the evidence
for the first penguin dominated charmless two-body baryonic mode, $B^-\to \Lambda\bar p$ decay,
at 4.1$\sigma$ level~\cite{Aaij:2016xfa}, giving
\be
{\cal B}(B^-\to\Lambda\bar p)=(2.4^{+1.0}_{-0.8}\pm 0.3)\times 10^{-7}.
\en
The result is consistent with the predictions given in a pole model calculation~\cite{Cheng:2001tr} and a topological amplitude approach~\cite{Chua:2013zga}. The latter work made use of the large $m_B$ mass,
the experimental rate of the tree-dominated $\overline B{}^0\to p\bar p$ decay~\cite{Aaij:2013fta},
\be
{\cal B}(\overline B {}^0\to p \overline{p})=
         (1.47{}^{+0.62+0.35}_{-0.51-0.14})\times 10^{-8},
\en
and a na\"{i}ve estimation on tree-penguin ratio.  
In fact, it was advocated in \cite{Chua:2013zga} that the $B^-\to\Lambda\bar p$ decay mode could be the second charmless two-body baryonic mode to be found experimentally.
With both the experimental evidences on $\overline B {}^0\to p \overline{p}$ and $B^-\to\Lambda\bar p$ decays, it is now possible to extract both tree and penguin amplitudes at the same time.

Most of the decay amplitudes of the two-body baryonic decays are non-factorizable.~\footnote{For a study on the factorization contributions, see Ref.~\cite{Hsiao:2014zza}.}
Various models, such as
pole
model, ~\cite{Deshpande:1987nc,Jarfi:1990ej,Cheng:2001tr,Cheng:2001ub},
sum rule, ~\cite{Chernyak:ag}, diquark
model, ~\cite{Ball:1990fw,Chang:2001jt} and approaches employing flavor symmetry
~\cite{Gronau:1987xq,He:re,Sheikholeslami:fa,Luo:2003pv}
were used (for recent reviews, see~\cite{review, review1})
to calculate the amplitudes.
Nevertheless all predictions on the $\overline B{}^0\to p\bar p$ rate are off by several orders of magnitude comparing to the LHCb result~\cite{Aaij:2013fta,review, review1}.  

Indeed, 
the $B^0\to p\bar p$ decay mode is more suppressed than expected. To see this one scales the
$\bar B^0\to\Lambda_c^+\bar p$ rate by the Cabibbo-Kobayashi-Maskawa (CKM) matrix elements $|V_{ub}/V_{cb}|^2$ and with a possible dynamical suppression factor $f_{dyn}$ included,~\cite{Cheng:2014qxa}
\be
\Br(\overline B^0\to p\bar p) = \Br(\overline B^0\to\Lambda_c^+\bar p)|V_{ub}/V_{cb}|^2\times f_{dyn} 
\sim 2\times 10^{-7}\times f_{dyn}.
\en
The data demands $f_{dyn}\sim 0.1$.
It was pointed out in Ref.~\cite{Cheng:2014qxa} that for a given tree operator $O_i$, one needs to consider additional contribution, which were missed in the literature, from its Fiertz transformed operator $O'_i$.
It was found that there are cancelations of Feynman diagrams induced by $O_i$ by that from $O'_i$
and, consequently, the smallness of the tree-dominated charmless two-body baryonic $B$ decays results from this partial cancellation.
Furthermore, as pointed out in Ref.~\cite{Cheng:2014qxa} the internal $W$-emission tree amplitude should be proportional to the Wilson coefficient combination $c_1+c_2$ rather than $c_1-c_2$, where the latter is usually claimed in the literature. We shall adapt this point in the present work.

We can make use of the newly measured $\overline B {}^0\to p\bar p$ and $B^-\to\Lambda \bar p$ rates to give information on other modes in a symmetry related approach.
The quark diagram or the so-called topological
approach has been extensively used in mesonic modes~\cite{Zeppenfeld:1980ex,Chau:tk,Chau:1990ay,Gronau:1994rj,Gronau:1995hn,Cheng:2014rfa}.
In fact, the approach is closely related to the flavor SU(3)
symmetry~\cite{Zeppenfeld:1980ex,Gronau:1994rj,Savage:ub}. 
In~\cite{Chua:2003it} the approach was extended to the charmless two-body baryonic case. 
It was further developed in \cite{Chua:2013zga}, where amplitudes for all low lying charmless two-body baryonic modes with full topological amplitudes are obtained.
Note that in general a typical amplitude has more than one tree and one penguin amplitudes. 
Asymptotic relations in the large $m_B$ limit~\cite{Brodsky:1980sx} can be used to relate various
topological amplitudes~\cite{Chua:2003it, Chua:2013zga}.

In this work,
using the experimental results on the $\overline B{}^0\to p\bar p$ and $B^-\to\Lambda \bar p$ decay rates, 
we will extract both tree and penguin amplitudes for the first time. 
Rates and direct $CP$ asymmetries of all low lying charmless two body baryonic decays can be explored.
Rates and $CP$ asymmetries of some modes can be checked experimentally in the near future in LHCb and Belle-II.
Note that $CP$ asymmetries of some modes can be added to the list of the tests of the Standard Model.
In particular, $\Delta S=-1$ pure penguin modes have small $CP$ asymmetries and they are expected to be sensitive to New Physics contributions. 
These modes are good candidates to be added to the lists of the tests of the Standard Model, especially for those with unsuppressed rates.

The layout of this paper is as following.
In Sec. II, we briefly review and update our formulation for charmless two-body baryonic decays modes.
In Sec. III, we present the numerical results on rates and direct $CP$ asymmetries of all low lying baryon modes.
Relations on rates and $CP$ asymmetries will also be studied.
Conclusion is given in Sec. IV, which is followed by two appendices.

\section{Two-body charmless baryonic $B$ decay amplitudes}

\begin{figure}[t]
\centering
 \subfigure[]{
  \includegraphics[width=0.4\textwidth]{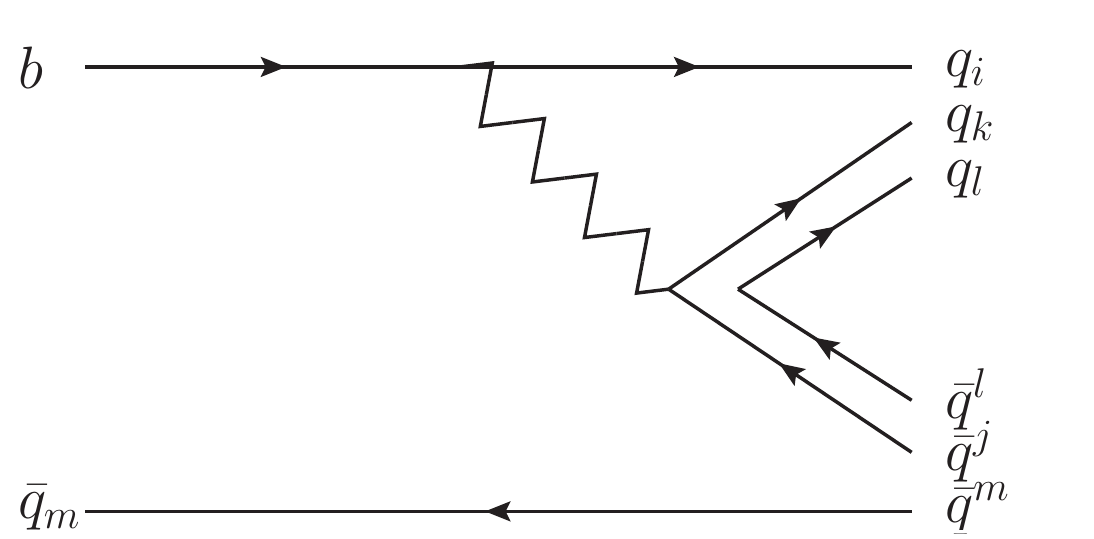}
}\subfigure[]{
  \includegraphics[width=0.4\textwidth]{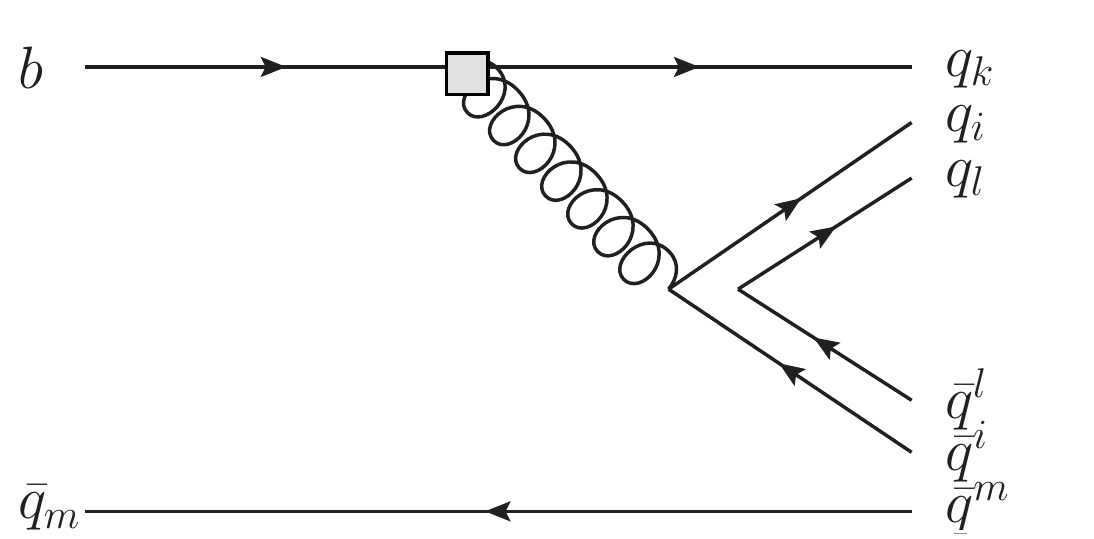}
}\\\subfigure[]{
  \includegraphics[width=0.4\textwidth]{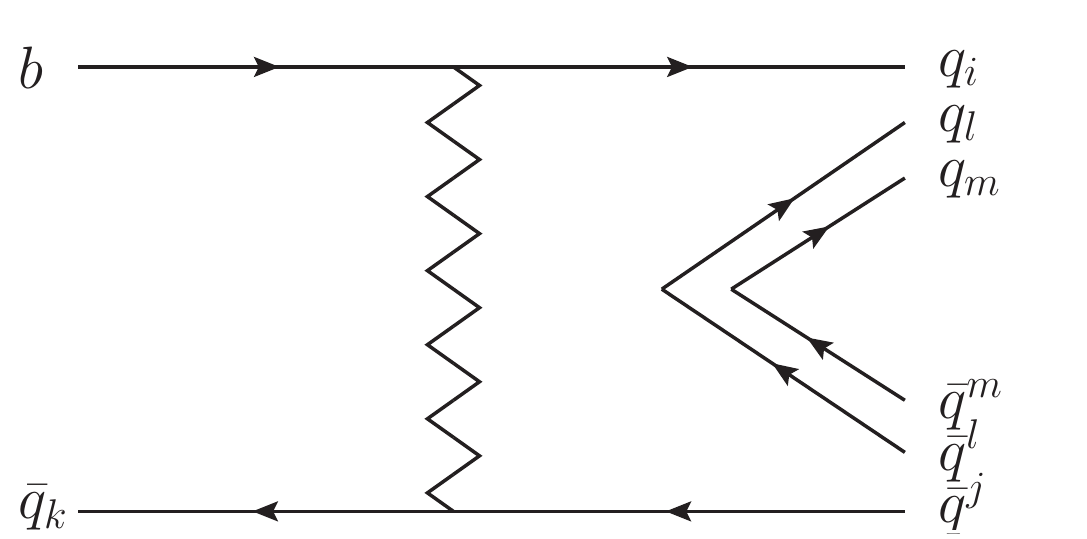}
}\subfigure[]{
  \includegraphics[width=0.4\textwidth]{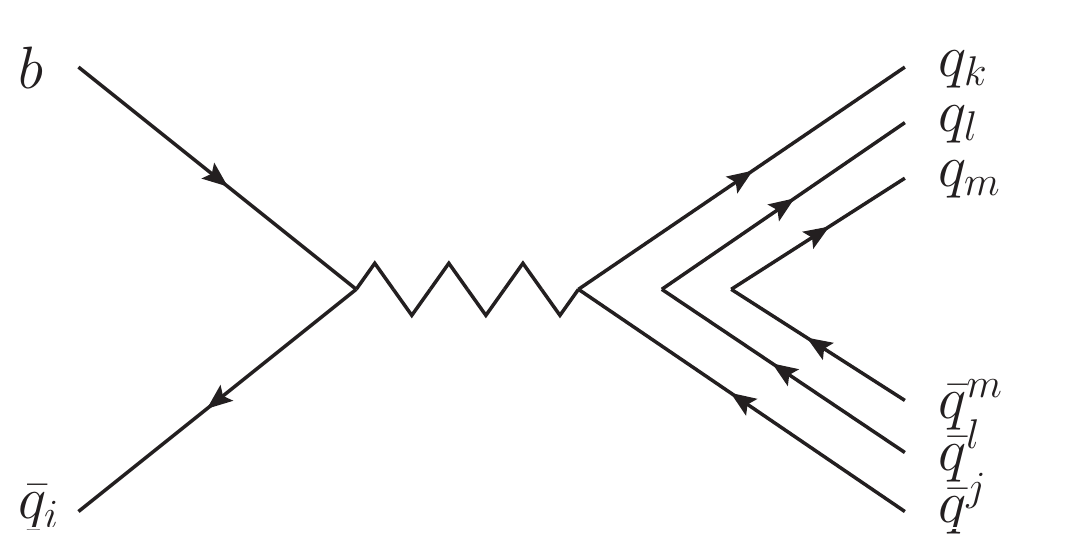}
}
\\\subfigure[]{
  \includegraphics[width=0.4\textwidth]{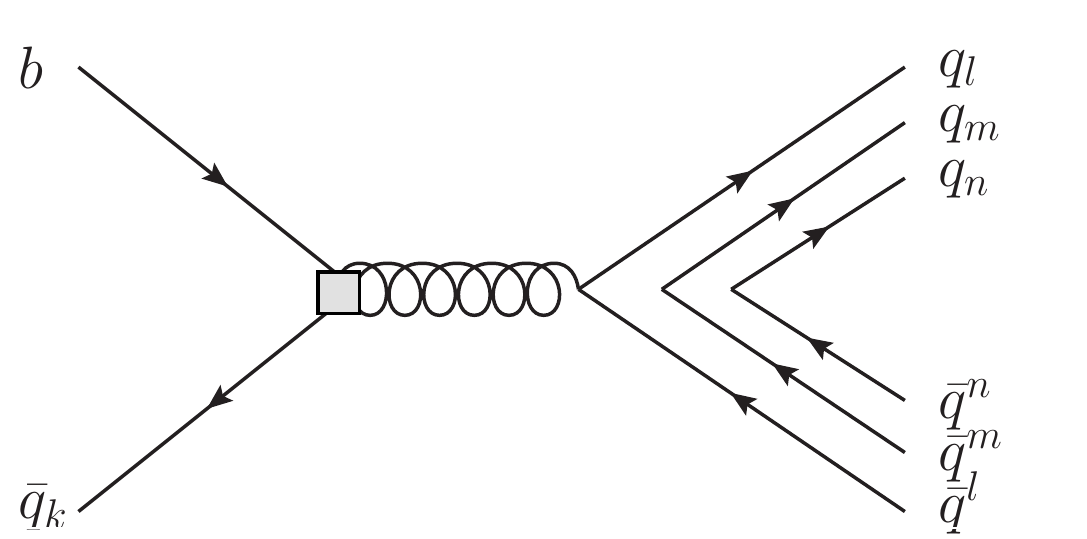}
}\subfigure[]{
  \includegraphics[width=0.4\textwidth]{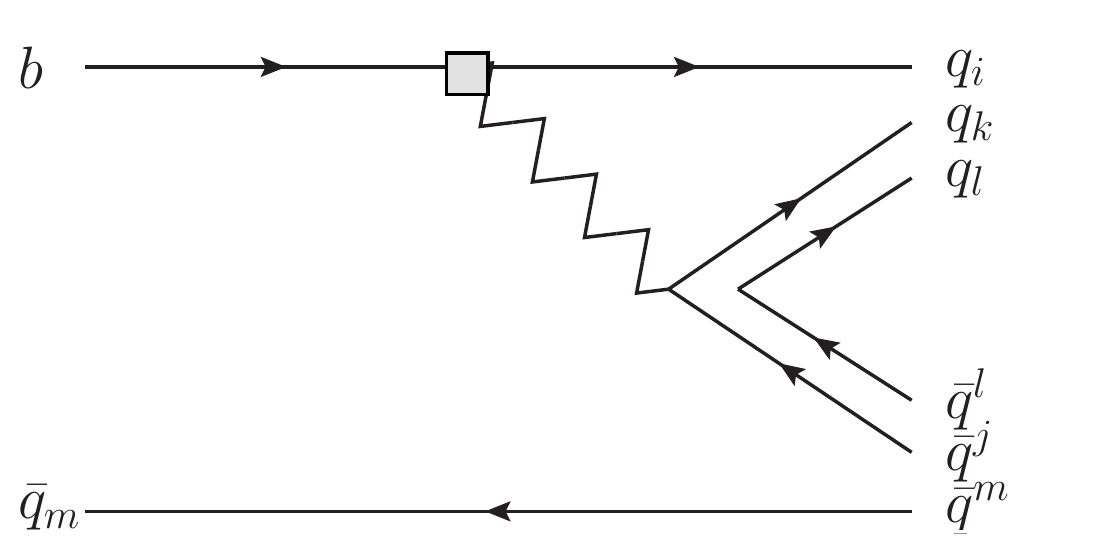}
}
\caption{Flavor flow or topological diagrams of
  (a) $T^{(\prime)}$ (tree), (b) $P^{(\prime)}$ (penguin), (c) $E^{(\prime)}$ ($W$-exchange),
  (d) $A^{(\prime)}$ (annihilation), (e) $PA^{(\prime)}$ (penguin annihilation) and (f) $P^{(\prime)}_{EW}$ (electroweak penguin)
  amplitudes in $\overline B$ to baryon pair decays for $\Delta S=0(-1)$ decays. 
} \label{fig:TA}
\end{figure}

There are more than 160 $\overline B\to\BB$, $\BD$, $\DB$, $\DD$ decay amplitudes with ${\mathcal B}$ and ${\cal D}$ the low lying octet and decuplet baryons, respectively.  Their decay amplitudes expressed in term of topological amplitudes can be found in~\cite{Chua:2013zga} and are collected in Appendix~\ref{appendix:amplitudes}, since they will be used extensively in this work. 
We show a few examples here, 
\be
A(B^-\to\Lambda\overline{p})
   &=&\frac{1}{\sq6}(T'_{1\BB}+2T'_{3\BB})
           +\frac{1}{\sqrt6}(10P'_{1\BB}-P'_{2\BB})
           -\frac{1}{3\sq6}(P'_{1EW\BB}-P'_{2EW\BB}
   \non\\
   &&-4P'_{3EW\BB}+4P'_{4EW\BB})
         +\frac{1}{\sq6}(10 A'_{1\BB}- A'_{2\BB}),   
  \non\\  
A(\bar B^0\to p\overline{p})
   &=&-T_{2\BB}+2T_{4\BB}
          +P_{2\BB}
           +\frac{2}{3}P_{2EW\BB}
          -5 E_{1\BB}+ E_{2\BB}
          -9PA_\BB,  
          \non\\
A(B^-\to\Xi^{-}\overline{\Xi^{0}})
   &=&-P_{2\BB}+\frac{1}{3}P_{2EW\BB}- A_{2\BB},
   \non\\               
A(\bar B^0_s\to p\overline{p})
   &=&-5E'_{1\BB}+ E'_{2\BB}-9 PA'_\BB,
   \label{eq:ampBB}
\en
\be
A(B^-\to p\overline{\Delta^{++}})
   &=&-\sq6 (T_{1\BD}-2T_{2\BD})+\sq6 P_\BD+2\sq{\frac{2}{3}}P_{1EW\BD}+\sq6 A_\BD,
   \non\\
A(\bar B^0\to\Xi^{-}\overline{\Sigma^{*-}})
   &=&\sq2P'_\BD
         -\frac{\sq2}{3}P'_{1EW\BD},
\en
\be
A(B^-\to\Delta^0\overline{p})
   &=&\sq2T_{1\DB}
          -\sq2 P_\DB
          +\frac{\sq2}{3}(3P_{1EW\DB}+P_{2EW\DB})
          -\sq2 A_\DB,
   \non\\
A(\bar B^0\to \Sigma^{*+}\overline{p})
   &=&\sq2 T'_{2\DB}
         +\sq2 P'_\DB
         +\frac{2\sq2}{3}P'_{2EW\DB},
\en
\be
A(\bar B^0\to\Delta^0\overline{\Delta^0})
   &=&2T_\DD+4 P_\DD+\frac{2}{3}P_{EW\DD}+2E_\DD+18PA_\DD,
   \non\\      
A(B^-\to \Sigma^{*+}\overline{\Delta^{++}})
   &=&2\sq3 T'_\DD+2\sq3 P'_\DD+\frac{4}{\sq3}P'_{EW\DD}+2{\sq3}A'_\DD,
\en
where $T^{(\prime)}$, $P^{(\prime)}$, $E^{(\prime)}$, $A^{(\prime)}$, $PA^{(\prime)}$ and $P^{(\prime)}_{EW}$ are
tree, penguin, $W$-exchange, annihilation, penguin annihilation and electroweak penguin amplitudes, respectively,
for $\Delta S=0(-1)$ decays (see Fig.~\ref{fig:TA}).
In most cases we needs more than one tree and one penguin amplitudes in
the baryonic decay amplitudes. 
Note that from the above amplitudes, $\overline B{}^0\to p\bar p$ decay is expected to be a tree dominated mode,
$B^-\to \Lambda\bar p$ decay a penguin dominated mode and $\overline B{}^0_s\to p\bar p$ decay a suppressed mode.

By considering the chirality nature of weak
interaction and asymptotic relations~\cite{Brodsky:1980sx} at the large $m_B$ limit ($m_q/m_B,m_{{\cal B},{\cal D}}/m_B\ll1$), the 
number of independent amplitudes is significantly reduced~\cite{Chua:2003it,Chua:2013zga}:
 \begin{eqnarray}
 T^{(\prime)}
 &=&T^{(\prime)}_{1\BB,2\BB,3\BB,4\BB}
 =T^{(\prime)}_{1\BD,2\BD}
 =T^{(\prime)}_{1\DB,2\DB}
 =T^{(\prime)}_{\DD},
 \non\\
 P^{(\prime)}
 &=&P^{(\prime)}_{1\BB,2\BB}
 =P^{(\prime)}_{{\cal B} \overline{\cal D}}
 =P^{(\prime)}_{{\cal D} \overline{\cal B}}
 =P^{(\prime)}_{{\cal D} \overline{\cal D}},
 \non\\
 P^{(\prime)}_{EW}
 &=&P^{(\prime)}_{1EW\BB,2EW\BB,3EW\BB,4EW\BB}
 =P^{(\prime)}_{1EW\BD,2EW\BD}
 =P^{(\prime)}_{1EW\DB,2EW\DB}
 =P^{(\prime)}_{EW\DD},
 \non\\
&&
\!\!\!\!\!\!\!\!\!\!\!\!\!\!\!\!\!\!\!\!\!\!
 E_{1\BB,2\BB,\BD,\DB,\DD},\,A_{1\BB,2\BB,\BD,\DB,\DD},
 \,PA_{\BB,\DD}\to0.
\label{eq:asymptoticrelations}
\end{eqnarray}
and there are only one tree, one penguin and one electroweak penguin amplitudes, which are estimated to be
\be
T^{(\prime)}&=&V_{ub} V^*_{ud(s)}\frac{G_f}{\sq2}(c_1+c_2) \chi \bar u'(1-\gamma_5)v,
\non\\
P^{(\prime)}&=&-V_{tb} V^*_{td(s)}\frac{G_f}{\sq2}[c_3+c_4+\kappa_1 c_5+\kappa_2 c_6]\chi \bar u'(1-\gamma_5)v,
\non\\
P^{(\prime)}_{EW}&=&-\frac{3}{2}V_{tb} V^*_{td(s)}\frac{G_f}{\sq2}[c_9+c_{10}+\kappa_1c_7+\kappa_2 c_8]\chi \bar u'(1-\gamma_5)v,
\label{eq:asymptotic1}
\en
where $c_i$ are the next-to-leading order Wilson coefficients, given by 
\be
c_1 = 1.081,\, 
c_2 = -0.190,\,
c_3 = 0.014,\,
c_4 = -0.036,\,
c_5 = 0.009,\, 
c_6 = -0.042,
\non\\ 
c_7 = -0.011\alpha_{EM},\, 
c_8 = 0.060\alpha_{EM},\, 
c_9 = -1.254\alpha_{EM},\,
c_{10} = 0.223\alpha_{EM},
\en
in the naive dimensional regularization scheme at scale $\mu=4.2$GeV~\cite{Beneke:2001ev}.
Note that the relative signs of $c_{1,3,9}$ and $c_{2,4,10}$ in Eq. (\ref{eq:asymptotic1}) are determined according to Ref.~\cite{Cheng:2014qxa}.
Since the Fierz transformation of $O_{5,6,7,8}$ are different from $O_{1,2,3,4}$, additional
coefficients $\kappa_i$ in front of $c_{5(7)}$ and $c_{6(8)}$ in Eq.~(\ref{eq:asymptotic1}) are assigned.
For simplicity we take $\kappa_1=\kappa_2=\kappa$.
The parameters $\chi$ and $\kappa$ will be extracted from $\overline B{}^0\to p\bar p$ and $B^-\to\Lambda\bar p$ data.
Note that $|\kappa|$ is expected to be of order ${\cal O}(1)$. 

In reality the $m_B$ mass is finite.
The decays amplitudes are not in the asymptotic forms. Some corrections are expected.
The correction on $T^{(\prime)}_i$, $P^{(\prime)}_i$ and $P^{(\prime)}_{EWi}$ are estimated to be of order $m_{\cal B}/m_B$ (the baryon and $B$ meson mass ratio). 
Hence, we have
\be
T^{(\prime)}_i=(1+r_{t,i}^{(\prime)}) T^{(\prime)},
\,
P^{(\prime)}_i=(1+r_{p,i}^{(\prime)}) P^{(\prime)}_,
\,
P^{(\prime)}_{EW i}=(1+r_{ewp,i}^{(\prime)}) P^{(\prime)}_{EW},
\label{eq:correction0}
\en
with
\be
 |r_{t,i}^{(\prime)}|, |r_{p,i}^{(\prime)}|, |r_{ewp,i}^{(\prime)}|\leq m_p/m_B,
\label{eq:correction1}
\en
parametrizing the correction to the asymptotic relations, Eq.~(\ref{eq:asymptoticrelations}).
Note that the phases of these $r^{(\prime)}$ can be any value.
For $P^{(\prime)}_i$, we replace the CKM factor $-V_{tb} V^*_{td(s)}$ by the sum of $V_{ub} V^*_{ud(s)}$ and $V_{cb} V^*_{cd(s)}$. The penguins with $V_{ub} V^*_{ud(s)}$ and $V_{cb} V^*_{cd(s)}$ are $u$-penguin ($P_i^{(\prime) u}$) and $c$-penguin ($P_i^{(\prime) c}$), respectively. The $r_{p,i}$ of the $u$-penguin and the $c$-penguin are independent.
Furthermore, although in the asymptotic limit the $\bar u' u$ and $-\bar u' \gamma_5 u$ terms have the same coefficients, see Eq.~(\ref{eq:asymptotic1}),
this will no longer be true in the finite $m_B$ case. 
In other words, the $r^{(\prime)}$ for $\bar u' u$ and $-\bar u' \gamma_5 u$ terms are independent. 
For subleading terms, such as exchange, penguin annihilation and annihilation amplitudes, we have
\be
E^{(\prime)}_i\equiv\eta_i \frac{f_B}{m_B} \frac{m_{\cal B}}{m_B} T^{(\prime)},\,
A^{(\prime)}_j\equiv\eta_j \frac{f_B}{m_B} \frac{m_{\cal B}}{m_B}T^{(\prime)},\,
PA^{(\prime)}_k\equiv\eta_k \frac{f_B}{m_B} \frac{m_{\cal B}}{m_B}P^{(\prime)},
\label{eq:correction2}
\en
where the ratio $f_B/m_B$ is from the usual estimation, the factor $m_{\cal B}/m_B$ is from the chirality structure.
Note that $|\eta_{i,j,k}|$ are estimated to be of order 1 and,
explicitly, we take
$
0\leq|\eta_{i,j,k}|\leq |\eta|=1,
$
with the bound $|\eta|$ set to 1. 

With the above amplitudes and formulas given in Appendices~\ref{appendix:amplitudes} and \ref{appendix:formulas}, we are ready to study the two-body baryonic decay rates. 
Note that for $\bar B\to \BD, \DB$ decays, we need to add a correction factor of $p_{cm}/(m_B/2)$ to the topological amplitudes shown in Eq.~(\ref{eq:asymptotic1}) with $p_{cm}$ the momentum of the final state baryons in the center of mass frame. The factor will further correct the amplitudes from their asymptotic forms.

\section{Numerical Results on Rates and Direct $CP$ Asymmetries}

\subsection{Tree and penguin amplitudes}

Using the recent data on the $\overline B{}^0\to p\bar p$ rate and the $B^-\to \Lambda\bar p$ rate, 
the unknown parameters $\chi$ and $\kappa$ in asymptotic amplitudes , Eq. (\ref{eq:asymptotic1}), are fitted to be
\be
\chi=(5.08^{+1.12}_{-1.02})\times 10^{-3}~{\rm GeV}^2,
\qquad
\kappa=1.92^{+0.39}_{-0.46}.
\label{eq:chikappa}
\en
Note that the value of $\kappa$ is indeed of order 1. The tree-penguin ratio is close to the na\"{i}ve estimation. 
The penguin-tree and tree-penguin ratios of the asymptotic amplitudes [see Eq. (\ref{eq:asymptoticrelations})] for $\Delta S=0$ and $-1$ transitions, respectively, can be extracted for the first time to be
\be
\left|\frac{P}{T}\right|=0.24\pm0.04,
\quad
\left|\frac{T'}{P'}\right|=0.21^{+0.05}_{-0.03}.
\label{eq:P/T}
\en
The above equation is one of the major results of this work.

\subsection{Numerical Results on Rates}

In the following we make use the $\overline B{}^0\to p\bar p$ and $B^-\to\Lambda\bar p$ data as inputs and predict rates of all other $\overline B\to\BB$, $\BD$, $\DB$, $\DD$ modes.  
Results are shown in Tables \ref{tab:BBDS=0}$\sim$\ref{tab:DDDS=-1}.
They are major results of this work.
A first glimpse on these tables reveals that all experimental upper bounds are satisfied.
This is a non-trivial check.
We will go through the discussions of the updated results on $\overline B\to\BB$, $\BD$, $\DB$, $\DD$ decay rates in below and make suggestions on future experimental searches.
We will give a summary of our suggestions at the end of this subsection.

\begin{table}[t!]
\caption{\label{tab:BBDS=0} Decay rates of $\Delta S=0$, $\overline B_q\to\BB$ modes.
The first uncertainty is from the uncertainties of the $\chi$ and $\kappa$ parameters, reflecting the uncertainties in the measurements of ${\cal B}(\overline B{}^0\to pp)$ and ${\cal B}(B^-\to \Lambda\bar p)$, 
the second uncertainty is obtained by varying the tree and penguin strong phase $\phi$,
the third uncertainty is from relaxing the asymptotic relations, by varying $r_{t,i}, r_{p,i}, r_{ewp,i}$ (see Eqs. (\ref{eq:correction0}) and 
(\ref{eq:correction1}))
and the last uncertainty is from sub-leading contributions, terms with $\eta_{i,j,k}$ (see Eq. (\ref{eq:correction2})). Occasionally the last uncertainty is shown to 
larger decimal place.
The latest experimental results are given in parentheses under the theoretical results.
The experimental $\overline B{}^0\to p\bar p$ rate is one of the inputs.}
\begin{ruledtabular}
\centering
{\begin{tabular}{llll}
Mode
          & ${\mathcal B}(10^{-8})$
          & Mode
          & ${\mathcal B}(10^{-8})$
          \\
\hline $B^-\to n\overline{p}$
          & $3.45^{+1.50}_{-1.23}{}^{+0.68}_{-0}{}^{+1.99}_{-1.50}\pm0.09$
          & $\overline B{}^0_s\to p\overline{\Sigma^{+}}$ 
          & $1.42^{+0.69}_{-0.51}{}^{+0.13}_{-0}{}^{+2.01}_{-1.12}\pm0$
           \\
$B^-\to \Sigma^{0}\overline{\Sigma^{+}}$ 
          & $3.29^{+1.56}_{-1.18}{}^{+0.52}_{-0}{}^{+2.09}_{-1.58}\pm0.11$
          & $\overline B{}^0_s\to n\overline{\Sigma^{0}}$
          & $0.75^{+0.36}_{-0.27}{}^{+0}_{-0.05}{}^{+0.32}_{-0.26}\pm0$
           \\ 
$B^-\to \Sigma^{-}\overline{\Sigma^{0}}$
          & $0.62^{+0.28}_{-0.22}\pm0{}^{+0.42}_{-0.31}{}^{+0.006}_{-0.004}$
          &$\overline B{}^0_s\to n\overline{\Lambda}$
          & $2.96^{+1.36}_{-1.06}{}^{+0.57}_{-0}{}^{+1.88}_{-1.41}\pm0$
           \\            
$B^-\to \Sigma^{-}\overline{\Lambda}$
          & $0.47^{+0.21}_{-0.17}\pm0{}^{+0.20}_{-0.17}{}^{+0.003}_{-0.002}$
          & $\overline B{}^0_s\to \Sigma^{0}\overline{\Xi^{0}}$ 
          & $10.85^{+5.23}_{-3.90}{}^{+1.22}_{-0}{}^{+5.23}_{-4.20}\pm0$
           \\ 
$B^-\to \Xi^{-}\overline{\Xi^{0}}$
          & $0.07^{+0.03}_{-0.03}\pm0\pm0.03{}^{+0.0005}_{-0.0003}$
          & $\overline B{}^0_s\to \Sigma^{-}\overline{\Xi^{-}}$ 
          & $1.76^{+0.78}_{-0.63}\pm0{}^{+0.76}_{-0.63}\pm0$
           \\ 
$B^-\to \Lambda\overline{\Sigma^+}$
          & $0.47^{+0.21}_{-0.17}\pm0{}^{+0.38}_{-0.17}{}^{+0.003}_{-0.002}$
          & $\overline B{}^0_s\to \Lambda\overline{\Xi^0}$
          & $0.11^{+0.05}_{-0.04}\pm0{}^{+0.63}_{-0.08}\pm0$
           \\ 
$\overline B{}^0\to p\overline{p}$ 
          & $1.47^{+0.71}_{-0.53}{}^{+0.14}_{-0}{}^{+2.07}_{-1.16}\pm0.12$
          & $\overline B{}^0\to \Sigma^{+}\overline{\Sigma^{+}}$
          & $0\pm0\pm0\pm0{}^{+0.003}_{-0}$
           \\
           &$1.47{}^{+0.62+0.35}_{-0.51-0.14}$~\cite{Aaij:2013fta}
           &
           &
           \\           
$\overline B{}^0\to n\overline{n}$
          & $6.66^{+3.15}_{-2.39}{}^{+1.05}_{-0}{}^{+4.25}_{-3.20}\pm0.07$
          & $\overline B{}^0\to \Sigma^{0}\overline{\Sigma^{0}}$ 
          & $1.52^{+0.72}_{-0.55}{}^{+0.24}_{-0}{}^{+0.97}_{-0.73}\pm0.07$
           \\
$\overline B{}^0\to \Xi^{0}\overline{\Xi^{0}}$
          & $0\pm0\pm0\pm0{}^{+0.0004}_{-0}$
          & $\overline B{}^0\to \Sigma^{-}\overline{\Sigma^{-}}$
          & $1.15^{+0.51}_{-0.41}\pm0{}^{+0.78}_{-0.58}\pm0.04$
           \\      
$\overline B{}^0\to \Xi^{-}\overline{\Xi^{-}}$
          & $0.07^{+0.03}_{-0.02}\pm0{}^{+0.03}_{-0.02}\pm0.01$
          & $\overline B{}^0\to \Sigma^{0}\overline{\Lambda}$ 
           & $4.10^{+1.98}_{-1.47}{}^{+0.39}_{-0}{}^{+1.90}_{-1.54}\pm0.05$
           \\
$\overline B{}^0\to \Lambda\overline{\Lambda}$
          & $0\pm0\pm0{}^{+0.23}_{-0}{}^{+0.0005}_{-0}$
          & $\overline B{}^0\to \Lambda\overline{\Sigma^{0}}$
           & $0.22^{+0.10}_{-0.08}\pm0{}^{+0.18}_{-0.08}\pm0.001$
           \\ 
           & ($<32$)~\cite{Tsai:2007pp}
           &
           &
          \\                                                                                                                                                                                                
\end{tabular}
}
\\
\end{ruledtabular}
\end{table}

\subsubsection{Rates of $\overline B_q\to \BB$ decays}

Predictions on $\Delta S=0$, $\overline B_q\to \BB$ decay rates are shown in Table \ref{tab:BBDS=0}. 
The first uncertainty is from the uncertainties of the $\chi$ and $\kappa$ parameters [see Eq. (\ref{eq:chikappa})], reflecting the uncertainties in the measurements of ${\cal B}(\overline B{}^0\to pp)$ and ${\cal B}(B^-\to \Lambda\bar p)$.
The second uncertainty is obtained by varying the tree and penguin strong phase $\phi$,
where we assign to the penguin amplitude:~\footnote{The strong phase of the tree amplitude is factored out and set to zero. Therefore, the strong phase $\phi$ is the relative strong phase of penguin and tree amplitudes.}
\be
P^{(\prime)}&=&-V_{tb} V^*_{td(s)}\frac{G_f}{\sq2}[c_3+c_4+\kappa c_5+\kappa c_6]\chi e^{i\phi} \bar u'(1-\gamma_5)v
\label{eq: penguin strong phase}
\en
and a similar expression for $P^{(\prime)}_{EW}$. We use a common strong phase for simplicity.
The third uncertainty is from relaxing the asymptotic relations by varying $r_{t,i}, r_{p,i}, r_{ewp,i}$ [see Eqs. (\ref{eq:correction0}) and 
(\ref{eq:correction1})]
and the last uncertainty is from sub-leading contributions, such as annihilation, penguin annihilation and exchange amplitudes 
[see Eq. (\ref{eq:correction2})].

From the tables we see that errors are reduced at least by a factor of two compared to the previous analysis in \cite{Chua:2013zga}.
These errors can provide useful informations:
(i) As noted before the first errors reflect the uncertainties in ${\cal B}(\overline B{}^0\to pp)$ and ${\cal B}(B^-\to \Lambda\bar p)$.
(ii) The second errors reflect the size of tree-penguin interferences. 
We can see from the table that in general the effects of tree-penguin interference on rates are not sizable. This is consistent with the tree-penguin ratios shown in Eq.~(\ref{eq:P/T}).
(iii)~The third errors, which correspond to corrections to amplitudes away from the asymptotic limit, are usually the largest ones. (iv)~Occasionally we show the last errors, which are from sub-leading contributions, to larger decimal place. 
These terms are with $\eta_{i,j,k}$ [see Eq. (\ref{eq:correction2})]. 
Note that for modes only have sub-leading contributions, the rates are proportional to $|\eta_{i,j,k}|^2$, while for modes having tree and/or penguin terms as well, these (sub-leading) contributions are roughly proportional to $\eta_{i,j,k}$.

For $\Delta S=0$, $\overline B_q\to\BB$ decays, there are three modes that can decay cascadely to all charged final states,
namely $\overline B{}^0\to p\bar p$, $\overline B{}^0\to\Xi^-\overline{\Xi^-}$ and $\Lambda\overline{\Lambda}$ decays.~\footnote{
To study the accessibilities of searching of the charmless two-body baryonic modes,
we note that 
(i)~$\Delta^{++,0}$, $\Lambda$, $\Xi^-$, $\Sigma^{*\pm}$, $\Xi^{*0}$ and $\Omega^-$ have non-suppressed decay modes of final states with all charged particles, 
(ii)~$\Delta^+$, $\Sigma^{+}$, $\Xi^0$, $\Sigma^{*0}$ and $\Xi^{*-}$ can be detected by detecting a $\pi^0$, 
(iii)~$\Sigma^{0}$ can be detected by detecting $\gamma$, 
(iv)~one needs $n$ in detecting $\Delta^-$ and $\Sigma^-$~\cite{Chua:2013zga,PDG}. 
Note that although some particles, such as $\Xi^-$ and $\Xi^{*0}$, can decay to final states with all charged particles, they may suffer from low reconstruction efficiencies. } 
We see from Table~\ref{tab:BBDS=0} that among these modes the $\overline B{}^0\to p\bar p$ decay has the highest rate and highest detectability. The rates of the other two modes are, in fact, one or two orders of magnitude smaller the $p\bar p$ rate.
In particular, the predicted $\overline B{}^0\to\Lambda\overline\Lambda$ rate is much smaller than the present experimental upper limit~\cite{Tsai:2007pp}. 
More modes can be searched for with $\pi^0$, $\gamma$, $\pi^0\pi^0$ and $\pi^0\gamma$ in the future experiments, such as in Belle II.
For example, with one $\pi^0$ or $\gamma$, one can search for 
$\overline B^0\to \Sigma^0\overline \Lambda$ and 
$\overline B{}^0_s\to p\overline{\Sigma^+}$ decays,
with $\pi^0\gamma$ 
one can also search for 
$\overline B{}^0_s\to\Sigma^0\overline{\Xi^0}$ and 
$B^-\to\Sigma^0\overline{\Sigma^+}$ decays, while with $\gamma\gamma$ one can search for
$\overline B{}^0\to\Sigma^0\overline{\Sigma^0}$ decays.
In fact, the $\overline B{}^0_s\to\Sigma^0\overline{\Xi^0}$ decay rate is of the order of $10^{-7}$, which is the highest rate in the table, but the mode is reconstructed through the cascade decay $\overline B{}^0_s\to\Sigma^0\overline{\Xi^0}\to\Lambda\gamma\,\bar\Lambda\pi^0$,
which requires both $\gamma$ and $\pi^0$ for the detection. 
It is understandable that why $\overline B{}^0\to p\bar p$ decay is the first mode with experimental evidence. 

From Eqs. (\ref{eq: BBBm, DS=0}), (\ref{eq: BBB0, DS=0}) and (\ref{eq: BBBs, DS=0}), we see that there are several modes without any tree ($T_{i\BB}$) contribution.
These include 
$B^-\to \Sigma^{-}\overline{\Sigma^{0}}$, 
$B^-\to \Sigma^{-}\overline{\Lambda}$,
$B^-\to \Xi^{-}\overline{\Xi^{0}}$,
$\overline B{}^0\to \Xi^{-}\overline{\Xi^{-}}$,
$\overline B{}^0\to \Sigma^{-}\overline{\Sigma^{-}}$,
$\overline B{}^0\to \Xi^{0}\overline{\Xi^{0}}$,
$\overline B{}^0\to \Sigma^{+}\overline{\Sigma^{+}}$
and 
$\overline B{}^0_s\to \Sigma^{-}\overline{\Xi^{-}}$ decays.
As shown in Table~\ref{tab:BBDS=0} the second uncertainties of the rates of these modes are vanishing.
Note that 
$\overline B{}^0\to \Xi^{-}\overline{\Xi^{-}}$, 
$\overline B{}^0\to \Sigma^{-}\overline{\Sigma^{-}}$
and $\overline B{}^0_s\to \Sigma^{-}\overline{\Xi^{-}}$ decays are pure penguin modes, which only have $P_{i\BB}$, $P_{iEW\BB}$ and $PA_{\BB}$ terms, 
while $\overline B{}^0\to \Xi^{0}\overline{\Xi^{0}}$ and
$\overline B{}^0\to \Sigma^{+}\overline{\Sigma^{+}}$ decays only have subleading contributions, namely $E_{i\BB}$ and $PA_\BB$. 
Note that although 
$B^-\to \Lambda\overline{\Sigma^+}$,
 $\overline B{}^0_s\to \Lambda\overline{\Xi^0}$,
$\overline B{}^0\to \Lambda\overline{\Lambda}$ and
$\overline B{}^0\to \Lambda\overline{\Sigma^{0}}$ decays
have tree amplitudes $T_{i\BB}$, these tree amplitudes cancel out in the asymptotic limit
[see Eqs (\ref{eq: BBBm, DS=0}), (\ref{eq: BBB0, DS=0}), (\ref{eq: BBBs, DS=0}) and (\ref{eq:asymptoticrelations})].
In particular, the tree, penguin, electroweak penguin and exchange amplitudes
in the $\overline B{}^0\to \Lambda\overline{\Lambda}$ amplitude
all cancel out in the asymptotic limit.
This mode is therefore sensitive to the correction to the asymptotic relations, Eqs. (\ref{eq:asymptoticrelations}),
(\ref{eq:correction0}) and (\ref{eq:correction1}). 
          
\begin{table}[t!]
\caption{\label{tab:BBDS=-1} Same as Table~\ref{tab:BBDS=0}, but with $\Delta S=-1$, $\overline B_q\to\BB$ modes.
The latest experimental result is given in the parenthesis under the theoretical results.
The experimental $B^-\to\Lambda\bar p$ rate is one of the inputs.}
\begin{ruledtabular}
\begin{tabular}{llll}
Mode
          & ${\mathcal B}(10^{-8})$
          & Mode
          & ${\mathcal B}(10^{-8})$
          \\
\hline $B^-\to \Sigma^{0}\overline{p}$
          & $0.76^{+0.35}_{-0.27}{}^{+0}_{-0.19}{}^{+0.50}_{-0.37}\pm0.001$
          & $\overline B{}^0\to \Sigma^{+}\overline{p}$ 
          & $1.83^{+0.76}_{-0.65}{}^{+0.47}_{-0}{}^{+0.82}_{-0.61}\pm0$
           \\
$B^-\to \Sigma^{-}\overline{n}$
          & $1.67^{+0.74}_{-0.59}\pm0{}^{+0.71}_{-0.58}{}^{+0.002}_{-0.002}$
          & $\overline B{}^0\to \Sigma^{0}\overline{n}$
          & $1.12^{+0.47}_{-0.40}{}^{+0.52}_{-0}{}^{+0.47}_{-0.36}\pm0$
           \\ 
$B^-\to \Xi^{0}\overline{\Sigma^{+}}$ 
          & $39.98^{+17.53}_{-14.23}{}^{+2.14}_{-0}{}^{+17.06}_{-14.02}\pm0.03$
          &$\overline B{}^0\to \Xi^{0}\overline{\Sigma^{0}}$ 
          & $18.50^{+8.11}_{-6.58}{}^{+0.99}_{-0}{}^{+7.89}_{-6.48}\pm0$
           \\            
$B^-\to\Xi^{-}\overline{\Sigma^{0}}$ 
          & $19.40^{+8.59}_{-6.90}\pm0{}^{+8.38}_{-6.88}\pm0.02$
          & $\overline B{}^0\to \Xi^{0}\overline{\Lambda}$ 
          & $2.59^{+1.07}_{-0.92}{}^{+0.67}_{-0}{}^{+2.80}_{-1.74}\pm0$
           \\ 
$B^-\to \Xi^{-}\overline{\Lambda}$ 
          & $2.36^{+1.05}_{-0.84}\pm0{}^{+2.65}_{-1.67}\pm0.005$
          & $\overline B{}^0\to  \Xi^{-}\overline{\Sigma^{-}}$
          & $35.88^{+15.89}_{-12.77}\pm0{}^{+15.50}_{-12.73}\pm0$
           \\ 
$B^-\to \Lambda\overline{p}$ 
          & $24.00^{+10.44}_{-8.54}{}^{+2.13}_{-0}{}^{+12.48}_{-9.85}\pm0.02$
          & $\overline B{}^0\to \Lambda\overline{n}$
          & $23.48^{+10.00}_{-8.36}{}^{+4.12}_{-0}{}^{+12.13}_{-9.50}\pm0$
           \\ 
           & $24^{+10}_{-8}\pm 3$~\cite{Aaij:2016xfa}
           &
           &
          \\                    
$\overline B{}^0_s\to p\overline{p}$
          & $0\pm0\pm0{}^{+0.007}_{-0}$
          & $\overline B{}^0_s\to \Sigma^{+}\overline{\Sigma^{+}}$ 
          & $1.76^{+0.73}_{-0.63}{}^{+0.45}_{-0}{}^{+0.79}_{-0.59}{}^{+0.21}_{-0.20}$
           \\
           & ($2.84{}^{+2.03+0.85}_{-1.68-0.18}$)~\cite{Aaij:2013fta}
           &
           &
          \\
$\overline B{}^0_s\to n\overline{n}$
          & $0\pm0\pm0{}^{+0.007}_{-0}$
          & $\overline B{}^0_s\to \Sigma^{0}\overline{\Sigma^{0}}$ 
          & $1.61^{+0.69}_{-0.57}{}^{+0.21}_{-0}{}^{+0.69}_{-0.55}{}^{+0.20}_{-0.19}$
           \\
$\overline B{}^0_s\to \Xi^{0}\overline{\Xi^{0}}$ 
          & $24.46^{+10.53}_{-8.71}{}^{+3.24}_{-0}{}^{+16.28}_{-12.07}{}^{+0.75}_{-0.74}$
          & $\overline B{}^0_s\to \Sigma^{-}\overline{\Sigma^{-}}$
          & $1.49^{+0.66}_{-0.53}\pm0{}^{+0.63}_{-0.52}{}^{+0.19}_{-0.18}$
           \\ 
$\overline B{}^0_s\to \Xi^{-}\overline{\Xi^{-}}$ 
          & $22.63^{+10.02}_{-8.05}\pm0{}^{+15.27}_{-11.36}{}^{+0.72}_{-0.71}$
          & $\overline B{}^0_s\to \Sigma^{0}\overline{\Lambda}$
          & $0.05^{+0.03}_{-0.02}{}^{+0.04}_{-0}{}^{+0.06}_{-0.04}\pm0.001$
           \\    
$\overline B{}^0_s\to \Lambda\overline{\Lambda}$ 
          & $14.90^{+6.42}_{-5.31}{}^{+1.97}_{-0}{}^{+7.58}_{-5.99}{}^{+0.61}_{-0.60}$
          & $\overline B{}^0_s\to \Lambda\overline{\Sigma^{0}}$
          & $0.05^{+0.03}_{-0.02}{}^{+0.04}_{-0}\pm0.02\pm0.001$
           \\                                                                                                                                                                       
\end{tabular}
\end{ruledtabular}
\end{table}

Predictions on $\Delta S=-1$, $\overline B_q\to \BB$ decay rates are shown in Table \ref{tab:BBDS=-1}. 
There are 9 modes that have rates of order $10^{-7}$, namely
$B^-\to\Xi^0\overline{\Sigma^+}$, 
$\Xi^-\overline{\Sigma^0}$, 
$\Lambda\bar p$, 
$\overline B{}^0\to \Xi^0\overline{\Sigma^0}$, 
$\Xi^-\overline{\Sigma^-}$, 
$\Lambda \bar n$,
$\overline B{}^0_s\to\Xi^0\overline{\Xi^0}$,
$\Xi^-\overline{\Xi^-}$
and $\Lambda\overline\Lambda$ decays.
On the other hand, there are 5 modes that can cascadely decay to all charged final states, namely
$B^-\to \Xi^-\overline \Lambda,\Lambda\bar p$, 
$\overline B{}^0_s\to p\bar p$, 
$\Xi^-\overline{\Xi^-}$
and
$\Lambda\overline\Lambda$. 
Comparing these two sets, we see that 
$B^-\to \Lambda\bar p$, 
$\overline B{}^0_s\to \Lambda\overline{\Lambda}$ 
and
$\Xi^-\overline{\Xi^-}$ decays are the only three modes that have rates of order $10^{-7}$ and can cascadely decay to all charge final states.
In fact, the $B^-\to \Lambda\bar p$ has the highest rate among these three modes and has the best detectability as the others need both $\Lambda$ and $\overline\Lambda$ for detections ($\overline B{}^0_s\to \Lambda\overline{\Lambda}$, 
$\overline B{}^0_s\to\Xi^-\overline{\Xi^-}\to \Lambda\pi^-\,\overline\Lambda \pi^+$).
It is understandable that why $B^-\to \Lambda\bar p$ is the first penguin mode with experimental evidence.
Nevertheless, it is interesting to search for $\overline B{}^0_s\to \Lambda\overline{\Lambda}$ and
$\Xi^-\overline{\Xi^-}$ decays as well.
One can also search for other modes.
For example, the 
$B^-\to \Xi^{-}\overline{\Lambda}$ 
mode can also cascadely decay to all charged final state and its rate is of order $10^{-8}$,
but this mode suffers from the low reconstruction efficiency of  $\Xi^-$.
With $\gamma$ one can search for  
$B^-\to\Xi^{-}\overline{\Sigma^{0}}$ 
decay, which has rate of order $10^{-7}$. 
With $\pi^0$ one can search for 
$\overline B{}^0\to \Xi^{0}\overline{\Lambda}$ 
and 
$\overline B{}^0\to \Sigma^{+}\overline{p}$ decays at $10^{-8}$ level.
With $\pi^0\pi^0$ one can search for 
$B^-\to \Xi^{0}\overline{\Sigma^{+}}$,  
  $\overline B{}^0_s\to \Xi^{0}\overline{\Xi^{0}}$, 
and
$\overline B{}^0_s\to \Sigma^{+}\overline{\Sigma^{+}}$ 
decays, where the first two modes have rate of order $10^{-7}$,
while with $\pi^0\gamma$ one can search for  $\overline B{}^0\to \Xi^{0}\overline{\Sigma^{0}}$, 
which has rate of order $10^{-7}$,
and finally with $\gamma\gamma$ one can search for 
$\overline B{}^0_s\to \Sigma^{0}\overline{\Sigma^{0}}$ 
decay at $10^{-8}$ level.
Note that with all charged final states, $\gamma$, $\pi^0\pi^0$ and $\pi^0\gamma$, 
most modes having rates of the order of $10^{-7}$ can be searched for in the future.

From Eqs. (\ref{eq: BBBm, DS=-1}), (\ref{eq: BBB0, DS=-1}) and (\ref{eq: BBBs, DS=-1}), 
we see that 
$B^-\to \Sigma^{-}\overline{n}$,
$B^-\to\Xi^{-}\overline{\Sigma^{0}}$, 
$B^-\to \Xi^{-}\overline{\Lambda}$, 
$\overline B{}^0\to  \Xi^{-}\overline{\Sigma^{-}}$,
$\overline B{}^0_s\to \Sigma^{-}\overline{\Sigma^{-}}$,
$\overline B{}^0_s\to \Xi^{-}\overline{\Xi^{-}}$, 
$\overline B{}^0_s\to p\overline{p}$ and
$\overline B{}^0_s\to n\overline{n}$ decays
do not have any tree ($T'_{i\BB}$) contribution.
As shown in Table~\ref{tab:BBDS=-1} the rates of these modes have vanishing second uncertainties.
In particular, we note that 
$\overline B{}^0\to  \Xi^{-}\overline{\Sigma^{-}}$,
$\overline B{}^0_s\to \Sigma^{-}\overline{\Sigma^{-}}$
and
$\overline B{}^0_s\to \Xi^{-}\overline{\Xi^{-}}$ are pure penguin modes,
which only have $P'_{i\BB}$, $P'_{iEW\BB}$ and $PA'_{\BB}$ terms,
while 
$\overline B{}^0_s\to p\overline{p}$ and
$\overline B{}^0_s\to n\overline{n}$ decays 
only have subleading contributions, the $E'_{i\BB}$ and $PA'_{\BB}$ terms.
We will return to these modes later.

The predicted $\overline B{}^0_s\to p\bar p$ rate is several order smaller than the present experimental result, which, however, has large uncertainty.
As noted this mode only receives contributions from sub-leading terms [see Eq. (\ref{eq:ampBB})]. 
One may enhance the subleading contributions, but will soon run into contradictions. For example, after enhancing the subleading contributions in the $\overline B{}^0\to p\bar p$ mode, the so-called ``subleading contributions" will oversize the leading tree contribution.
Note that, some enhancement on subleading contributions is possible in the presence of final state rescattering (see for example, \cite{FSI, FSI1}),
but it is unlikely that the enhancement to be so significant.
Note that in Ref.~\cite{Hsiao:2014zza}, when partial conservation of axial-vector current is relaxed, the calculated $B_s\to p\bar p$ rate can be close to data, but the predicted $B^-\to\Lambda\bar p$ rate is of the order $10^{-8}$ and is in tension with the data~\cite{Aaij:2016xfa}. 
We need a more precise measurement on the $B_s\to p\bar p$ rate to settle the issue.

\begin{table}[t!]
\caption{\label{tab:BDDS=0} Same as Table~\ref{tab:BBDS=0}, but with $\Delta S=0$, $\overline B_q\to\BD$ modes.
The latest experimental result is given in the parenthesis. 
}
\begin{ruledtabular}
\centering
\begin{tabular}{llll}
Mode
          & ${\mathcal B}(10^{-8})$
          & Mode
          & ${\mathcal B}(10^{-8})$
          \\
\hline $B^-\to p\overline{\Delta^{++}}$
          & $6.21^{+3.01}_{-2.23}{}^{+0.58}_{-0}{}^{+8.77}_{-4.89}\pm0.08$
          & $\overline B{}^0_s\to p\overline{\Sigma^{*+}}$
          & $1.84^{+0.89}_{-0.66}{}^{+0.17}_{-0}{}^{+2.60}_{-1.45}\pm0$
           \\
           & ($<14$)~\cite{Wei:2007fg}
           &
           &
          \\
$B^-\to n\overline{\Delta^+}$
          & $2.18^{+1.05}_{-0.79}{}^{+0}_{-0.15}{}^{+0.93}_{-0.75}\pm0.03$
          & $\overline B{}^0_s\to n\overline{\Sigma^{*0}}$
          & $0.97^{+0.47}_{-0.35}{}^{+0}_{-0.07}{}^{+0.41}_{-0.33}\pm0$
           \\ 
$B^-\to\Sigma^0\overline{\Sigma^{*+}}$
          & $3.14^{+1.53}_{-1.13}{}^{+0.17}_{-0}{}^{+1.34}_{-1.10}\pm0.02$
          &$\overline B{}^0_s\to \Sigma^{0}\overline{\Xi^{*0}}$
          & $2.77^{+1.35}_{-1.00}{}^{+0.15}_{-0}{}^{+1.19}_{-0.97}\pm0$
           \\            
$B^-\to\Sigma^-\overline{\Sigma^{*0}}$
          & $0.04\pm0.02\pm0\pm0.02^{+0.0003}_{-0.0002}$
          & $\overline B{}^0_s\to \Sigma^{-}\overline{\Xi^{*-}}$
          & $0.08\pm0.03\pm0\pm0.03\pm0$
           \\ 
$B^-\to\Xi^{-}\overline{\Xi^{*0}}$
          & $0.07^{+0.03}_{-0.02}\pm0{}^{+0.03}_{-0.02}{}^{+0.0004}_{-0.0003}$
          & $\overline B{}^0_s\to \Xi^{-}\overline{\Omega^-}$
          & $0.18^{+0.08}_{-0.07}\pm0{}^{+0.08}_{-0.06}\pm0$
           \\ 
$B^-\to\Lambda\overline{\Sigma^{*+}}$
          & $0.14^{+0.06}_{-0.05}\pm0{}^{+0.21}_{-0.05}\pm{+0.001}$
          & $\overline B{}^0_s\to \Lambda\overline{\Xi^{*0}}$
          & $0.12^{+0.05}_{-0.04}\pm0{}^{+0.19}_{-0.05}\pm0$
           \\ 
$\overline B{}^0\to p\overline{\Delta^+}$
          & $1.92^{+0.93}_{-0.69}{}^{+0.18}_{-0}{}^{+2.71}_{-1.51}\pm0.02$
          & $\overline B{}^0\to \Sigma^{+}\overline{\Sigma^{*+}}$
          & $0\pm0\pm0{}^{+0.0001}_{-0}$
           \\
$\overline B{}^0\to n\overline{\Delta^0}$
          & $2.01^{+0.97}_{-0.73}{}^{+0}_{-0.14}{}^{+0.86}_{-0.69}\pm0.03$
          & $\overline B{}^0\to \Sigma^{0}\overline{\Sigma^{*0}}$
          & $1.45^{+0.71}_{-0.52}{}^{+0.08}_{-0}{}^{+0.62}_{-0.51}\pm0.01$
           \\
$\overline B{}^0\to \Xi^{0}\overline{\Xi^{*0}}$
          & $0\pm0\pm0{}^{+0.0001}_{-0}$
          & $\overline B{}^0\to \Sigma^{-}\overline{\Sigma^{*-}}$
          & $0.08^{+0.04}_{-0.03}\pm0{}\pm0.03\pm0$
           \\      
$\overline B{}^0\to \Xi^{-}\overline{\Xi^{*-}}$
          & $0.06^{+0.03}_{-0.02}\pm0{}^{+0.03}_{-0.02}\pm0$
          & $\overline B{}^0\to \Lambda\overline{\Sigma^{*0}}$
          & $0.06^{+0.03}_{-0.02}\pm0{}^{+0.10}_{-0.02}{}^{+0.0004}_{-0.0003}$
           \\                                                                                                                                                               
\end{tabular}
\end{ruledtabular}
\end{table}

\subsubsection{Rates of $\overline B_q\to \BD$ decays}

Predictions on rates of  $\Delta S=0$, $\overline B_q\to\BD$ decays are shown in Table~\ref{tab:BDDS=0}.
Modes that can cascadely decay to all charged final states and with unsuppressed rates are
$B^-\to p\overline{\Delta^{++}}$ and $\overline B{}^0_s\to p\overline{\Sigma^{*+}}$ decays. 
In particular, the predicted rate of $B^-\to p\overline{\Delta^{++}}$ decay is the highest one in the table and is just roughly a factor of 2 smaller than the present experimental upper bound \cite{Wei:2007fg}, which, however, has not been updated for quite a while.
This mode could be just around the corner. 
On the other hand, four other modes that can cascadely decay to all charged final states, namely 
$B^-\to \Xi^-\overline{\Xi^{*0}}$, 
$\Lambda\overline{\Sigma^{*+}}$,
$\overline B{}^0_s\to \Xi^-\overline{\Omega^-}$
and
$\Lambda\overline{\Xi^{*0}}$ decays,
are one or two order of magnitude more suppressed. 
Note that
with $\gamma$ one can search for 
$B^-\to\Sigma^0\overline{\Sigma^{*+}}$ 
and
$\overline B{}^0_s\to \Sigma^{0}\overline{\Xi^{*0}}$ 
 decays, 
with $\pi^0$ one can search for   
$\overline B{}^0\to p\overline{\Delta^+}$ 
decay,
while with $\gamma\pi^0$ one can search for
$\overline B{}^0\to \Sigma^{0}\overline{\Sigma^{*0}}$ 
decay in the future.
All of these modes have rates of orders $10^{-8}$.

From Eqs. (\ref{eq: BDBm, DS=0}), (\ref{eq: BDB0, DS=0}) and (\ref{eq: BDBs, DS=0}),
we see that       
$B^-\to\Sigma^-\overline{\Sigma^{*0}}$,
$B^-\to\Xi^{-}\overline{\Xi^{*0}}$,
$\overline B{}^0_s\to \Sigma^{-}\overline{\Xi^{*-}}$,
$\overline B{}^0\to \Xi^{-}\overline{\Xi^{*-}}$,
$\overline B{}^0\to \Sigma^{-}\overline{\Sigma^{*-}}$,
$\overline B{}^0_s\to \Xi^{-}\overline{\Omega^-}$,
$\overline B{}^0\to \Sigma^{+}\overline{\Sigma^{*+}}$ and
 $\overline B{}^0\to \Xi^{0}\overline{\Xi^{*0}}$ decay amplitudes,
do not have any tree ($T_{i\BD}$) contribution.
As shown in Table~\ref{tab:BDDS=0}, the rates of these modes have vanishing second uncertainties.
We note that
$\overline B{}^0\to \Xi^{-}\overline{\Xi^{*-}}$ and
$\overline B{}^0\to \Sigma^{-}\overline{\Sigma^{*-}}$
decays are pure penguin modes with only
$P_{\BD}$ and $P_{iEW\BD}$ contributions,
while
$\overline B{}^0\to \Sigma^{+}\overline{\Sigma^{*+}}$ and
$\overline B{}^0\to \Xi^{0}\overline{\Xi^{*0}}$ decays
are pure exchange modes with only $E_{\BD}$ contributions. 
Although 
$B^-\to\Lambda\overline{\Sigma^{*+}}$,
$\overline B{}^0_s\to \Lambda\overline{\Xi^{*0}}$ and
$\overline B{}^0\to \Lambda\overline{\Sigma^{*0}}$ decay amplitudes
have tree amplitudes, these tree amplitudes cancel out in the asymptotic limit.         

There are relations among rates. 
Using formulas in Appendices~\ref{appendix:amplitudes} and \ref{appendix:formulas},
we have~\cite{Chua:2013zga} 
\be
\Br(B^-\to\Xi^{-}\overline{\Xi^{*0}})
&=&2 \Br(B^-\to\Sigma^-\overline{\Sigma^{*0}})
\left(\frac{p_{cm}(B^-\to\Xi^{-}\overline{\Xi^{*0}})}{p_{cm}(B^-\to\Sigma^-\overline{\Sigma^{*0}})}\right)^3
\non\\
   &\simeq&2 \Br(B^-\to\Sigma^-\overline{\Sigma^{*0}}),
   \non\\
   3\tau_{B_d} \Br(\bar B^0\to\Sigma^{-}\overline{\Sigma^{*-}})
   &\simeq&3\tau_{B_d}\Br(\bar B^0\to\Xi^{-}\overline{\Xi^{*-}})
   \simeq
3\tau_{B_s} \Br(\bar B^0_s\to\Sigma^{-}\overline{\Xi^{*-}})
\non\\
   &\simeq&\tau_{B_s}\Br(\bar B^0_s\to\Xi^{-}\overline{\Omega^-}),
\non\\   
\Br(\bar B^0\to\Sigma^{+}\overline{\Sigma^{*+}})
    &\simeq&\Br(\bar B^0\to\Xi^{0}\overline{\Xi^{*0}}).       
 \en 
One can check from Table~\ref{tab:BDDS=0} that the rates of these modes roughly satisfy the above relations and the agreement will be improved when the SU(3) breaking effects are taken into account.
Note that these relations do not relay on the asymptotic relations.

Predictions on $\Delta S=-1$, $\overline B_q\to\BD$ decay rates are shown in Table \ref{tab:BDDS=-1}. 
Both $\overline B{}^0\to \Xi^-\overline{\Sigma^{*-}}$ and $\Lambda\overline{\Delta^0}$ modes can cascadely decay to all charge final states,
where only the $\overline B{}^0\to \Xi^-\overline{\Sigma^{*-}}$ rate can reach $10^{-8}$ level. 
In principle, this mode can be detected through $\overline B{}^0\to \Xi^-\overline{\Sigma^{*-}}\to \Lambda\pi^-\,\overline\Lambda\pi^+$ decay, but may be restricted by the low reconstruction efficiency of $\Xi^-$.
Note that with $\pi^0$ one can search for
$\overline B{}^0_s\to \Sigma^{+}\overline{\Sigma^{*+}}$, 
               $B^-\to \Xi^{0}\overline{\Sigma^{*+}}$, 
               $B^-\to \Sigma^+\overline{\Delta^{++}}$, 
              $\overline B{}^0_s\to \Xi^{-}\overline{\Xi^{*-}}$, 
              $\overline B{}^0_s\to \Xi^{0}\overline{\Xi^{*0}}$ 
              and 
              $B^-\to \Xi^{-}\overline{\Sigma^{*0}}$ 
decays, which have rates of orders $10^{-8}$,
while with $\gamma$ one can search for
             $\overline B{}^0\to \Sigma^{0}\overline{\Delta^0}$ 
decay in the future.
In particular, the $B^-\to \Sigma^+\overline{\Delta^{++}}$ decay rate is the highest one in the table. 
With $\gamma\pi^0$, $\pi^0\pi^0$ or $\gamma\gamma$, one can search for
$B^-\to \Sigma^0\overline{\Delta^+}$, 
              $\overline B{}^0\to \Sigma^{+}\overline{\Delta^+}$, 
             $\overline B{}^0_s\to \Sigma^{0}\overline{\Sigma^{*0}}$ 
and
             $\overline B{}^0\to \Xi^{0}\overline{\Sigma^{*0}}$ 
             decays, which have rates of orders (or close to) $10^{-8}$, in the future.
Note that the predicted $\overline B{}^0\to \Lambda\overline{\Delta^0}$ and $B^-\to \Lambda\overline{\Delta^+}$ rates
are both roughly two orders of magnitude below the present experimental bounds~\cite{Wang:2007as}.

From Eqs. (\ref{eq: BDBm, DS=-1}), (\ref{eq: BDB0, DS=-1}) and (\ref{eq: BDBs, DS=-1}), we see that  
$B^-\to \Sigma^-\overline{\Delta^0}$,
$B^-\to \Xi^{-}\overline{\Sigma^{*0}}$,
$\overline B{}^0\to \Sigma^{-}\overline{\Delta^-}$,
$\overline B{}^0\to \Xi^{-}\overline{\Sigma^{*-}}$, 
$\overline B{}^0_s\to \Sigma^{-}\overline{\Sigma^{*-}}$,
$\overline B{}^0_s\to \Xi^{-}\overline{\Xi^{*-}}$, 
$\overline B{}^0_s\to p\overline{\Delta^+}$ and
$\overline B{}^0_s\to n\overline{\Delta^0}$ decays,
do not have any tree ($T'_{i\BD}$) contribution.
As shown in Table~\ref{tab:BDDS=-1},  the rates of these modes have vanishing second uncertainties.
Note that
$\overline B{}^0\to \Sigma^{-}\overline{\Delta^-}$,
$\overline B{}^0\to \Xi^{-}\overline{\Sigma^{*-}}$, 
$\overline B{}^0_s\to \Sigma^{-}\overline{\Sigma^{*-}}$ and
$\overline B{}^0_s\to \Xi^{-}\overline{\Xi^{*-}}$ 
decays are pure penguin modes with only
$P'_{\BD}$ and $P'_{iEW\BD}$ contributions,
while
$\overline B{}^0_s\to p\overline{\Delta^+}$ and
$\overline B{}^0_s\to n\overline{\Delta^0}$
are pure exchange modes with only $E'_{\BD}$ contributions. 

\begin{table}[t!]
\caption{\label{tab:BDDS=-1} Same as Table~\ref{tab:BBDS=0}, but with $\Delta S=-1$, $\overline B_q\to\BD$ modes.
The latest experimental results are given in parentheses. 
}
\begin{ruledtabular}
\centering
\begin{tabular}{llll}
Mode
          & ${\mathcal B}(10^{-8})$
          & Mode
          & ${\mathcal B}(10^{-8})$
          \\
\hline $B^-\to \Sigma^+\overline{\Delta^{++}}$
          & $6.99^{+2.90}_{-2.49}{}^{+1.80}_{-0}{}^{+3.13}_{-2.33}\pm0.002$
          & $\overline B{}^0\to \Sigma^{+}\overline{\Delta^+}$
          & $2.16^{+0.90}_{-0.77}{}^{+0.56}_{-0}{}^{+0.97}_{-0.72}\pm0$
           \\
$B^-\to \Sigma^0\overline{\Delta^+}$
          & $4.25^{+1.83}_{-1.51}{}^{+0.56}_{-0}{}^{+1.83}_{-1.46}\pm0.003$
          & $\overline B{}^0\to \Sigma^{0}\overline{\Delta^0}$
          & $3.93^{+1.69}_{-1.40}{}^{+0.52}_{-0}{}^{+1.69}_{-1.35}\pm0$
           \\ 
$B^-\to \Sigma^-\overline{\Delta^0}$
          & $1.96^{+0.87}_{-0.70}\pm0{}^{+0.83}_{-0.69}\pm0.002$
          &$\overline B{}^0\to \Sigma^{-}\overline{\Delta^-}$
          & $5.46^{+2.42}_{-1.94}\pm0{}^{+2.32}_{-1.91}\pm0$
           \\            
$B^-\to \Xi^{0}\overline{\Sigma^{*+}}$ 
          & $1.49^{+0.67}_{-0.53}{}^{+0}_{-0.38}{}^{+0.93}_{-0.70}\pm0.002$
          & $\overline B{}^0\to \Xi^{0}\overline{\Sigma^{*0}}$ 
          & $0.69^{+0.31}_{-0.24}{}^{+0}_{-0.17}{}^{+0.43}_{-0.33}\pm0$
           \\ 
$B^-\to \Xi^{-}\overline{\Sigma^{*0}}$
          & $0.81^{+0.36}_{-0.29}\pm0{}^{+0.34}_{-0.28}\pm0.001$
          & $\overline B{}^0\to \Xi^{-}\overline{\Sigma^{*-}}$
          & $1.49^{+0.66}_{-0.53}\pm0{}^{+0.63}_{-0.52}\pm0$
           \\ 
$B^-\to \Lambda\overline{\Delta^{+}}$
          & $0.15^{+0.07}_{-0.05}{}^{+0.11}_{-0}{}^{+0.07}_{-0.05}\pm0$
          & $\overline B{}^0\to \Lambda\overline{\Delta^0}$
          & $0.14^{+0.07}_{-0.05}{}^{+0.10}_{-0}{}^{+0.06}_{-0.05}\pm0$
           \\
           & ($<82$)~\cite{Wang:2007as}
           &
           & ($<93$)~\cite{Wang:2007as} 
          \\ 
$\overline B{}^0_s\to p\overline{\Delta^+}$
          & $0\pm0\pm0{}^{+0.000005}_{-0}$
          & $\overline B{}^0_s\to \Sigma^{+}\overline{\Sigma^{*+}}$
          & $2.07^{+0.86}_{-0.74}{}^{+0.53}_{-0}{}^{+0.93}_{-0.69}\pm0.001$
           \\
$\overline B{}^0_s\to n\overline{\Delta^0}$
          & $0\pm0\pm0{}^{+0.000005}_{-0}$
          & $\overline B{}^0_s\to \Sigma^{0}\overline{\Sigma^{*0}}$
          & $1.89^{+0.81}_{-0.67}{}^{+0.25}_{-0}{}^{+0.81}_{-0.65}\pm0.001$
           \\
$\overline B{}^0_s\to \Xi^{0}\overline{\Xi^{*0}}$
          & $1.31^{+0.59}_{-0.47}{}^{+0}_{-0.33}{}^{+0.82}_{-0.62}\pm0.002$
          & $\overline B{}^0_s\to \Sigma^{-}\overline{\Sigma^{*-}}$ 
          & $1.74^{+0.77}_{-0.62}\pm0{}^{+0.74}_{-0.61}\pm0$
           \\ 
$\overline B{}^0_s\to \Xi^{-}\overline{\Xi^{*-}}$
          & $1.42^{+0.63}_{-0.51}\pm0{}^{+0.60}_{-0.50}\pm0$
          & $\overline B{}^0_s\to \Lambda\overline{\Sigma^{*0}}$
          & $0.07^{+0.03}_{-0.02}{}^{+0.05}_{-0}{}^{+0.03}_{-0.02}\pm0.001$
           \\                                                                                                                                                       
\end{tabular}
\end{ruledtabular}
\end{table}

The rates of some modes are related.
From Appendices~\ref{appendix:amplitudes} and \ref{appendix:formulas},
we have~\cite{Chua:2013zga}
\be
2\Br(B^-\to\Xi^{-}\overline{\Sigma^{*0}})
   &=&\Br(B^-\to\Sigma^-\overline{\Delta^0}),
\non\\ 
3\tau_{B_s}\Br(\bar B^0_s\to\Sigma^{-}\overline{\Sigma^{*-}})
   &=&3\tau_{B_s}\Br(\bar B^0_s\to\Xi^{-}\overline{\Xi^{*-}})
     =3 \tau_{B_d}\Br(\bar B^0\to\Xi^{-}\overline{\Sigma^{*-}})
\non\\
   &=&\tau_{B_d}\Br(\bar B^0\to\Sigma^{-}\overline{\Delta^-}), 
\non\\
\Br(\bar B^0_s\to p\overline{\Delta^+})
   &=&\Br(\bar B^0_s\to n\overline{\Delta^0}),
\en
where these relations are subjected to corrections from SU(3) breaking in $|p_{cm}|^3$.
These relations do not relay on the asymptotic relations. 
As shown in Table~\ref{tab:BDDS=0} the rates of these modes roughly satisfy the above relations and the agreement will be improved when the SU(3) breaking effects are taken into account.

\begin{table}[t!]
\caption{\label{tab:DBDS=0} Same as Table~\ref{tab:BBDS=0}, but with $\Delta S=0$, $\overline B_q\to\DB$ modes.
The latest experimental result is given in the parenthesis. 
}
\begin{ruledtabular}
\centering
\begin{tabular}{llll}
Mode
          & ${\mathcal B}(10^{-8})$
          & Mode
          & ${\mathcal B}(10^{-8})$
          \\
\hline $B^-\to \Delta^0\overline{p}$
          & $2.19^{+1.05}_{-0.79}{}^{+0}_{-0.15}{}^{+0.92}_{-0.74}\pm0.03$
          & $\overline B{}^0_s\to \Delta^+\overline{\Sigma^{+}}$
          & $1.70^{+0.82}_{-0.61}{}^{+0.16}_{-0}{}^{+0.79}_{-0.64}\pm0$
           \\
           & ($<138$)~\cite{Wei:2007fg}
           &
           &
          \\
$B^-\to \Delta^-\overline{n}$
          & $0.33^{+0.15}_{-0.12}\pm0{}^{+0.14}_{-0.12}{}^{+0.002}_{-0.001}$
          & $\overline B{}^0_s\to \Delta^0\overline{\Sigma^{0}}$
          & $0.96^{+0.46}_{-0.34}{}^{+0.15}_{-0}{}^{+0.50}_{-0.40}\pm0$
           \\ 
$B^-\to \Sigma^{*0}\overline{\Sigma^{+}}$ 
          & $0.85^{+0.41}_{-0.31}{}^{+0}_{-0.06}{}^{+0.36}_{-0.29}\pm0.01$
          &$\overline B{}^0_s\to \Delta^-\overline{\Sigma^{-}}$
          & $0.27^{+0.12}_{-0.10}\pm0{}^{+0.11}_{-0.09}\pm0$
           \\            
$B^-\to \Sigma^{*-}\overline{\Sigma^{0}}$
          & $0.04\pm0.02\pm0\pm0.02{}^{+0.0003}_{-0.0002}$
          & $\overline B{}^0_s\to \Sigma^{*0}\overline{\Xi^{0}}$
          & $2.73^{+1.33}_{-0.98}{}^{+0.15}_{-0}{}^{+1.17}_{-0.96}\pm0$
           \\ 
$B^-\to \Xi^{*-}\overline{\Xi^{0}}$
          & $0.07^{+0.03}_{-0.02}\pm0{}^{+0.03}_{-0.02}{}^{+0.0004}_{-0.0003}$
          & $\overline B{}^0_s\to \Sigma^{*-}\overline{\Xi^{-}}$
          & $0.07\pm0.03\pm0\pm0.03\pm0$
           \\ 
$B^-\to \Sigma^{*-}\overline{\Lambda}$
          & $0.14^{+0.06}_{-0.05}\pm0{}^{+0.06}_{-0.05}\pm0.001$
          & $\overline B{}^0_s\to \Delta^0\overline{\Lambda}$
          & $2.59^{+1.27}_{-0.93}{}^{+0.03}_{-0}{}^{+1.01}_{-0.84}\pm0$
           \\ 
$\overline B{}^0\to \Delta^+\overline{p}$
          & $1.92^{+0.93}_{-0.69}{}^{+0.18}_{-0}{}^{+0.89}_{-0.72}\pm0.02$
          & $\overline B{}^0\to \Sigma^{*+}\overline{\Sigma^{+}}$
          & $0\pm0\pm0\pm0{}^{+0.0001}_{-0}$
           \\
$\overline B{}^0\to \Delta^0\overline{n}$
          & $7.46^{+3.64}_{-2.69}{}^{+0.40}_{-0}{}^{+3.19}_{-2.62}\pm0.05$
          & $\overline B{}^0\to \Sigma^{*0}\overline{\Sigma^{0}}$
          & $0.39^{+0.19}_{-0.14}{}^{+0}_{-0.03}{}^{+0.17}_{-0.13}\pm0.005$
           \\
$\overline B{}^0\to \Xi^{*0}\overline{\Xi^{0}}$
          & $0\pm0\pm0\pm0{}^{+0.0001}_{-0}$
          & $\overline B{}^0\to \Sigma^{*-}\overline{\Sigma^{-}}$
          & $0.08^{+0.04}_{-0.03}\pm0\pm0.03\pm0$
           \\      
$\overline B{}^0\to \Xi^{*-}\overline{\Xi^{-}}$
          & $0.06^{+0.03}_{-0.02}\pm0{}^{+0.03}_{-0.02}\pm0$
          & $\overline B{}^0\to \Sigma^{*0}\overline{\Lambda}$
           & $1.18^{+0.57}_{-0.42}{}^{+0.11}_{-0}{}^{+0.55}_{-0.44}\pm0.02$
           \\                                                                                                                                                               
\end{tabular}
\end{ruledtabular}
\end{table}

\subsubsection{Rates of $\overline B_q\to \DB$ decays}

Predictions on $\Delta S=0$, $\overline B_q\to \DB$ decay rates are shown in Table \ref{tab:DBDS=0}. 
The mode with the highest decay rate is $\overline B{}^0\to\Delta^0\bar n$ decay, which is unfortunately hard to detect.
Both $B^-\to \Delta^0\overline p$ and $\overline B{}^0_s\to\Delta^0\overline\Lambda$ decays can cascadely decay to all charge final states. 
These two modes both have rates at the order of $10^{-8}$, but the $B^-\to \Delta^0\overline p$ decay has better detectability.
Note that the predicted rate of this mode is roughly two orders of magnitude below the present experimental limit~\cite{Wei:2007fg}, which, however, has not been updated since 2008.
With $\pi^0$, one can search for 
$\overline B{}^0\to \Delta^+\overline{p}$ 
and       
$\overline B{}^0\to \Sigma^{*0}\overline{\Lambda}$ 
decays, 
while $\overline B{}^0_s\to \Delta^0\overline{\Sigma^{0}}$ decay, 
which has a slightly smaller rate, can be searched for with $\pi^0$ in the future.
With $\pi^0\pi^0$, one can search for
$\overline B{}^0_s\to \Sigma^{*0}\overline{\Xi^{0}}$, 
$\overline B{}^0_s\to \Delta^+\overline{\Sigma^{+}}$, 
and
$B^-\to \Sigma^{*0}\overline{\Sigma^{+}}$ 
decays, in the future. 
These modes all have rates at the level of $10^{-8}$.     

Note that
$B^-\to \Delta^-\overline{n}$,
$B^-\to \Sigma^{*-}\overline{\Sigma^{0}}$,
$B^-\to \Xi^{*-}\overline{\Xi^{0}}$,
$B^-\to \Sigma^{*-}\overline{\Lambda}$,
$\overline B{}^0\to \Xi^{*-}\overline{\Xi^{-}}$,
$\overline B{}^0_s\to \Delta^-\overline{\Sigma^{-}}$,
$\overline B{}^0_s\to \Sigma^{*-}\overline{\Xi^{-}}$,
$\overline B{}^0\to \Sigma^{*-}\overline{\Sigma^{-}}$,
$\overline B{}^0\to \Xi^{*0}\overline{\Xi^{0}}$ and
$\overline B{}^0\to \Sigma^{*+}\overline{\Sigma^{+}}$ decays
do not have any tree ($T_{i\DB}$) contribution 
[see Eqs. (\ref{eq: DBBm, DS=0}), (\ref{eq: DBB0, DS=0}) and (\ref{eq: DBBs, DS=0})],
and, consequently, their rates in Table~\ref{tab:DBDS=0} have vanishing second uncertainties.
In particular, the 
$\overline B{}^0\to \Xi^{*-}\overline{\Xi^{-}}$,
$\overline B{}^0_s\to \Delta^-\overline{\Sigma^{-}}$,
$\overline B{}^0_s\to \Sigma^{*-}\overline{\Xi^{-}}$ and
$\overline B{}^0\to \Sigma^{*-}\overline{\Sigma^{-}}$ decays
are pure penguin modes, which only have $P_\DB$ and $P_{i EW\DB}$ contributions,
while 
$\overline B{}^0\to \Xi^{*0}\overline{\Xi^{0}}$ and
$\overline B{}^0\to \Sigma^{*+}\overline{\Sigma^{+}}$ decays
are pure exchange ($E_\DB$) modes. 

The rates of some modes are related.
Using the formulas in Appendices~\ref{appendix:amplitudes} and \ref{appendix:formulas},
we have~\cite{Chua:2013zga}
\be
\Br(B^-\to\Delta^0\overline{p})
   &=&
     2\Br(B^-\to\Sigma^{*0}\overline{\Sigma^{+}}),
   \non\\
\Br(B^-\to\Delta^-\overline{n})
   &=& 
   3 \Br(B^-\to\Xi^{*-}\overline{\Xi^{0}})
  =6 \Br(B^-\to\Sigma^{*-}\overline{\Sigma^{0}})
\non\\
   &=&2 \Br(B^-\to\Sigma^{*-}\overline{\Lambda}),
\non\\
3\tau_{B_d}\Br(\bar B^0\to\Sigma^{*-}\overline{\Sigma^{-}})
  &=&3\tau_{B_d}\Br(\bar B^0\to\Xi^{*-}\overline{\Xi^{-}})
   = \tau_{B_s}\Br(\bar B^0_s\to\Delta^-\overline{\Sigma^{-}})
   \non\\
   &=&3\tau_{B_s}\Br(\bar B^0_s\to\Sigma^{*-}\overline{\Xi^{-}}),
\non\\ 
\Br(\bar B^0\to\Sigma^{*+}\overline{\Sigma^{+}})
   &=&\Br(\bar B^0\to\Xi^{*0}\overline{\Xi^{0}}),
\en
where these relations are subjected to corrections from SU(3) breaking in $|p_{cm}|^3$.
These relations do not relay on the asymptotic relations. 
Note that the relations on $B^-$ decay rates are new compared to those in \cite{Chua:2013zga}. 
As shown in Table~\ref{tab:DBDS=0} the rates of these modes roughly satisfy the above relations and the agreement will be improved when the SU(3) breaking effects are taken into account.

\begin{table}[t!]
\caption{\label{tab:DBDS=-1} Same as Table~\ref{tab:BBDS=0}, but with $\Delta S=-1$, $\overline B_q\to\DB$ modes.}
\begin{ruledtabular}
\centering
\begin{tabular}{llll}
Mode
          & ${\mathcal B}(10^{-8})$
          & Mode
          & ${\mathcal B}(10^{-8})$
          \\
\hline $B^-\to \Sigma^{*0}\overline{p}$
          & $0.94^{+0.43}_{-0.33}{}^{+0}_{-0.24}{}^{+0.51}_{-0.40}\pm0.001$
          & $\overline B{}^0\to \Sigma^{*+}\overline{p}$
          & $2.25^{+0.93}_{-0.80}{}^{+0.58}_{-0}{}^{+0.93}_{-0.75}\pm0$
           \\
          &$<47$~\cite{Wang:2007as}
          &
          &$<26$~\cite{Wang:2007as}
          \\ 
$B^-\to \Sigma^{*-}\overline{n}$
          & $2.05^{+0.91}_{-0.73}\pm0{}^{+0.87}_{-0.72}\pm0.002$
          & $\overline B{}^0\to \Sigma^{*0}\overline{n}$
          & $1.39^{+0.58}_{-0.49}{}^{+0.65}_{-0}{}^{+0.54}_{-0.45}\pm0$
           \\ 
$B^-\to \Xi^{*0}\overline{\Sigma^{+}}$
          & $1.44^{+0.65}_{-0.51}{}^{+0}_{-0.37}{}^{+0.78}_{-0.61}\pm0.002$
          &$\overline B{}^0\to \Xi^{*0}\overline{\Sigma^{0}}$
          & $0.67^{+0.30}_{-0.24}{}^{+0}_{-0.17}{}^{+0.36}_{-0.28}\pm0$
           \\            
$B^-\to \Xi^{*-}\overline{\Sigma^{0}}$
          & $0.78^{+0.35}_{-0.28}\pm0{}^{+0.33}_{-0.27}\pm0.001$
          & $\overline B{}^0\to \Xi^{*-}\overline{\Sigma^{-}}$
          & $1.45^{+0.64}_{-0.51}\pm0{}^{+0.61}_{-0.51}\pm0$
           \\ 
$B^-\to \Omega^-\overline{\Xi^{0}}$
          & $3.77^{+1.67}_{-1.34}\pm0{}^{+1.60}_{-1.32}\pm0.003$
          & $\overline B{}^0\to \Omega^-\overline{\Xi^{-}}$
          & $3.47^{+1.54}_{-1.24}\pm0{}^{+1.47}_{-1.21}\pm0$
           \\ 
$B^-\to \Xi^{*-}\overline{\Lambda}$
          & $2.48^{+1.10}_{-0.88}\pm0{}^{+1.05}_{-0.87}\pm0.002$
          & $\overline B{}^0\to \Xi^{*0}\overline{\Lambda}$ 
          & $2.71^{+1.13}_{-0.97}{}^{+0.70}_{-0}{}^{+1.12}_{-0.90}\pm0$
           \\ 
$\overline B{}^0_s\to \Delta^+\overline{p}$
          & $0\pm0\pm0\pm0{}^{+0.000005}_{-0}$
          & $\overline B{}^0_s\to \Sigma^{*+}\overline{\Sigma^{+}}$
          & $1.98^{+0.82}_{-0.71}{}^{+0.51}_{-0}{}^{+0.82}_{-0.66}\pm0.001$
           \\
$\overline B{}^0_s\to \Delta^0\overline{n}$
          & $0\pm0\pm0\pm0{}^{+0.000005}_{-0}$
          & $\overline B{}^0_s\to \Sigma^{*0}\overline{\Sigma^{0}}$
          & $1.81^{+0.78}_{-0.64}{}^{+0.24}_{-0}{}^{+0.74}_{-0.61}\pm0.001$
           \\
$\overline B{}^0_s\to \Xi^{*0}\overline{\Xi^{0}}$
          & $1.99^{+0.83}_{-0.71}{}^{+0.93}_{-0}{}^{+0.78}_{-0.64}\pm0.0002$
          & $\overline B{}^0_s\to \Sigma^{*-}\overline{\Sigma^{-}}$
          & $1.67^{+0.74}_{-0.59}\pm0{}^{+0.71}_{-0.58}\pm0$
           \\ 
$\overline B{}^0_s\to \Xi^{*-}\overline{\Xi^{-}}$
          & $1.36^{+0.60}_{-0.48}\pm0{}^{+0.58}_{-0.47}\pm0$
          & $\overline B{}^0_s\to \Sigma^{*0}\overline{\Lambda}$
          & $0.06^{+0.03}_{-0.02}{}^{+0.05}_{-0}\pm0.02\pm0.001$
           \\                                                                                                                                                  
\end{tabular}
\end{ruledtabular}
\end{table}

Predictions on $\Delta S=-1$, $\overline B_q\to \DB$ decay rates are shown in Table \ref{tab:DBDS=-1}. 
Note  that 
$\overline B{}^0\to \Omega^-\overline{\Xi^{-}}$, 
$\overline B{}^0\to \Xi^{*0}\overline{\Lambda}$ 
and            
$\overline B{}^0\to \Sigma^{*+}\overline{p}$ 
decays are the only three modes that can cascadely decay to all charged final states. 
All of them have rates of order $10^{-8}$.
We need more data to search for them as the reconstruction efficiencies of the final states of the first two modes are low,
while the predicted $\overline B{}^0\to \Sigma^{*+}\overline{p}$ rate is one order of magnitude below the experimental limit,
which has not been updated since 2007~\cite{Wang:2007as}.
Note that with one $\pi^0$ one can search for
$B^-\to \Omega^-\overline{\Xi^{0}}$, 
$B^-\to \Xi^{*-}\overline{\Lambda}$, 
$\overline B{}^0_s\to \Xi^{*0}\overline{\Xi^{0}}$, 
$\overline B{}^0_s\to \Sigma^{*+}\overline{\Sigma^{+}}$, 
$B^-\to \Xi^{*0}\overline{\Sigma^{+}}$, 
$\overline B{}^0_s\to \Xi^{*-}\overline{\Xi^{-}}$ 
and         
$B^-\to \Sigma^{*0}\overline{p}$ 
decays, which have rates of order $10^{-8}$, in the future.
In particular,
the $B^-\to\Omega^-\overline{\Xi^0}$ decay has the highest rate in the table
and the predicted $B^-\to \Sigma^{*0}\overline{p}$ rate is more than one order of magnitude below the experimental limit,
which has not been updated since 2007~\cite{Wang:2007as}.
With $\pi^0\gamma$ one can search for
$\overline B{}^0_s\to \Sigma^{*0}\overline{\Sigma^{0}}$ 
and
$B^-\to \Xi^{*-}\overline{\Sigma^{0}}$ 
decays, which have rates of order (or close to) $10^{-8}$, in the future.

Note that
$B^-\to \Sigma^{*-}\overline{n}$, 
$B^-\to \Xi^{*-}\overline{\Sigma^{0}}$, 
$B^-\to \Omega^-\overline{\Xi^{0}}$, 
$B^-\to \Xi^{*-}\overline{\Lambda}$, 
$\overline B{}^0\to \Omega^-\overline{\Xi^{-}}$, 
$\overline B{}^0_s\to \Sigma^{*-}\overline{\Sigma^{-}}$,
$\overline B{}^0\to \Xi^{*-}\overline{\Sigma^{-}}$,
$\overline B{}^0_s\to \Xi^{*-}\overline{\Xi^{-}}$ 
$\overline B{}^0_s\to \Delta^+\overline{p}$ and
$\overline B{}^0_s\to \Delta^0\overline{n}$ decays
do not have any tree ($T'_{i\DB}$) contribution
[see Eqs. (\ref{eq: DBBm, DS=-1}), (\ref{eq: DBB0, DS=-1}) and (\ref{eq: DBBs, DS=-1})],
and, consequently, their rates in Table~\ref{tab:DBDS=-1} have vanishing second uncertainties.
In particular, the 
$\overline B{}^0\to \Xi^{*-}\overline{\Sigma^{-}}$ and
$\overline B{}^0_s\to \Xi^{*-}\overline{\Xi^{-}}$ 
decays
are pure penguin modes, which only have $P'_\DB$ and $P'_{i EW\DB}$ contributions,
while 
$\overline B{}^0_s\to \Delta^+\overline{p}$ and
$\overline B{}^0_s\to \Delta^0\overline{n}$ decays
are pure exchange ($E'_\DB$) modes.

The rates of some modes are related.
Using the formulas in Appendices~\ref{appendix:amplitudes} and \ref{appendix:formulas},
we have~\cite{Chua:2013zga}
\be
   2 \Br(B^-\to\Sigma^{*0}\overline{p})
   &=&\Br(B^-\to\Xi^{*0}\overline{\Sigma^{+}}),
   \non\\
3\Br(B^-\to\Sigma^{*-}\overline{n})
   &=&
6 \Br(B^-\to\Xi^{*-}\overline{\Sigma^{0}})
   =  
\Br(B^-\to\Omega^-\overline{\Xi^{0}})
\non\\
    &=&
2 \Br(B^-\to\Xi^{*-}\overline{\Lambda}),   
\non\\
3\tau_{B_s}\Br(\bar B^0_s\to\Sigma^{*-}\overline{\Sigma^{-}})
   &=&3\tau_{B_s}\Br(\bar B^0_s\to\Xi^{*-}\overline{\Xi^{-}})
   =3 \tau_{B_d}\Br(\bar B^0\to\Xi^{*-}\overline{\Sigma^{-}})
\non\\   
   &=& \tau_{B_d}\Br(\bar B^0\to\Omega^-\overline{\Xi^{-}}),
\non\\
\Br(\bar B^0_s\to\Delta^+\overline{p})
   &=&\Br(\bar B^0_s\to\Delta^0\overline{n}),
\en
where these relations are subjected to corrections from SU(3) breaking in $|p_{cm}|^3$.
Note that the relations on $B^-$ decay rates are new compared to those in \cite{Chua:2013zga}. 
These relations do not relay on the asymptotic relations. 
The rates in Table~\ref{tab:BDDS=-1} roughly satisfy the above relations and the agreement will be improved when the SU(3) breaking effects are taken into account.

\begin{table}[t!]
\caption{\label{tab:DDDS=0} Same as Table~\ref{tab:BBDS=0}, but with $\Delta S=0$,
$\overline B_q\to\DD$ modes. 
}
\centering
\begin{ruledtabular}
\begin{tabular}{llll}
Mode
          & ${\mathcal B}(10^{-8})$
          & Mode
          & ${\mathcal B}(10^{-8})$
          \\
\hline $B^-\to \Delta^+ \overline{\Delta^{++}}$ 
          & $17.15^{+8.31}_{-6.17}{}^{+1.61}_{-0}{}^{+7.97}_{-6.45}\pm0.22$
          & $\overline B{}^0_s\to \Delta^{+} \overline{\Sigma^{*+}}$ 
          & $5.18^{+2.51}_{-1.86}{}^{+0.49}_{-0}{}^{+2.41}_{-1.95}\pm0$
           \\
$B^-\to \Delta^0 \overline{\Delta^{+}}$ 
          & $6.47^{+3.07}_{-2.33}{}^{+1.02}_{-0}{}^{+3.39}_{-2.67}\pm0.14$
          & $\overline B{}^0_s\to \Delta^{0} \overline{\Sigma^{*0}}$
          & $2.93^{+1.39}_{-1.05}{}^{+0.46}_{-0}{}^{+1.53}_{-1.21}\pm0$
           \\ 
$B^-\to \Delta^- \overline{\Delta^{0}}$
          & $0.92^{+0.41}_{-0.33}\pm0{}^{+0.39}_{-0.32}{}^{+0.006}_{-0.004}$
          &$\overline B{}^0_s\to \Delta^{-} \overline{\Sigma^{*-}}$
          & $0.83^{+0.37}_{-0.29}\pm0{}^{+0.35}_{-0.29}\pm0$
           \\            
$B^-\to \Sigma^{*0} \overline{\Sigma^{*+}}$ 
          & $3.02^{+1.43}_{-1.08}{}^{+0.47}_{-0}{}^{+1.58}_{-1.24}\pm0.07$
          & $\overline B{}^0_s\to \Sigma^{*0} \overline{\Xi^{*0}}$ 
          & $2.73^{+1.29}_{-0.98}{}^{+0.43}_{-0}{}^{+1.43}_{-1.12}\pm0$
           \\ 
$B^-\to \Sigma^{*-} \overline{\Sigma^{*0}}$
          & $0.57^{+0.25}_{-0.20}\pm0{}^{+0.24}_{-0.20}\pm0.003$
          & $\overline B{}^0_s\to \Sigma^{*-} \overline{\Xi^{*-}}$ 
          & $1.03^{+0.46}_{-0.37}\pm0{}^{+0.44}_{-0.36}\pm0$
           \\ 
$B^-\to \Xi^{*-} \overline{\Xi^{*0}}$
          & $0.26^{+0.12}_{-0.09}\pm0{}^{+0.11}_{-0.09}{}^{+0.002}_{-0.001}$
          & $\overline B{}^0_s\to \Xi^{*-} \overline{\Omega^{-}}$
          & $0.71^{+0.31}_{-0.25}\pm0{}^{+0.30}_{-0.25}\pm0$
           \\ 
$\overline B{}^0\to \Delta^{++} \overline{\Delta^{++}}$
          & $0\pm0\pm0\pm0{}^{+0.004}_{-0}$
          & $\overline B{}^0\to \Sigma^{*+} \overline{\Sigma^{*+}}$
          & $0\pm0\pm0\pm0{}^{+0.002}_{-0}$
           \\
           & $<1.1\times10^4$~\cite{Bortoletto:1989mu}
           \\
$\overline B{}^0\to \Delta^{+} \overline{\Delta^{+}}$
          & $5.29^{+2.56}_{-1.90}{}^{+0.50}_{-0}{}^{+2.46}_{-2.00}\pm0.16$
          & $\overline B{}^0\to \Sigma^{*0} \overline{\Sigma^{*0}}$ 
          & $1.40^{+0.66}_{-0.50}{}^{+0.22}_{-0}{}^{+0.73}_{-0.58}\pm0.06$
           \\
$\overline B{}^0\to \Delta^{0} \overline{\Delta^{0}}$
          & $5.99^{+2.83}_{-2.15}{}^{+0.94}_{-0}{}^{+3.14}_{-2.47}\pm0.13$
          & $\overline B{}^0\to \Sigma^{*-} \overline{\Sigma^{*-}}$
          & $1.05^{+0.47}_{-0.37}\pm0{}^{+0.45}_{-0.37}\pm0.07$
           \\ 
           &$<1.5\times10^5$~\cite{Bortoletto:1989mu}
           \\
$\overline B{}^0\to \Delta^{-} \overline{\Delta^{-}}$
          & $2.55^{+1.13}_{-0.91}\pm0{}^{+1.08}_{-0.89}\pm0.11$
          & $\overline B{}^0\to \Xi^{*0} \overline{\Xi^{*0}}$
          & $0\pm0\pm0\pm0{}^{+0.001}_{-0}$ 
           \\ 
$\overline B{}^0\to \Omega^{-} \overline{\Omega^{-}}$
          & $0\pm0\pm0\pm0{}^{+0.001}_{-0}$%
          & $\overline B{}^0\to \Xi^{*-} \overline{\Xi^{*-}}$
          & $0.24^{+0.11}_{-0.09}\pm0{}^{+0.10}_{-0.08}\pm0.03$
           \\                                                                                                                                                         
\end{tabular}
\end{ruledtabular}
\end{table}

\subsubsection{Rates of $\overline B_q\to \DD$ decays}

Predictions on $\Delta S=0$, $\overline B_q\to\DD$ decay rates are shown in Table~\ref{tab:DDDS=0}.
There are six modes that can cascadely decay to all charged final states.
They are
$\overline B{}^0\to\Delta^{++}\overline{\Delta^{++}}$,
$\Delta^0\overline{\Delta^0}$,
$\Omega^-\overline{\Omega^-}$,
$\Sigma^{*+}\overline{\Sigma^{*+}}$,
$\Sigma^{*-}\overline{\Sigma^{*-}}$
and $\Xi^{*0}\overline{\Xi^{*0}}$ decays.
However, most of them have highly suppressed rates, except
$\overline{B}{}^0\to \Delta^0\overline{\Delta^0}$ 
and 
$\overline{B}{}^0\to \Sigma^{*-}\overline{\Sigma^{*-}}$ decays, which have rates of order $10^{-8}$ and should be searchable.
In particular, the bound on $\overline B{}^0\to \Delta^{0} \overline{\Delta^{0}}$ rate has not been updated for almost three decades~\cite{Bortoletto:1989mu}.
Note that with $\pi^0$, one can search for plenty of unsuppressed modes.
These modes include
$B^-\to \Delta^+ \overline{\Delta^{++}}$, 
             $B^-\to \Delta^0 \overline{\Delta^{+}}$, 
          $\overline B{}^0_s\to \Delta^{+} \overline{\Sigma^{*+}}$, 
             $B^-\to \Sigma^{*0} \overline{\Sigma^{*+}}$, 
          $\overline B{}^0_s\to \Delta^{0} \overline{\Sigma^{*0}}$, 
             $\overline B{}^0_s\to \Sigma^{*0} \overline{\Xi^{*0}}$, 
          $\overline B{}^0_s\to \Sigma^{*-} \overline{\Xi^{*-}}$, 
and          
             $\overline B{}^0_s\to \Xi^{*-} \overline{\Omega^{-}}$ 
decays.             
In particular, the mode with the highest rate (the only one at order $10^{-7}$) can be searched through the
$B^-\to \Delta^+\overline{\Delta^{++}}\to p\pi^0\bar p\pi^-$ decay.
With $\pi^0\pi^0$, one can also search for
          $\overline B{}^0\to \Delta^{+} \overline{\Delta^{+}}$ 
and          
             $\overline B{}^0\to \Sigma^{*0} \overline{\Sigma^{*0}}$ 
decays.

Note that
$B^-\to \Delta^- \overline{\Delta^{0}}$,
$B^-\to \Sigma^{*-} \overline{\Sigma^{*0}}$,
$B^-\to \Xi^{*-} \overline{\Xi^{*0}}$,
$\overline B{}^0\to \Delta^{-} \overline{\Delta^{-}}$,
$\overline B{}^0\to \Sigma^{*-} \overline{\Sigma^{*-}}$, 
$\overline B{}^0\to \Xi^{*-} \overline{\Xi^{*-}}$,
$\overline B{}^0_s\to \Delta^{-} \overline{\Sigma^{*-}}$,
$\overline B{}^0_s\to \Sigma^{*-} \overline{\Xi^{*-}}$, 
$\overline B{}^0_s\to \Xi^{*-} \overline{\Omega^{-}}$, 
$\overline B{}^0\to \Delta^{++} \overline{\Delta^{++}}$,
$\overline B{}^0\to \Sigma^{*+} \overline{\Sigma^{*+}}$,
$\overline B{}^0\to \Xi^{*0} \overline{\Xi^{*0}}$ and
$\overline B{}^0\to \Omega^{-} \overline{\Omega^{-}}$ decays
do not have any tree ($T_{\DD}$) contribution 
[see Eqs. (\ref{eq: DDBm, DS=0}), (\ref{eq: DDB0, DS=0}) and (\ref{eq: DDBs, DS=0})],
and, consequently, their rates in Table~\ref{tab:DDDS=0} have vanishing second uncertainties.
In particular, the 
$\overline B{}^0\to \Delta^{-} \overline{\Delta^{-}}$,
$\overline B{}^0\to \Sigma^{*-} \overline{\Sigma^{*-}}$, 
$\overline B{}^0\to \Xi^{*-} \overline{\Xi^{*-}}$,
$\overline B{}^0_s\to \Delta^{-} \overline{\Sigma^{*-}}$,
$\overline B{}^0_s\to \Sigma^{*-} \overline{\Xi^{*-}}$ and 
$\overline B{}^0_s\to \Xi^{*-} \overline{\Omega^{-}}$ 
decays
are pure penguin modes, which only have $P_\DD$, $P_{EW\DD}$ and $PA_\DD$ contributions,
the 
$\overline B{}^0\to \Delta^{++} \overline{\Delta^{++}}$,
$\overline B{}^0\to \Sigma^{*+} \overline{\Sigma^{*+}}$ and
$\overline B{}^0\to \Xi^{*0} \overline{\Xi^{*0}}$ decays
are subleading modes,
which only have $E_\DD$ and $P_\DD$ contributions,
while the
$\overline B{}^0\to \Omega^{-} \overline{\Omega^{-}}$ decay
is a pure penguin annihilation ($PA_\DD$) mode.

The rates of some modes are related.
Using formulas in Appendices~\ref{appendix:amplitudes} and \ref{appendix:formulas},
we have~\cite{Chua:2013zga}
\be
2 \Br(B^-\to\Delta^-\overline{\Delta^{0}})
    &=&3 \Br(B^-\to\Sigma^{*-}\overline{\Sigma^{*0}})
      =6 \Br(B^-\to\Xi^{*-}\overline{\Xi^{*0}}),
\non\\
\Br(B^-\to\Delta^0\overline{\Delta^{+}})
      &=&2 \Br(B^-\to\Sigma^{*0}\overline{\Sigma^{*+}}),
\non\\
\Br(\overline{B^0_s}\to\Delta^{0}\overline{\Sigma^{*0}})
     &=&\Br(\overline{B^0_s}\to \Sigma^{*0}\overline{\Xi^{*0}}),
\non\\
4 \Br(\overline{B^0_s}\to\Delta^{-}\overline{\Sigma^{*-}})
     &=&4 \Br(\overline{B^0_s}\to\Xi^{*-}\overline{\Omega^-})
      =3 \Br(\overline{B^0_s}\to\Sigma^{*-}\overline{\Xi^{*-}}),
\en
where these relations are subjected to SU(3) breaking from the phase space factors.
These relations do not relay on the asymptotic relations. 
As shown in Table~\ref{tab:DDDS=0} the rates of these modes roughly satisfy the above relations and the agreement will be improved when the SU(3) breaking effects are taken into account.

\begin{table}[t!]
\caption{\label{tab:DDDS=-1} Same as Table~\ref{tab:BBDS=0}, but with $\Delta S=-1$, $\overline B_q\to\DD$ modes.}
\begin{ruledtabular}
\centering
\begin{tabular}{llll}
Mode
          & ${\mathcal B}(10^{-8})$
          & Mode
          & ${\mathcal B}(10^{-8})$
          \\
\hline $B^-\to \Sigma^{*+} \overline{\Delta^{++}}$ 
          & $21.48^{+8.92}_{-7.65}{}^{+5.53}_{-0}{}^{+8.86}_{-7.15}\pm0.007$
          & $\overline B{}^0\to \Sigma^{*+} \overline{\Delta^{+}}$ 
          & $6.63^{+2.75}_{-2.36}{}^{+1.71}_{-0}{}^{+2.73}_{-2.21}\pm0$
           \\
$B^-\to \Sigma^{*0} \overline{\Delta^{+}}$ 
          & $13.08^{+5.63}_{-4.66}{}^{+1.73}_{-0}{}^{+5.32}_{-4.39}\pm0.008$
          & $\overline B{}^0\to \Sigma^{*0} \overline{\Delta^{0}}$ 
          & $12.11^{+5.21}_{-4.31}{}^{+1.60}_{-0}{}^{+4.93}_{-4.06}\pm0$
           \\ 
$B^-\to \Sigma^{*-} \overline{\Delta^{0}}$ 
          & $6.06^{+2.68}_{-2.16}\pm0{}^{+2.57}_{-2.12}\pm0.005$
          &$\overline B{}^0\to \Sigma^{*-} \overline{\Delta^{-}}$
          & $16.83^{+7.46}_{-5.99}\pm0{}^{+7.15}_{-5.88}\pm0$ 
           \\            
$B^-\to \Xi^{*0} \overline{\Sigma^{*+}}$ 
          & $24.26^{+10.44}_{-8.64}{}^{+3.21}_{-0}{}^{+9.87}_{-8.13}\pm0.01$
          & $\overline B{}^0\to \Xi^{*0} \overline{\Sigma^{*0}}$ 
          & $11.23^{+4.83}_{-4.00}{}^{+1.49}_{-0}{}^{+4.57}_{-3.77}\pm0$
           \\ 
$B^-\to \Xi^{*-} \overline{\Sigma^{*0}}$ 
          & $11.24^{+4.98}_{-4.00}\pm0{}^{+4.77}_{-3.93}\pm0.01$
          & $\overline B{}^0\to \Xi^{*-} \overline{\Sigma^{*-}}$ 
          & $20.79^{+9.21}_{-7.40}\pm0{}^{+8.83}_{-7.27}\pm0$ 
           \\ 
$B^-\to \Omega^{-} \overline{\Xi^{*0}}$ 
          & $15.49^{+6.86}_{-5.51}\pm0{}^{+6.57}_{-5.41}\pm0.01$
          & $\overline B{}^0\to \Omega^{-} \overline{\Xi^{*-}}$ 
          & $14.32^{+6.34}_{-5.10}\pm0{}^{+6.08}_{-5.01}\pm0$ 
           \\ 
$\overline B{}^0_s\to \Delta^{++} \overline{\Delta^{++}}$
          & $0\pm0\pm0\pm0{}^{+0.02}_{-0}$ 
          & $\overline B{}^0_s\to \Sigma^{*+} \overline{\Sigma^{*+}}$ 
          & $6.49^{+2.69}_{-2.31}{}^{+1.67}_{-0}{}^{+2.67}_{-2.16}{}^{+0.77}_{-0.72}$ 
           \\
$\overline B{}^0_s\to \Delta^{+} \overline{\Delta^{+}}$
          & $0\pm0\pm0\pm0{}^{+0.02}_{-0}$ 
          & $\overline B{}^0_s\to \Sigma^{*0} \overline{\Sigma^{*0}}$ 
          & $5.93^{+2.55}_{-2.11}{}^{+0.79}_{-0}{}^{+2.41}_{-1.99}{}^{+0.74}_{-0.70}$ 
           \\
$\overline B{}^0_s\to \Delta^{0} \overline{\Delta^{0}}$
          & $0\pm0\pm0\pm0{}^{+0.02}_{-0}$ 
          & $\overline B{}^0_s\to \Sigma^{*-} \overline{\Sigma^{*-}}$ 
          & $5.48^{+2.43}_{-1.95}\pm0{}^{+2.33}_{-1.92}{}^{+0.72}_{-0.67}$ 
           \\ 
$\overline B{}^0_s\to \Delta^{-} \overline{\Delta^{-}}$
          & $0\pm0\pm0\pm0{}^{+0.02}_{-0}$ 
          & $\overline B{}^0_s\to \Xi^{*0} \overline{\Xi^{*0}}$ 
          & $21.92^{+9.44}_{-7.81}{}^{+2.90}_{-0}{}^{+8.92}_{-7.35}{}^{+1.36}_{-1.32}$
           \\ 
$\overline B{}^0_s\to \Omega^{-} \overline{\Omega^{-}}$ 
          & $41.95^{+18.58}_{-14.93}\pm0{}^{+17.81}_{-14.67}{}^{+1.79}_{-1.75}$ 
          & $\overline B{}^0_s\to \Xi^{*-} \overline{\Xi^{*-}}$ 
          & $20.29^{+8.99}_{-7.22}\pm0{}^{+8.62}_{-7.10}{}^{+1.30}_{-1.26}$ 
           \\                                                                                                                                                        
\end{tabular}
\end{ruledtabular}
\end{table}

Predictions on $\Delta S=-1$, $\overline B_q\to\DD$ decay rates are shown in Table~\ref{tab:DDDS=-1}.
There are eight modes that can cascadely decay to all charged final states with unsuppressed rates.
They are
$\overline B{}^0_s\to \Omega^{-} \overline{\Omega^{-}}$, 
$B^-\to \Xi^{*0} \overline{\Sigma^{*+}}$, 
$\overline B{}^0_s\to \Xi^{*0} \overline{\Xi^{*0}}$, 
$B^-\to \Sigma^{*+} \overline{\Delta^{++}}$, 
$B^-\to \Omega^{-} \overline{\Xi^{*0}}$, 
$\overline B{}^0_s\to \Sigma^{*+} \overline{\Sigma^{*+}}$, 
$B^-\to \Sigma^{*-} \overline{\Delta^{0}}$ 
and              
$\overline B{}^0_s\to \Sigma^{*-} \overline{\Sigma^{*-}}$ 
decays.
It is interesting that many (the first five) of them have rates of order $10^{-7}$.
In particular, the $\overline B{}^0_s\to\Omega^-\overline{\Omega^-}$ decay has the highest rate and good detectability. 
Note that with one $\pi^0$ one can search for
          $\overline B{}^0\to \Xi^{*-} \overline{\Sigma^{*-}}$, 
              $\overline B{}^0\to \Omega^{-} \overline{\Xi^{*-}}$, 
              $\overline B{}^0_s\to \Xi^{*-} \overline{\Xi^{*-}}$, 
             $B^-\to \Sigma^{*0} \overline{\Delta^{+}}$, 
             $\overline B{}^0\to \Sigma^{*0} \overline{\Delta^{0}}$ 
and
             $\overline B{}^0\to \Xi^{*0} \overline{\Sigma^{*0}}$ 
decays in the future. 
All of them have rates of order $10^{-7}$.             
With $\pi^0\pi^0$ one can search for
            $B^-\to \Xi^{*-} \overline{\Sigma^{*0}}$, 
            $\overline B{}^0_s\to \Sigma^{*0} \overline{\Sigma^{*0}}$ 
and            
            $\overline B{}^0\to \Sigma^{*+} \overline{\Delta^{+}}$ 
decays,
where the first one has rate of order $10^{-7}$.

Note that
$B^-\to \Sigma^{*-} \overline{\Delta^{0}}$, 
$B^-\to \Xi^{*-} \overline{\Sigma^{*0}}$, 
$B^-\to \Omega^{-} \overline{\Xi^{*0}}$, 
$\overline B{}^0\to \Sigma^{*-} \overline{\Delta^{-}}$,
$\overline B{}^0\to \Xi^{*-} \overline{\Sigma^{*-}}$, 
$\overline B{}^0\to \Omega^{-} \overline{\Xi^{*-}}$, 
$\overline B{}^0_s\to \Sigma^{*-} \overline{\Sigma^{*-}}$, 
$\overline B{}^0_s\to \Omega^{-} \overline{\Omega^{-}}$, 
$\overline B{}^0_s\to \Xi^{*-} \overline{\Xi^{*-}}$, 
$\overline B{}^0_s\to \Delta^{++} \overline{\Delta^{++}}$,
$\overline B{}^0_s\to \Delta^{+} \overline{\Delta^{+}}$,
$\overline B{}^0_s\to \Delta^{0} \overline{\Delta^{0}}$ and
$\overline B{}^0_s\to \Delta^{-} \overline{\Delta^{-}}$ decays
do not have any tree ($T'_{\DD}$) contribution
[see Eqs. (\ref{eq: DDBm, DS=-1}), (\ref{eq: DDB0, DS=-1}) and (\ref{eq: DDBs, DS=-1})],
and, consequently, their rates in Table~\ref{tab:DDDS=-1} have vanishing second uncertainties.
The 
$\overline B{}^0\to \Sigma^{*-} \overline{\Delta^{-}}$,
$\overline B{}^0\to \Xi^{*-} \overline{\Sigma^{*-}}$, 
$\overline B{}^0\to \Omega^{-} \overline{\Xi^{*-}}$, 
$\overline B{}^0_s\to \Sigma^{*-} \overline{\Sigma^{*-}}$, 
$\overline B{}^0_s\to \Omega^{-} \overline{\Omega^{-}}$ and 
$\overline B{}^0_s\to \Xi^{*-} \overline{\Xi^{*-}}$ 
decays
are pure penguin modes, which only have $P'_\DD$, $P'_{EW\DD}$ and $PA'_\DD$ contributions,
the 
$\overline B{}^0_s\to \Delta^{++} \overline{\Delta^{++}}$
$\overline B{}^0_s\to \Delta^{+} \overline{\Delta^{+}}$ and
$\overline B{}^0_s\to \Delta^{0} \overline{\Delta^{0}}$ decays
are subleading modes,
which only have $E'_\DD$ and $P'_\DD$ contributions,
while the
$\overline B{}^0_s\to \Delta^{-} \overline{\Delta^{-}}$ decay
is a pure penguin annihilation ($PA'_\DD$) mode.

The rates of some modes are related.
Using formulas in Appendices~\ref{appendix:amplitudes} and \ref{appendix:formulas},
we have~\cite{Chua:2013zga}~\footnote{A typo in the last relation is corrected.}
\be
6 \Br(B^-\to\Sigma^{*-}\overline{\Delta^0})
     &=&   
2 \Br(B^-\to\Omega^-\overline{\Xi^{*0}})
      =3 \Br(B^-\to\Xi^{*-}\overline{\Sigma^{*0}}),
\non\\
2 \Br(B^-\to\Sigma^{*0}\overline{\Delta^+})
     &=&\Br(B^-\to\Xi^{*0}\overline{\Sigma^{*+}}),
\non\\
\Br(\bar B^0\to\Sigma^{*0}\overline{\Delta^0})
   &=&\Br(\bar B^0\to\Xi^{*0}\overline{\Sigma^{*0}}),
\non\\
4 \Br(\bar B^0\to\Sigma^{*-}\overline{\Delta^-})
   &=& 3 \Br(\bar B^0\to\Xi^{*-}\overline{\Sigma^{*-}})
   = 4 \Br(\bar B^0\to\Omega^-\overline{\Xi^{*-}}).
\en
where these relations are subjected to SU(3) breaking from the phase space factors.
These relations do not relay on the asymptotic relations. 
The rates in Table~\ref{tab:DDDS=-1} roughly satisfy the above relations and the agreement will be improved when the SU(3) breaking effects are taken into account.

\begin{table}[t!]
\caption{\label{tab:rate}  Modes with (relatively) unsuppressed rates and (relatively) good detectability are summarized.}
\begin{ruledtabular}
\centering
{\footnotesize\begin{tabular}{llll}
          & Group I
          & Group II
          & Group III
\\
          & All charged final states
          & with single $\pi^0/\gamma$
          & with $\pi^0\pi^0$, $\pi^0\gamma$ or $\gamma\gamma$
          \\
          \hline
$\BB, \Delta S=0$        
          & $\overline B{}^0\to p\overline{p}$; 
          & $\overline B{}^0\to \Sigma^{0}\overline{\Lambda}$, 
              $\overline B{}^0_s\to p\overline{\Sigma^{+}}$; 
              
          & $\overline B{}^0_s\to \Sigma^{0}\overline{\Xi^{0}}$, 
              $B^-\to \Sigma^{0}\overline{\Sigma^{+}}$, 
          \\
          &   
          & 
          & $\overline B{}^0\to \Sigma^{0}\overline{\Sigma^{0}}$ 
          \\ 
\hline
$\BB, \Delta S=-1$
          & $B^-\to \Lambda\overline{p}$, 
              $\overline B{}^0_s\to \Xi^{-}\overline{\Xi^{-}}$; 
          & $B^-\to\Xi^{-}\overline{\Sigma^{0}}$, 
              $\overline B{}^0\to \Xi^{0}\overline{\Lambda}$; 
          & $B^-\to \Xi^{0}\overline{\Sigma^{+}}$, 
              $\overline B{}^0_s\to \Xi^{0}\overline{\Xi^{0}}$, 
          \\
          & $\overline B{}^0_s\to \Lambda\overline{\Lambda}$, 
              $B^-\to \Xi^{-}\overline{\Lambda}$; 
          & $\overline B{}^0\to \Sigma^{+}\overline{p}$ 
          & $\overline B{}^0\to \Xi^{0}\overline{\Sigma^{0}}$, 
              $\overline B{}^0_s\to \Sigma^{+}\overline{\Sigma^{+}}$, 
          \\
          &
          &
          & $\overline B{}^0_s\to \Sigma^{0}\overline{\Sigma^{0}}$ 
          \\
\hline  
$\BD,\Delta S=0$
          & $B^-\to p\overline{\Delta^{++}}$, 
             $\overline B{}^0_s\to p\overline{\Sigma^{*+}}$; 
          & $B^-\to\Sigma^0\overline{\Sigma^{*+}}$, 
              $\overline B{}^0_s\to \Sigma^{0}\overline{\Xi^{*0}}$; 
          & $\overline B{}^0\to \Sigma^{0}\overline{\Sigma^{*0}}$
          \\
          &  
          &   $\overline B{}^0\to p\overline{\Delta^+}$ 
          &
          \\
\hline  
$\BD,\Delta S=-1$
          & $\overline B{}^0\to \Xi^{-}\overline{\Sigma^{*-}}$; 
              
          & $B^-\to \Sigma^+\overline{\Delta^{++}}$, 
              $\overline B{}^0\to \Sigma^{0}\overline{\Delta^0}$; 
          & $B^-\to \Sigma^0\overline{\Delta^+}$, 
              $\overline B{}^0\to \Sigma^{+}\overline{\Delta^+}$, 
          \\
          &
          &  $\overline B{}^0_s\to \Sigma^{+}\overline{\Sigma^{*+}}$, 
               $B^-\to \Xi^{0}\overline{\Sigma^{*+}}$; 

          &   $\overline B{}^0_s\to \Sigma^{0}\overline{\Sigma^{*0}}$, 
               $\overline B{}^0\to \Xi^{0}\overline{\Sigma^{*0}}$ 
          \\       
          &
          &   $\overline B{}^0_s\to \Xi^{-}\overline{\Xi^{*-}}$, 
                $\overline B{}^0_s\to \Xi^{0}\overline{\Xi^{*0}}$,
          &
          \\
          &
          &  $B^-\to \Xi^{-}\overline{\Sigma^{*0}}$ 
          &
          \\  
\hline  
$\DB,\Delta S=0$
          & $B^-\to \Delta^0\overline{p}$, 
              $\overline B{}^0_s\to \Delta^0\overline{\Lambda}$; 
          & $\overline B{}^0\to \Delta^+\overline{p}$, 
             $\overline B{}^0\to \Sigma^{*0}\overline{\Lambda}$; 
          & $\overline B{}^0_s\to \Sigma^{*0}\overline{\Xi^{0}}$, 
              $\overline B{}^0_s\to \Delta^+\overline{\Sigma^{+}}$, 
          \\
          &
          & $\overline B{}^0_s\to \Delta^0\overline{\Sigma^{0}}$
          & $B^-\to \Sigma^{*0}\overline{\Sigma^{+}}$ 
          \\ 
\hline
$\DB,\Delta S=-1$
         & $\overline B{}^0\to \Omega^-\overline{\Xi^{-}}$, 
            $\overline B{}^0\to \Xi^{*0}\overline{\Lambda}$ ;
         & $B^-\to \Omega^-\overline{\Xi^{0}}$, 
             $B^-\to \Xi^{*-}\overline{\Lambda}$; 
         & $\overline B{}^0_s\to \Sigma^{*0}\overline{\Sigma^{0}}$, 
             $B^-\to \Xi^{*-}\overline{\Sigma^{0}}$ 
         \\
         & $\overline B{}^0\to \Sigma^{*+}\overline{p}$ 
         & 
            $\overline B{}^0_s\to \Xi^{*0}\overline{\Xi^{0}}$, 
            $\overline B{}^0_s\to \Sigma^{*+}\overline{\Sigma^{+}}$, 
         &
         \\
         &
         &  $B^-\to \Xi^{*0}\overline{\Sigma^{+}}$, 
             $\overline B{}^0_s\to \Xi^{*-}\overline{\Xi^{-}}$, 
         &
         \\ 
         &
         & $B^-\to \Sigma^{*0}\overline{p}$
         &
         \\ 
\hline
$\DD,\Delta S=0$
         & $\overline B{}^0\to \Delta^{0} \overline{\Delta^{0}}$, 
             $\overline B{}^0\to \Sigma^{*-} \overline{\Sigma^{*-}}$; 
         & $B^-\to \Delta^+ \overline{\Delta^{++}}$, 
             $B^-\to \Delta^0 \overline{\Delta^{+}}$; 
         & $\overline B{}^0\to \Delta^{+} \overline{\Delta^{+}}$, 
             $\overline B{}^0\to \Sigma^{*0} \overline{\Sigma^{*0}}$ 
         \\
         & 
         & $\overline B{}^0_s\to \Delta^{+} \overline{\Sigma^{*+}}$, 
             $B^-\to \Sigma^{*0} \overline{\Sigma^{*+}}$, 
         &
         \\
         &
         & $\overline B{}^0_s\to \Delta^{0} \overline{\Sigma^{*0}}$, 
             $\overline B{}^0_s\to \Sigma^{*0} \overline{\Xi^{*0}}$, 
         &
         \\
         &
         & $\overline B{}^0_s\to \Sigma^{*-} \overline{\Xi^{*-}}$, 
             $\overline B{}^0_s\to \Xi^{*-} \overline{\Omega^{-}}$ 
         &
         \\    
\hline
$\DD,\Delta S=-1$
         & $\overline B{}^0_s\to \Omega^{-} \overline{\Omega^{-}}$, 
              $B^-\to \Xi^{*0} \overline{\Sigma^{*+}}$; 
         & $\overline B{}^0\to \Xi^{*-} \overline{\Sigma^{*-}}$, 
              $\overline B{}^0\to \Omega^{-} \overline{\Xi^{*-}}$; 
         & $\overline B{}^0_s\to \Xi^{*-} \overline{\Xi^{*-}}$, 
             $B^-\to \Sigma^{*0} \overline{\Delta^{+}}$, 
         \\
         & $\overline B{}^0_s\to \Xi^{*0} \overline{\Xi^{*0}}$, 
              $B^-\to \Sigma^{*+} \overline{\Delta^{++}}$; 
         &  $\overline B{}^0\to \Sigma^{*0} \overline{\Delta^{0}}$, 
             $\overline B{}^0\to \Xi^{*0} \overline{\Sigma^{*0}}$; 
         &   $B^-\to \Xi^{*-} \overline{\Sigma^{*0}}$, 
              $\overline B{}^0_s\to \Sigma^{*0} \overline{\Sigma^{*0}}$ 
         \\
         & $B^-\to \Omega^{-} \overline{\Xi^{*0}}$, 
              $\overline B{}^0_s\to \Sigma^{*+} \overline{\Sigma^{*+}}$; 
         &  $\overline B{}^0\to \Sigma^{*+} \overline{\Delta^{+}}$ 
         &
         \\
         & $B^-\to \Sigma^{*-} \overline{\Delta^{0}}$, 
              $\overline B{}^0_s\to \Sigma^{*-} \overline{\Sigma^{*-}}$ 
         &
         &
         \\      
  \end{tabular}
}
\\
\end{ruledtabular}
\end{table}

We have shown that $\overline B{}^0\to p\bar p$ and $B^-\to\Lambda\bar p$ decays have better chance to be found experimentally.
We have identified several modes that can cascadely decay to all charged final states with (relatively) unsuppressed rates.
They are searchable in near future. 
In particular, we note that the predicted $B^-\to p\overline{\Delta^{++}}$ rate is close to the experimental bound,
which has not been updated in the last ten years~\cite{Wei:2007fg}.
Furthermore, the bounds on $B^-\to \Delta^0\overline{p}$ and $\overline B{}^0\to \Sigma^{*+}\overline{p}$ rates have not been updated in the last ten years~\cite{Wei:2007fg, Wang:2007as} and the bound on $\overline B{}^0\to \Delta^{0} \overline{\Delta^{0}}$ rate was last given in 1989~\cite{Bortoletto:1989mu},
while their rates are predicted to be of the order of $10^{-8}$.
Also note that the $\overline B{}^0_s\to \Omega^{-} \overline{\Omega^{-}}$ rate is predicted to be the highest rate.
It will also be interesting for Belle-II, which will be turned on soon, to search for baryonic modes, as there are many modes having unsuppressed rates but require $\pi^0$ or $\gamma$ for detection.  
We pointed out several modes without tree amplitudes. Some of them are pure penguin modes. 
As we shall see in the next subsection, these will affect their $CP$ asymmetries.
We summarize our suggestions for experimental searches in Table~\ref{tab:rate}. These modes have (relatively) unsuppressed rates and better detectability. Modes are arranged according to their quantum numbers, detection and rates (in descending order) and they are assigned into Groups I, II and III accordingly, where Group I modes are modes that have unsuppressed rates and can cascadely decay to all charged final states, Group II modes are modes that can be searched with $\pi^0$ or $\gamma$ and Group III modes are modes that can be searched with $\pi^0\pi^0$, $\pi^0\gamma$ or $\gamma\gamma$.

\subsection{Numerical Results on Direct $CP$ Asymmetries}

In this subsection, 
we will give results of direct $CP$ asymmetries of all modes and plot asymmetries of several interesting modes.
Note that this study becomes possible as we now have the information of the tree-penguin ratio, Eq. (\ref{eq:P/T}).

\begin{table}[t!]
\caption{\label{tab:AcpBBDS=0} Direct $CP$ asymmetries ($\A$ in $\%$) of $\Delta S=0$, $\overline B_q\to\BB$ modes
for $\phi=0$, $\pm\pi/4$ and $\pm\pi/2$.
The uncertainties are from varying the strong phases of $r^{(\prime)}_{t,i}$, $r^{(\prime)}_{p,i}$, $r^{(\prime)}_{ewp,i}$ and $\eta_{i,j,k}$ (see Eqs. (\ref{eq:correction0}) and (\ref{eq:correction2})).}
\begin{ruledtabular}
\centering
{
\footnotesize
\begin{tabular}{lccclccc}
Mode
          & $\phi=0$
          & $\phi=\pm\pi/4$
          & $\phi=\pm\pi/2$
          & Mode
          & $\phi=0$
          & $\phi=\pm\pi/4$
          & $\phi=\pm\pi/2$
          \\
\hline $B^-\to n\overline{p}$
          & $0\pm32.1$%
          & $\mp(74.0_{-25.6}^{+19.7})$%
          & $\mp(97.9_{-15.6}^{+2.1})$%
          & $\overline B{}^0_s\to p\overline{\Sigma^{+}}$ 
          & $0\pm21.2$%
          & $\mp(36.0_{-17.7}^{+21.2})$%
          & $\mp(49.3^{+20.2}_{-15.2})$%
           \\
$B^-\to \Sigma^{0}\overline{\Sigma^{+}}$ 
          & $0\pm31.2$%
          & $\mp(59.3^{+23.4}_{-27.2})$%
          & $\mp(79.5^{+16.1}_{-22.4})$%
          & $\overline B{}^0_s\to n\overline{\Sigma^{0}}$ 
          & $0\pm14.9$%
          & $\pm(26.7^{+14.6}_{-13.8})$%
          & $\pm(38.8^{+13.9}_{-12.8})$%
          \\ 
$B^-\to \Sigma^{-}\overline{\Sigma^{0}}$
          & $0\pm54.6$%
          & $0\pm54.6$%
          & $0\pm54.6$%
          &$\overline B{}^0_s\to n\overline{\Lambda}$
          & $0\pm31.0$%
          & $\mp(71.6^{+19.3}_{-25.4})$%
          & $\mp(94.9^{+5.1}_{-16.3})$%
           \\            
$B^-\to \Sigma^{-}\overline{\Lambda}$
          & $0\pm35.7$%
          & $0\pm35.7$%
          & $0\pm35.7$%
          & $\overline B{}^0_s\to \Sigma^{0}\overline{\Xi^{0}}$ 
          & $0\pm15.4$%
          & $\mp(42.8^{+14.1}_{-13.9})$%
          & $\mp(58.3^{+12.9}_{-12.6})$%
           \\ 
$B^-\to \Xi^{-}\overline{\Xi^{0}}$
          & $0\pm35.7$%
          & $0\pm35.7$%
          & $0\pm35.7$%
          & $\overline B{}^0_s\to \Sigma^{-}\overline{\Xi^{-}}$ 
          & $0\pm30.0$%
          & $0\pm30.0$%
          & $0\pm30.0$%
           \\ 
$B^-\to \Lambda\overline{\Sigma^+}$
          & $0\pm70.7$%
          & $0\pm70.7$%
          & $0\pm70.7$%
          & $\overline B{}^0_s\to \Lambda\overline{\Xi^0}$
          & $0\pm100$%
          & $0\pm100$%
          & $0\pm100$%
           \\ 
$\overline B{}^0\to p\overline{p}$ 
          & $0\pm26.9$%
          & $\mp(36.0^{+26.8}_{-22.0})$%
          & $\mp(49.3^{+25.2}_{-18.3})$%
          & $\overline B{}^0\to \Sigma^{+}\overline{\Sigma^{+}}$
          & $-100\sim100$%
          & $-100\sim100$%
          & $-100\sim100$%
           \\           
$\overline B{}^0\to n\overline{n}$
          & $0\pm31.6$%
          & $\mp(59.3^{+23.5}_{-27.9})$%
          & $\mp(79.5^{+16.0}_{-23.2})$%
          & $\overline B{}^0\to \Sigma^{0}\overline{\Sigma^{0}}$ 
          & $0\pm33.6$%
          & $\mp(59.3^{+24.8}_{-29.6})$%
          & $\mp(79.5^{+16.7}_{-24.4})$%
           \\
$\overline B{}^0\to \Xi^{0}\overline{\Xi^{0}}$
          & $-100\sim100$%
          & $-100\sim100$%
          & $-100\sim100$%
          & $\overline B{}^0\to \Sigma^{-}\overline{\Sigma^{-}}$ 
          & $0\pm47.1$%
          & $0\pm47.1$%
          & $0\pm47.1$%
           \\      
$\overline B{}^0\to \Xi^{-}\overline{\Xi^{-}}$
          & $0\pm32.1$%
          & $0\pm32.1$%
          & $0\pm32.1$%
          & $\overline B{}^0\to \Sigma^{0}\overline{\Lambda}$ 
          & $0\pm 13.3$%
          & $\mp(36.0^{+12.8}_{-12.1})$%
          & $\mp(49.3^{+12.1}_{-11.2})$%
           \\
$\overline B{}^0\to \Lambda\overline{\Lambda}$
          & $-100\sim100$%
          & $-100\sim100$%
          & $-100\sim100$%
          & $\overline B{}^0\to \Lambda\overline{\Sigma^{0}}$ 
          & $0\pm70.7$%
          & $0\pm70.7$%
          & $0\pm70.7$%
          \\                                                                                                                                                                                                
\end{tabular}
}
\\
\end{ruledtabular}
\end{table}

\subsubsection{$CP$ asymmetries of $\overline B_q\to \BB$ decays}

In Table~\ref{tab:AcpBBDS=0} we give results of direct $CP$ asymmetries of $\Delta S=0$, $\overline B_q\to\BB$ modes. 
The central values are the asymmetries generated from tree-penguin interference
where only the asymptotic amplitudes and Eq.~(\ref{eq: penguin strong phase}) are used.
We show results for $\phi=0$, $\pm\pi/4$ and $\pm\pi/2$.
The uncertainties are from  relaxing the asymptotic relation by using Eq.~(\ref{eq:correction0}) and varying the strong phases of $r^{(\prime)}_{t,i}$, $r^{(\prime)}_{p,i}$, $r^{(\prime)}_{ewp,i}$ and from sub-leading terms Eq.~(\ref{eq:correction2}) with strong phases from $\eta_{i,j,k}$.
Note that to satisfy the experimental $\overline B{}^0\to p\overline p$ and $B^-\to\Lambda\overline p$ rates, 
the sizes of $|r^{(\prime)}|$ are reduced by 60\% in all $\overline B_q\to\BB$ modes.

Now we discuss the $CP$ asymmetries of Group I, II, III modes, according to their rates and detectability as noted in Table~\ref{tab:rate}.
For the Group I mode, the $CP$ asymmetry of the  
$\overline B{}^0\to p\overline{p}$ decay 
can be as large as $\mp 49\%$.
For the Group II modes, 
$\A(\overline B{}^0\to \Sigma^{0}\overline{\Lambda})$ 
and $\A(\overline B{}^0_s\to p\overline{\Sigma^{+}})$ 
are similar to $\A(\overline B{}^0\to p\overline{p})$.
For the Group III modes, the $CP$ asymmetries of               
$B^-\to \Sigma^{0}\overline{\Sigma^{+}}$ 
and
$\overline B{}^0\to \Sigma^{0}\overline{\Sigma^{0}}$ decays 
are similar and can reach $\mp 80\%$, 
while the $CP$ asymmetry of the
$\overline B{}^0_s\to \Sigma^{0}\overline{\Xi^{0}}$ 
decay
is smaller then these two modes, but can still reach $\mp58\%$.
The $CP$ asymmetries of these modes basically all have the same sign when $|\phi|$ is large enough.

As noted in the previous subsection, 
$B^-\to \Sigma^{-}\overline{\Sigma^{0}}$, 
$B^-\to \Sigma^{-}\overline{\Lambda}$,
$B^-\to \Xi^{-}\overline{\Xi^{0}}$,
$\overline B{}^0\to \Xi^{-}\overline{\Xi^{-}}$,
$\overline B{}^0_s\to \Sigma^{-}\overline{\Xi^{-}}$ and
$\overline B{}^0\to \Sigma^{-}\overline{\Sigma^{-}}$ decays
do not have any tree ($T_{i\BB}$) contribution.
Although the $\overline B{}^0\to \Xi^{-}\overline{\Xi^{-}}$, 
$\overline B{}^0\to \Sigma^{-}\overline{\Sigma^{-}}$
and $\overline B{}^0_s\to \Sigma^{-}\overline{\Xi^{-}}$ decays are pure penguins modes, 
which only have $P_{i\BB}$, $P_{iEW\BB}$ and $PA_{\BB}$ contributions,
it is still possible for these modes to have sizable $CP$ asymmetries.
Since the sizes of $u$-penguin ($P^u$) and $c$-penguin ($P^c$) in $\Delta S=0$ modes are not very different, as their ratio can be estimated as
\be
\left|\frac{P^u}{P^c}\right|\simeq \left|\frac{V_{ub}V^*_{ud}}{V_{cb}V^*_{cd}}\right|\simeq 0.38,
\label{eq: PuPc}
\en
and they can have different strong phases to produce $CP$ asymmetries.~\footnote{Similarly the sizable $CP$ asymmetry of a pure penguin mode $\A(\overline B^0\to K^0\overline K^0)=-16.7^{+4.7+4.5+1.5+4.6}_{-3.7-5.1-1.7-3.6}\,\%$ as predicted in a QCD factorization (QCDF) calculation \cite{Beneke:2003zv} can be understood.}
For subleading modes, $\overline B{}^0\to \Xi^{0}\overline{\Xi^{0}}$ and
$\overline B{}^0\to \Sigma^{+}\overline{\Sigma^{+}}$ decays, which only have $E_{i\BB}$ and $PA_\BB$ contributions,
their $CP$ asymmetries can be any value.
Note that
$B^-\to \Lambda\overline{\Sigma^+}$,
 $\overline B{}^0_s\to \Lambda\overline{\Xi^0}$,
$\overline B{}^0\to \Lambda\overline{\Lambda}$ and
$\overline B{}^0\to \Lambda\overline{\Sigma^{0}}$ decay
with tree amplitudes $T_{i\BB}$ canceled in the asymptotic limits,
have have large uncertainties on $CP$ asymmetries, 
which mostly come from the corrections to the asymptotic relations.
Measuring $CP$ asymmetries of these modes can give information on the corrections to the asymptotic relations.

\begin{table}[t!]
\caption{\label{tab:AcpBBDS=-1} Same as Table~\ref{tab:AcpBBDS=0}, but with $\Delta S=-1$, $\overline B_q\to\BB$ modes.}
\begin{ruledtabular}
\centering
{
\footnotesize
\begin{tabular}{lccclccc}
Mode
          & $\phi=0$
          & $\phi=\pm\pi/4$
          & $\phi=\pm\pi/2$
          & Mode
          & $\phi=0$
          & $\phi=\pm\pi/4$
          & $\phi=\pm\pi/2$
          \\
\hline $B^-\to \Sigma^{0}\overline{p}$
          & $0\pm20.9$%
          & $\mp(28.9^{+22.1}_{-18.6})$%
          & $\mp(45.0^{+23.0}_{-16.7})$%
          & $\overline B{}^0\to \Sigma^{+}\overline{p}$ 
          & $0\pm16.4$%
          & $\pm(27.1^{+16.6}_{-13.6})$%
          & $\pm(35.3^{+15.5}_{-11.9})$%
           \\
$B^-\to \Sigma^{-}\overline{n}$
          & $0\pm1.8$%
          & $0\pm1.8$%
          & $0\pm1.8$%
          & $\overline B{}^0\to \Sigma^{0}\overline{n}$
          & $0\pm29.0$%
          & $\pm(47.8^{+23.8}_{-23.1})$%
          & $\pm(58.6^{+19.5}_{-17.8})$%
           \\ 
$B^-\to \Xi^{0}\overline{\Sigma^{+}}$ 
          & $0\pm2.6$%
          & $\pm(5.8^{+3.4}_{-2.3})$%
          & $\pm(8.1^{+3.9}_{-2.8})$%
          &$\overline B{}^0\to \Xi^{0}\overline{\Sigma^{0}}$ 
          & $0\pm2.3$%
          & $\pm(5.8^{+3.0}_{-2.1})$%
          & $\pm(8.1^{+3.6}_{-2.5})$%
           \\            
$B^-\to\Xi^{-}\overline{\Sigma^{0}}$ 
          & $0\pm1.8$%
          & $0\pm1.8$%
          & $0\pm1.8$%
          & $\overline B{}^0\to \Xi^{0}\overline{\Lambda}$ 
          & $0\pm24.5$%
          & $\pm(27.1^{+28.3}_{-16.2})$%
          & $\pm(35.3^{+27.2}_{-13.0})$%
           \\ 
$B^-\to \Xi^{-}\overline{\Lambda}$ 
          & $0\pm5.5$%
          & $0\pm5.5$%
          & $0\pm5.5$%
          & $\overline B{}^0\to  \Xi^{-}\overline{\Sigma^{-}}$ 
          & $0\pm1.5$%
          & $0\pm1.5$%
          & $0\pm1.5$%
           \\ 
$B^-\to \Lambda\overline{p}$ 
          & $0\pm4.8$%
          & $\pm(9.6^{+5.9}_{-4.0})$%
          & $\pm(13.2^{+6.5}_{-4.3})$%
          & $\overline B{}^0\to \Lambda\overline{n}$
          & $0\pm8.4$%
          & $\pm(18.7^{+9.6}_{-6.9})$%
          & $\pm(24.9^{+9.8}_{-6.6})$%
          \\                    
$\overline B{}^0_s\to p\overline{p}$
          & $-100\sim100$%
          & $-100\sim100$%
          & $-100\sim100$%
          & $\overline B{}^0_s\to \Sigma^{+}\overline{\Sigma^{+}}$ 
          & $0\pm21.0$%
          & $\pm(27.1^{+21.6}_{-16.7})$%
          & $\pm(35.3^{+20.1}_{-14.1})$%
          \\
$\overline B{}^0_s\to n\overline{n}$
          & $-100\sim100$%
          & $-100\sim100$%
          & $-100\sim100$%
          & $\overline B{}^0_s\to \Sigma^{0}\overline{\Sigma^{0}}$ 
          & $0\pm11.1$%
          & $\pm(14.2^{+12.5}_{-8.9})$%
          & $\pm(19.2^{+12.7}_{-8.2})$%
           \\
$\overline B{}^0_s\to \Xi^{0}\overline{\Xi^{0}}$ 
          & $0\pm8.4$%
          & $\pm(14.2^{+10.4}_{-6.5})$%
          & $\pm(19.2^{+11.1}_{-6.3})$%
          & $\overline B{}^0_s\to \Sigma^{-}\overline{\Sigma^{-}}$
          & $0\pm1.6$%
          & $0\pm1.6$%
          & $0\pm1.6$%
           \\ 
$\overline B{}^0_s\to \Xi^{-}\overline{\Xi^{-}}$ 
          & $0\pm2.5$%
          & $0\pm2.5$%
          & $0\pm2.5$%
          & $\overline B{}^0_s\to \Sigma^{0}\overline{\Lambda}$
          & $0\pm46.4$%
          & $\pm(74.3^{+21.7}_{-33.6})$%
          & $\pm(84.7^{+13.8}_{-19.7})$%
           \\    
$\overline B{}^0_s\to \Lambda\overline{\Lambda}$ 
          & $0\pm7.0$%
          & $\pm(14.2^{+8.3}_{-5.7})$%
          & $\pm(19.2^{+8.8}_{-5.6})$%
          & $\overline B{}^0_s\to \Lambda\overline{\Sigma^{0}}$
          & $0\pm42.8$%
          & $\pm(74.3^{+19.7}_{-33.2})$%
          & $\pm(84.7^{+12.5}_{-20.9})$%
           \\                                                                                                                                                                        
\end{tabular}
}
\end{ruledtabular}
\end{table}

In Table~\ref{tab:AcpBBDS=-1} we give results of direct $CP$ asymmetries of $\Delta S=-1$, $\overline B_q\to\BB$ modes. 
The $CP$ asymmetries of the Group I modes are as following.
The $CP$ asymmetries of the $B^-\to \Lambda\overline{p}$ 
and
$\overline B{}^0_s\to \Lambda\overline{\Lambda}$ 
decays are similar reaching $\pm 13\%$ and $\pm19\%$, respectively,
and their signs are opposite to 
the sign of $\A(\overline B{}^0\to p\overline p)$.
The $CP$ asymmetries of 
$\overline B{}^0_s\to \Xi^{-}\overline{\Xi^{-}}$ 
and
$B^-\to \Xi^{-}\overline{\Lambda}$ decays 
are vanishingly small. 
We will return to these modes later.  
From the table we see that for the Group II modes,
$\A(B^-\to\Xi^{-}\overline{\Sigma^{0}})$ 
is vanishingly, 
while    
$\A(\overline B{}^0\to \Xi^{0}\overline{\Lambda})$ 
and
$\A(\overline B{}^0\to \Sigma^{+}\overline{p})$ 
are similar and can be as large as $\pm35\%$.
For the Group III modes,  
$\A(\overline B{}^0_s\to \Sigma^{+}\overline{\Sigma^{+}})$ 
is the largest one reaching $\pm35\%$,    
$\A(\overline B{}^0_s\to \Xi^{0}\overline{\Xi^{0}})$ 
and
$\A(\overline B{}^0_s\to \Sigma^{0}\overline{\Sigma^{0}})$ 
are similar and can be as large as $\pm 19\%$,
while   
$\A(B^-\to \Xi^{0}\overline{\Sigma^{+}})$ 
and
$\A(\overline B{}^0\to \Xi^{0}\overline{\Sigma^{0}})$ 
are similar and are not sizable, but can reach $\pm8\%$,
Some of the above modes have rates of orders $10^{-7}$ (see Table~\ref{tab:BBDS=-1})
and with unsuppressed $CP$ asymmetries.

From Table~\ref{tab:AcpBBDS=-1}, we see that
the $CP$ asymmetries of 
$B^-\to \Sigma^{-}\overline{n}$,
$B^-\to\Xi^{-}\overline{\Sigma^{0}}$, 
$B^-\to \Xi^{-}\overline{\Lambda}$, 
$\overline B{}^0\to  \Xi^{-}\overline{\Sigma^{-}}$,
$\overline B{}^0_s\to \Sigma^{-}\overline{\Sigma^{-}}$ and
$\overline B{}^0_s\to \Xi^{-}\overline{\Xi^{-}}$ 
decays
have vanishing central values as they
do not have any tree ($T'_{i\BB}$) contribution.
In particular,
$\overline B{}^0\to  \Xi^{-}\overline{\Sigma^{-}}$,
$\overline B{}^0_s\to \Sigma^{-}\overline{\Sigma^{-}}$
and
$\overline B{}^0_s\to \Xi^{-}\overline{\Xi^{-}}$ decays are pure penguin modes,
which only have $P'_{i\BB}$, $P'_{iEW\BB}$ and $PA'_{\BB}$ terms.
Their $CP$ asymmetries are small. 
This can be understood as $P^{\prime u}$ is much smaller than $P^{\prime c}$. We can estimate the $CP$ asymmetry as following: 
\be
|\A|\simeq 
2 \left|\frac{P^{\prime u}}{P^{\prime c}}\right|
\sin\gamma\,
|\sin\delta|
\lesssim 
2\left|\frac{V_{ub}V^*_{us}}{V_{cb}V^*_{cs}}\right|
\sin\gamma
\simeq 3.8\%, 
\label{eq: A DeltaS penguin}
\en
where $\delta$ is the strong phase different of the penguins.~\footnote{It is useful to compare to the $CP$ asymmetry of the $\Delta S=1$ pure penguin $PP$ mode mode: 
$\A(\overline B{}^0_s\to K^0\overline K^0)=0.9^{+0.2+0.2+0.1+0.2}_{-0.2-0.2-0.2-0.3}\,\%$ in a QCDF calculation~\cite{Beneke:2003zv}.}
The smallness of the direct CP reflects the fact that in $\Delta S=-1$ decays, the $c$-penguin is much larger than the $u$-penguin as 
their CKM factors are off by about a factor of 50.
They can be tests of the Standard Model.
In particular, the $\overline B{}^0_s\to \Xi^{-}\overline{\Xi^{-}}$ decay being a Group I mode, 
can cascadely decays to all charged final states and with unsuppressed rate ($\sim 2\times 10^{-7}$, see Table~\ref{tab:BBDS=-1}).
The $CP$ asymmetry of this mode can be searched for.
Nevertheless the search is quite demanding as it requires tagging of $B_s$ and suffers from low efficiency of $\Xi^-$ recontruction. 
Subleading modes,
$\overline B{}^0_s\to p\overline{p}$ and
$\overline B{}^0_s\to n\overline{n}$ decays with $E'_{i\BB}$ and $PA'_{\BB}$ contributions
can have any value on their $CP$ asymmetries.

There are relations on 
direct $CP$ asymmetries between $\Delta S=0$ and $\Delta S=-1$ modes by using 
the so-called $U$-spin symmetry~\cite{Uspin, Uspin1}.
With
\be
\Delta_{CP}(\overline B_q\to f)\equiv \Gamma(\overline B_q\to f)-\Gamma(B_q\to \bar f)
=\frac{2}{\tau(B_q)}{\cal A}(\overline B_q\to f){\cal B}(\overline B_q\to f),
\en
we have~\cite{Chua:2013zga}
\be
\Delta_{CP}(B^-\to n\overline{p})
   &=&-\Delta_{CP}(B^-\to\Xi^{0}\overline{\Sigma^{+}}), 
\non\\
\Delta_{CP}(B^-\to\Xi^{-}\overline{\Xi^{0}})
   &=&-\Delta_{CP}(B^-\to\Sigma^{-}\overline{n}), 
\non\\
\Delta_{CP}(\bar B^0\to p\overline{p})
   &=&
-\Delta_{CP}(\bar B^0_s\to\Sigma^{+}\overline{\Sigma^{+}}), 
\non\\   
\Delta_{CP}(\bar B^0\to n\overline{n})
   &=&
-\Delta_{CP}(\bar B^0_s\to\Xi^{0}\overline{\Xi^{0}}), 
\non\\ 
\Delta_{CP}(\bar B^0\to\Sigma^{+}\overline{\Sigma^{+}})
   &=&-\Delta_{CP}(\bar B^0_s\to p\overline{p}), 
\non\\
\Delta_{CP}(\bar B^0\to\Sigma^{-}\overline{\Sigma^{-}})
   &=&-\Delta_{CP}(\bar B^0_s\to\Xi^{-}\overline{\Xi^{-}}),
\non\\      
\Delta_{CP}(\bar B^0\to\Xi^{0}\overline{\Xi^{0}})
   &=&-\Delta_{CP}(\bar B^0_s\to n\overline{n}), 
\non\\ 
\Delta_{CP}(\bar B^0\to\Xi^{-}\overline{\Xi^{-}})
   &=&-\Delta_{CP}(\bar B^0_s\to\Sigma^{-}\overline{\Sigma^{-}}), 
\non\\
\Delta_{CP}(\bar B^0_s\to p\overline{\Sigma^{+}})
   &=&-\Delta_{CP}(\bar B^0\to \Sigma^{+}\overline{p}), 
   \non\\
\Delta_{CP}(\bar B^0_s\to\Sigma^{-}\overline{\Xi^{-}})
   &=&-\Delta_{CP}(\bar B^0\to\Xi^{-}\overline{\Sigma^{-}}). 
   \label{eq: DCPBB}
\en
The minus signs in the above relations are from
$Im(V_{ub}V^*_{ud} V^*_{tb}V_{td})
=-Im(V_{ub}V^*_{us} V^*_{tb}V_{ts})$.
Note that these relations do not rely on the large $m_B$ limit, but are subjected to corrections from SU(3) breaking in the phase space factors and topological amplitudes. 
Some relations are satisfied trivially as all the related $CP$ asymmetries are always vanishing.
We can checked that the results shown in Tables~\ref{tab:AcpBBDS=0} and \ref{tab:AcpBBDS=-1} roughly satisfy these relations and the agreement can be improved when SU(3) breaking effects are taken into account.~\footnote{
Note that the values and signs of
$\Delta_{CP}(\bar B^0\to\Sigma^{+}\overline{\Sigma^{+}})$,
$\Delta_{CP}(\bar B^0\to\Xi^{0}\overline{\Xi^{0}})$,
$\Delta_{CP}(\bar B^0_s\to p\overline{p})$
and
$\Delta_{CP}(\bar B^0_s\to n\overline{n})$
cannot be read out from the tables. The relative signs of modes in the sixth, eighth and tenth relations cannot be read out from the tables.}
For example, using the first three relations of the above equation and the corresponding rates from Tables~\ref{tab:BBDS=0} and \ref{tab:BBDS=-1}, we have
\be
{\cal A}(B^-\to\Xi^{0}\overline{\Sigma^{+}})
&=&-{\cal A}(B^-\to n\overline{p}) 
   \frac{{\cal B}(B^-\to n\overline{p})}{{\cal B}(B^-\to\Xi^{0}\overline{\Sigma^{+}})}
\non\\
&\simeq&-(0.087) {\cal A}(B^-\to n\overline{p}),
\non\\
{\cal A}(B^-\to\Xi^{-}\overline{\Xi^{0}})
&=&-{\cal A}(B^-\to\Sigma^{-}\overline{n}) 
   \frac{{\cal B}(B^-\to\Sigma^{-}\overline{n})}{{\cal B}(B^-\to\Xi^{-}\overline{\Xi^{0}})}
\non\\
&\simeq&-(23.9) {\cal A}(B^-\to\Sigma^{-}\overline{n}).
\non\\
{\cal A}(\overline B{}^0_s\to\Sigma^+\overline{\Sigma^+})
&=&-{\cal A}(\overline B{}^0\to p\bar p)\frac{\tau(B^0_s)}{\tau(B^0)}
   \frac{{\cal B}(\overline B{}^0\to p\bar p)}{{\cal B}(\overline B{}^0_s\to\Sigma^+\overline{\Sigma^+})}
\non\\
&\simeq&-(0.8) {\cal A}(\overline B{}^0\to p\bar p),
\en
which are roughly satisfied compared to results of the $CP$ asymmetries in Tables~\ref{tab:AcpBBDS=0} and \ref{tab:AcpBBDS=-1}. Note that the rate ratios in the above relations are not fixed. For example, they can change with $\phi$. The values of the rate ratios used are just rough estimations using the center values of rates in Tables~\ref{tab:BBDS=0} and \ref{tab:BBDS=-1}. 

Some of these relations are useful to constrain the sizes of $CP$ asymmetries of $\Delta S=-1$ pure penguin modes model independently. From the sixth, eighth and tenth relations we have
\be
|\A(\bar B^0_s\to\Xi^{-}\overline{\Xi^{-}})|
&=&
|\A(\bar B^0\to\Sigma^{-}\overline{\Sigma^{-}})|
\frac{\tau(B^0_s)}{\tau(B^0)}
\frac{{\cal B}(\bar B^0\to\Sigma^{-}\overline{\Sigma^{-}})}{{\cal B}(\bar B^0_s\to\Xi^{-}\overline{\Xi^{-}})}
\non\\
&\leq& \frac{\tau(B^0_s)}{\tau(B^0)}
\frac{{\cal B}(\bar B^0\to\Sigma^{-}\overline{\Sigma^{-}})}{{\cal B}(\bar B^0_s\to\Xi^{-}\overline{\Xi^{-}})} 
\simeq 6.5\%,
\non\\  
|\A(\bar B^0_s\to\Sigma^{-}\overline{\Sigma^{-}})|
&=&
|\A(\bar B^0\to\Xi^{-}\overline{\Xi^{-}})|
\frac{\tau(B^0_s)}{\tau(B^0)}
\frac{{\cal B}(\bar B^0\to\Xi^{-}\overline{\Xi^{-}})}{{\cal B}(\bar B^0_s\to\Sigma^{-}\overline{\Sigma^{-}})}
\non\\
&\leq& \frac{\tau(B^0_s)}{\tau(B^0)}
\frac{{\cal B}(\bar B^0\to\Xi^{-}\overline{\Xi^{-}})}{{\cal B}(\bar B^0_s\to\Sigma^{-}\overline{\Sigma^{-}})}
\simeq 4.6\%,
\non\\  
|\A(\bar B^0\to\Xi^{-}\overline{\Sigma^{-}})|
&=&
|\A(\bar B^0_s\to\Sigma^{-}\overline{\Xi^{-}})|
\frac{\tau(B^0)}{\tau(B^0_s)}
\frac{{\cal B}(\bar B^0_s\to\Sigma^{-}\overline{\Xi^{-}})}{{\cal B}(\bar B^0\to\Xi^{-}\overline{\Sigma^{-}})}
\non\\
&\leq& \frac{\tau(B^0)}{\tau(B^0_s)}
\frac{{\cal B}(\bar B^0_s\to\Sigma^{-}\overline{\Xi^{-}})}{{\cal B}(\bar B^0\to\Xi^{-}\overline{\Sigma^{-}})}
\simeq 5.0\%.
\en 
We see from Table~\ref{tab:AcpBBDS=-1}  that the above inequalities are all satisfied.
Note that the constraining powers of these inequalities are similar to the one in Eq.~(\ref{eq: A DeltaS penguin}).

\begin{table}[t!]
\caption{\label{tab:AcpBDDS=0} Same as Table~\ref{tab:AcpBBDS=0}, but with $\Delta S=0$, $\overline B_q\to\BD$ modes.
}
\begin{ruledtabular}
\centering
{
\footnotesize
\begin{tabular}{lccclccc}
Mode
          & $\phi=0$
          & $\phi=\pm\pi/4$
          & $\phi=\pm\pi/2$
          & Mode
          & $\phi=0$
          & $\phi=\pm\pi/4$
          & $\phi=\pm\pi/2$
          \\
\hline $B^-\to p\overline{\Delta^{++}}$ 
          & $0\pm48.7$%
          & $\mp(36.0^{+49.3}_{-33.9})$%
          & $\mp(49.3^{+44.8}_{-22.0})$%
          & $\overline B{}^0_s\to p\overline{\Sigma^{*+}}$ 
          & $0\pm47.9$%
          & $\mp(36.0^{+48.6}_{-33.4})$%
          & $\mp(49.3^{+44.3}_{-21.9})$%
          \\
$B^-\to n\overline{\Delta^+}$
          & $0\pm19.9$%
          & $\pm(26.7^{+20.5}_{-17.3})$%
          & $\pm(38.8^{+20.1}_{-15.2})$%
          & $\overline B{}^0_s\to n\overline{\Sigma^{*0}}$
          & $0\pm19.6$%
          & $\pm(26.7^{+20.1}_{-17.1})$%
          & $\pm(38.8^{+19.7}_{-15.0})$%
           \\ 
$B^-\to\Sigma^0\overline{\Sigma^{*+}}$ 
          & $0\pm11.5$%
          & $\mp(20.8^{+12.4}_{-9.9})$%
          & $\mp(28.9^{+12.5}_{-9.0})$%
          &$\overline B{}^0_s\to \Sigma^{0}\overline{\Xi^{*0}}$ 
          & $0\pm11.4$%
          & $\mp(20.8^{+12.3}_{-9.8})$%
          & $\mp(28.9^{+12.4}_{-8.9})$%
           \\            
$B^-\to\Sigma^-\overline{\Sigma^{*0}}$
          & $0\pm35.7$%
          & $0\pm35.7$%
          & $0\pm35.7$%
          & $\overline B{}^0_s\to \Sigma^{-}\overline{\Xi^{*-}}$ 
          & $0\pm29.9$%
          & $0\pm29.9$%
          & $0\pm29.9$%
           \\ 
$B^-\to\Xi^{-}\overline{\Xi^{*0}}$
          & $0\pm35.7$%
          & $0\pm35.7$%
          & $0\pm35.7$%
          & $\overline B{}^0_s\to \Xi^{-}\overline{\Omega^-}$
          & $0\pm29.9$%
          & $0\pm29.9$%
          & $0\pm29.9$%
           \\ 
$B^-\to\Lambda\overline{\Sigma^{*+}}$
          & $0\pm100$%
          & $0\pm100$%
          & $0\pm100$%
          & $\overline B{}^0_s\to \Lambda\overline{\Xi^{*0}}$
          & $0\pm100$%
          & $0\pm100$%
          & $0\pm100$%
           \\ 
$\overline B{}^0\to p\overline{\Delta^+}$ 
          & $0\pm48.7$%
          & $\mp(36.0^{+49.3}_{-33.9})$%
          & $\mp(49.4^{+44.8}_{-22.0})$%
          & $\overline B{}^0\to \Sigma^{+}\overline{\Sigma^{*+}}$
          & $0$%
          & $0$%
          & $0$%
           \\
$\overline B{}^0\to n\overline{\Delta^0}$
          & $0\pm19.9$%
          & $\pm(26.7^{+20.5}_{-17.3})$%
          & $\pm(38.8^{+20.1}_{-15.2})$%
          & $\overline B{}^0\to \Sigma^{0}\overline{\Sigma^{*0}}$ 
          & $0\pm11.5$%
          & $\mp(20.8^{+12.4}_{-9.9})$%
          & $\mp(28.9^{+12.5}_{-9.0})$%
           \\
$\overline B{}^0\to \Xi^{0}\overline{\Xi^{*0}}$
          & $0$%
          & $0$%
          & $0$%
          & $\overline B{}^0\to \Sigma^{-}\overline{\Sigma^{*-}}$ 
          & $0\pm29.9$%
          & $0\pm29.9$%
          & $0\pm29.9$%
           \\      
$\overline B{}^0\to \Xi^{-}\overline{\Xi^{*-}}$
          & $0\pm29.9$%
          & $0\pm29.9$%
          & $0\pm29.9$%
          & $\overline B{}^0\to \Lambda\overline{\Sigma^{*0}}$
          & $0\pm100$%
          & $0\pm100$%
          & $0\pm100$%
           \\
\end{tabular}
}
\end{ruledtabular}
\end{table}

\subsubsection{$CP$ asymmetries of $\overline B_q\to \BD$ decays}

In Table~\ref{tab:AcpBDDS=0} we give results of direct $CP$ asymmetries of $\Delta S=0$, $\overline B_q\to\BD$ modes. 
The $CP$ asymmetries of Group I modes,
$B^-\to p\overline{\Delta^{++}}$ 
and
$\overline B{}^0_s\to p\overline{\Sigma^{*+}}$ 
decays are similar and can be as large as $\mp49\%$.
For Group II modes,
$\A(B^-\to\Sigma^0\overline{\Sigma^{*+}})$ 
and
$\A(\overline B{}^0_s\to \Sigma^{0}\overline{\Xi^{*0}})$ 
are similar and can be as large as $\mp29\%$,
while
$\A(\overline B{}^0\to p\overline{\Delta^+})$ 
is more sizable and is similar to the $CP$ asymmetries of 
$B^-\to p\overline{\Delta^{++}}$ 
and
$\overline B{}^0_s\to p\overline{\Sigma^{*+}}$ 
decays, reaching $\mp49\%$.
The $CP$ asymmetry of the Group III mode,
$\A(\overline B{}^0\to \Sigma^{0}\overline{\Sigma^{*0}})$ 
is similar to $\A(B^-\to\Sigma^0\overline{\Sigma^{*+}})$ 
and
$\A(\overline B{}^0_s\to \Sigma^{0}\overline{\Xi^{*0}})$, 
reaching $\mp29\%$.   
The $CP$ asymmetries of these modes basically all share the sign of $\A(\overline B{}^0\to p\overline p)$
for large enough $|\phi|$.

From Eqs. (\ref{eq: BDBm, DS=0}), (\ref{eq: BDB0, DS=0}) and (\ref{eq: BDBs, DS=0}),
we can easily identify the following relations
\be
\A(B^-\to\Xi^{-}\overline{\Xi^{*0}})
&=&\A(B^-\to\Sigma^-\overline{\Sigma^{*0}}),
   \non\\
\A(\bar B^0\to\Sigma^{-}\overline{\Sigma^{*-}})
   &=&\A(\bar B^0\to\Xi^{-}\overline{\Xi^{*-}})
   =\A(\bar B^0_s\to\Sigma^{-}\overline{\Xi^{*-}})
\non\\
   &=&\A(\bar B^0_s\to\Xi^{-}\overline{\Omega^-}), 
\non\\     
\A(\bar B^0\to\Sigma^{+}\overline{\Sigma^{*+}})
    &=&\A(\bar B^0\to\Xi^{0}\overline{\Xi^{*0}})=0.    
 \en  
We see from Table~\ref{tab:AcpBDDS=0} that these relations are satisfied. 
Note that these relations do not relay on the asymptotic limit, while the first relation is subjected to corrections from SU(3) breaking  in the topological amplitudes.

As shown in Table~\ref{tab:AcpBDDS=0}, 
the $CP$ asymmetries of               
$B^-\to\Sigma^-\overline{\Sigma^{*0}}$,
$B^-\to\Xi^{-}\overline{\Xi^{*0}}$,
$\overline B{}^0_s\to \Sigma^{-}\overline{\Xi^{*-}}$,
$\overline B{}^0\to \Xi^{-}\overline{\Xi^{*-}}$,
$\overline B{}^0\to \Sigma^{-}\overline{\Sigma^{*-}}$,
$\overline B{}^0_s\to \Xi^{-}\overline{\Omega^-}$,
$\overline B{}^0\to \Sigma^{+}\overline{\Sigma^{*+}}$ and
 $\overline B{}^0\to \Xi^{0}\overline{\Xi^{*0}}$ decays,
have vanishing central values,
as they do not have any tree ($T_{i\BD}$) contribution. 
In particular,
$\overline B{}^0\to \Xi^{-}\overline{\Xi^{*-}}$ and
$\overline B{}^0\to \Sigma^{-}\overline{\Sigma^{*-}}$
decays are $\Delta S=0$ pure penguin modes with only
$P_{\BD}$ and $P_{iEW\BD}$ contributions,
while
$\overline B{}^0\to \Sigma^{+}\overline{\Sigma^{*+}}$ and
$\overline B{}^0\to \Xi^{0}\overline{\Xi^{*0}}$ decays
are pure exchange modes with only $E_{\BD}$ contributions. 
The $CP$ asymmetries of the former two modes need not be vanishing [see discussion around Eq.~(\ref{eq: PuPc})], while the $CP$ asymmetries of the latter two modes are always vanishing.
Note that although
$B^-\to\Lambda\overline{\Sigma^{*+}}$,
$\overline B{}^0_s\to \Lambda\overline{\Xi^{*0}}$ and
$\overline B{}^0\to \Lambda\overline{\Sigma^{*0}}$ decays
have vanishing tree contributions in the asymptotic limit, resulting their $CP$ asymmetries to have vanishing center values, 
their uncertainties, mostly from corrections to the asymptotic relations, are sizable, however.  
Measuring the $CP$ asymmetries of these modes can give information on the breaking of the asymptotic relations.

\begin{table}[t!]
\caption{\label{tab:AcpBDDS=-1} Same as Table~\ref{tab:AcpBBDS=0}, but with $\Delta S=-1$, $\overline B_q\to\BD$ modes.
}
\begin{ruledtabular}
\centering
{
\footnotesize
\begin{tabular}{lccclccc}
Mode
          & $\phi=0$
          & $\phi=\pm\pi/4$
          & $\phi=\pm\pi/2$
          & Mode
          & $\phi=0$
          & $\phi=\pm\pi/4$
          & $\phi=\pm\pi/2$
          \\
\hline $B^-\to \Sigma^+\overline{\Delta^{++}}$ 
          & $0\pm30.3$%
          & $\pm(27.1^{+29.0}_{-25.9})$%
          & $\pm(35.3^{+26.4}_{-21.8})$%
          & $\overline B{}^0\to \Sigma^{+}\overline{\Delta^+}$ 
          & $0\pm30.1$%
          & $\pm(27.1^{+28.8}_{-25.7})$%
          & $\pm(35.3^{+26.2}_{-21.6})$%
          \\
$B^-\to \Sigma^0\overline{\Delta^+}$ 
          & $0\pm16.1$%
          & $\pm(14.2^{+17.1}_{-13.7})$%
          & $\pm(19.2^{+16.9}_{-12.6})$%
          & $\overline B{}^0\to \Sigma^{0}\overline{\Delta^0}$ 
          & $0\pm15.8$%
          & $\pm(14.2^{+16.8}_{-13.5})$%
          & $\pm(19.2^{+16.6}_{-12.3})$%
           \\ 
$B^-\to \Sigma^-\overline{\Delta^0}$
          & $0\pm1.8$%
          & $0\pm1.8$%
          & $0\pm1.8$%
          &$\overline B{}^0\to \Sigma^{-}\overline{\Delta^-}$ 
          & $0\pm1.5$%
          & $0\pm1.5$%
          & $0\pm1.5$%
           \\            
$B^-\to \Xi^{0}\overline{\Sigma^{*+}}$ 
          & $0\pm21.3$%
          & $\mp(28.9^{+22.9}_{-18.5})$%
          & $\mp(45.0^{+24.0}_{-16.3})$%
          & $\overline B{}^0\to \Xi^{0}\overline{\Sigma^{*0}}$ 
          & $0\pm21.0$%
          & $\mp(28.9^{+22.6}_{-18.2})$%
          & $\mp(45.0^{+23.7}_{-16.0})$%
           \\ 
$B^-\to \Xi^{-}\overline{\Sigma^{*0}}$ 
          & $0\pm1.8$%
          & $0\pm1.8$%
          & $0\pm1.8$%
          & $\overline B{}^0\to \Xi^{-}\overline{\Sigma^{*-}}$ 
          & $0\pm1.5$%
          & $0\pm1.5$%
          & $0\pm1.5$%
           \\ 
$B^-\to \Lambda\overline{\Delta^{+}}$
          & $0\pm52.9$%
          & $\pm(74.3^{+22.9}_{-41.7})$%
          & $\pm(84.7^{+14.4}_{-25.1})$%
          & $\overline B{}^0\to \Lambda\overline{\Delta^0}$
          & $0\pm52.9$%
          & $\pm(74.3^{+22.9}_{-41.7})$%
          & $\pm(84.7^{+14.4}_{-25.1})$%
          \\ 
$\overline B{}^0_s\to p\overline{\Delta^+}$
          & $0$%
          & $0$%
          & $0$%
          & $\overline B{}^0_s\to \Sigma^{+}\overline{\Sigma^{*+}}$ 
          & $0\pm30.3$%
          & $\pm(27.1^{+29.0}_{-25.9})$%
          & $\pm(35.3^{+26.4}_{-21.8})$%
           \\
$\overline B{}^0_s\to n\overline{\Delta^0}$
          & $0$%
          & $0$%
          & $0$%
          & $\overline B{}^0_s\to \Sigma^{0}\overline{\Sigma^{*0}}$ 
          & $0\pm16.0$%
          & $\pm(14.2^{+17.0}_{-13.6})$%
          & $\pm(19.2^{+16.7}_{-12.4})$%
           \\
$\overline B{}^0_s\to \Xi^{0}\overline{\Xi^{*0}}$ 
          & $0\pm21.3$%
          & $\mp(28.9^{+22.9}_{-18.5})$%
          & $\mp(45.0^{+24.0}_{-16.3})$%
          & $\overline B{}^0_s\to \Sigma^{-}\overline{\Sigma^{*-}}$ 
          & $0\pm1.5$%
          & $0\pm1.5$%
          & $0\pm1.5$%
           \\ 
$\overline B{}^0_s\to \Xi^{-}\overline{\Xi^{*-}}$ 
          & $0\pm1.5$%
          & $0\pm1.5$%
          & $0\pm1.5$%
          & $\overline B{}^0_s\to \Lambda\overline{\Sigma^{*0}}$
          & $0\pm53.6$%
          & $\pm(74.3^{+23.1}_{-42.3})$%
          & $\pm(84.7^{+14.5}_{-25.4})$%
           \\ 
\end{tabular}
}
\end{ruledtabular}
\end{table}

In Table~\ref{tab:AcpBDDS=-1} we give results of direct $CP$ asymmetries of $\Delta S=-1$, $\overline B_q\to\BD$ modes.  
The $CP$ asymmetry of the Group I mode
 $\overline B{}^0\to \Xi^{-}\overline{\Sigma^{*-}}$ 
is vanishing. We will return to this later.
For Group II modes,
$\A(B^-\to \Sigma^+\overline{\Delta^{++}})$ 
and
$\A(\overline B{}^0_s\to \Sigma^{+}\overline{\Sigma^{*+}})$ 
are similar reaching $\pm35\%$,
$\A(B^-\to \Xi^{0}\overline{\Sigma^{*+}})$ 
and
$\A(\overline B{}^0_s\to \Xi^{0}\overline{\Xi^{*0}})$ 
are similar and sizable reaching $\mp45\%$, 
but with sign opposite to most modes in Table~\ref{tab:AcpBDDS=-1},
$\A(\overline B{}^0\to \Sigma^{0}\overline{\Delta^0})$ 
can reach $\pm19\%$,
while
$\A(\overline B{}^0_s\to \Xi^{-}\overline{\Xi^{*-}})$ 
and
$\A(B^-\to \Xi^{-}\overline{\Sigma^{*0}})$ 
are vanishingly small. 
In the Group III modes,
$\A(\overline B{}^0\to \Xi^{0}\overline{\Sigma^{*0}})$ 
is the largest one reaching $\mp45\%$ and is similar to 
$\A(B^-\to \Xi^{0}\overline{\Sigma^{*+}})$ 
and
$\A(\overline B{}^0_s\to \Xi^{0}\overline{\Xi^{*0}})$, 
$\A(\overline B{}^0\to \Sigma^{+}\overline{\Delta^+})$ 
can be as large as $\pm35\%$,
while
$\A(B^-\to \Sigma^0\overline{\Delta^+})$ 
and
$\A(\overline B{}^0_s\to \Sigma^{0}\overline{\Sigma^{*0}})$ 
are similar and can be as large as $\pm19\%$.
Note that for large enough $|\phi|$ most modes in the Table have signs basically equal to the sign of $\A(B^-\to\Lambda \overline p)$.

From Eqs. (\ref{eq: BDBm, DS=-1}), (\ref{eq: BDB0, DS=-1}) and (\ref{eq: BDBs, DS=-1}),
we can easily identify the following relations
\be
      \A(B^-\to\Xi^{-}\overline{\Sigma^{*0}})   
   &=&\A(B^-\to\Sigma^-\overline{\Delta^0}), 
\non\\
   \A(\bar B^0_s\to\Sigma^{-}\overline{\Sigma^{*-}})   
   &=&\A(\bar B^0_s\to\Xi^{-}\overline{\Xi^{*-}})
     =\A(\bar B^0\to\Xi^{-}\overline{\Sigma^{*-}})
\non\\
   &=&\A(\bar B^0\to\Sigma^{-}\overline{\Delta^-}), 
   \non\\      
       \A(\bar B^0_s\to p\overline{\Delta^+})
   &=&\A(\bar B^0_s\to n\overline{\Delta^0})=0.
\en
We see from Table~\ref{tab:AcpBDDS=-1} that these relations are satisfied. 
Note that these relations do not relay on the asymptotic limit, but are subjected to SU(3) breaking in the topological amplitudes.  

As shown in Table~\ref{tab:AcpBDDS=-1}, 
the $CP$ asymmetries of
$B^-\to \Sigma^-\overline{\Delta^0}$,
$B^-\to \Xi^{-}\overline{\Sigma^{*0}}$,
$\overline B{}^0\to \Sigma^{-}\overline{\Delta^-}$,
$\overline B{}^0\to \Xi^{-}\overline{\Sigma^{*-}}$, 
$\overline B{}^0_s\to \Sigma^{-}\overline{\Sigma^{*-}}$,
$\overline B{}^0_s\to \Xi^{-}\overline{\Xi^{*-}}$, 
$\overline B{}^0_s\to p\overline{\Delta^+}$ and
$\overline B{}^0_s\to n\overline{\Delta^0}$ decays
have vanishing central values,
as they do not have any tree ($T'_{i\BD}$) contribution. 
In particular,
$\overline B{}^0\to \Sigma^{-}\overline{\Delta^-}$,
$\overline B{}^0\to \Xi^{-}\overline{\Sigma^{*-}}$, 
$\overline B{}^0_s\to \Sigma^{-}\overline{\Sigma^{*-}}$ and
$\overline B{}^0_s\to \Xi^{-}\overline{\Xi^{*-}}$ 
decays are pure penguin modes with only
$P'_{\BD}$ and $P'_{iEW\BD}$ contributions,
while
$\overline B{}^0_s\to p\overline{\Delta^+}$ and
$\overline B{}^0_s\to n\overline{\Delta^0}$
are pure exchange modes with only $E'_{\BD}$ contributions.
The $CP$ asymmetries of the $\Delta S=-1$ pure penguin modes are small [see Eq.~(\ref{eq: A DeltaS penguin})], while the CP asymmetries of pure exchange modes are always vanishing.
These can be tests of the Standard Model.
In particular, the $\overline B{}^0\to \Xi^{-}\overline{\Sigma^{*-}}$ decay
can cascadely decay to all charged final states and have unsuppressed decay rate (see Table~\ref{tab:BDDS=-1}),
but may suffer from low reconstruction efficiencies of the final state particles.
Nevertheless, it is still interesting to search for this modes and its $CP$ asymmetry.

The $U$-spin relations for octet-antidecuplet modes are given by~\cite{Chua:2013zga}
\be
\Delta_{CP}(B^-\to n\overline{\Delta^+})
   &=&-\Delta_{CP}(B^-\to\Xi^{0}\overline{\Sigma^{*+}}),
\non\\      
\Delta_{CP}(B^-\to\Xi^{-}\overline{\Xi^{*0}})
   &=&2 \Delta_{CP}(B^-\to\Sigma^-\overline{\Sigma^{*0}})
      =-2\Delta_{CP}(B^-\to\Xi^{-}\overline{\Sigma^{*0}})
\non\\   
   &=&-\Delta_{CP}(B^-\to\Sigma^-\overline{\Delta^0}), 
\non\\
\Delta_{CP}(B^-\to p\overline{\Delta^{++}})
   &=&-
\Delta_{CP}(B^-\to \Sigma^+\overline{\Delta^{++}}),
   \non\\   
\Delta_{CP}(\bar B^0\to n\overline{\Delta^0})
   &=&-\Delta_{CP}(\bar B^0_s\to\Xi^{0}\overline{\Xi^{*0}}),
\non\\   
3\Delta_{CP}(\bar B^0\to\Sigma^{-}\overline{\Sigma^{*-}})
   &=&3 \Delta_{CP}(\bar B^0\to\Xi^{-}\overline{\Xi^{*-}})
   =3 \Delta_{CP}(\bar B^0_s\to\Sigma^{-}\overline{\Xi^{*-}})
\non\\   
   &=&\Delta_{CP}(\bar B^0_s\to\Xi^{-}\overline{\Omega^-})
   = -3 \Delta_{CP}(\bar B^0_s\to\Sigma^{-}\overline{\Sigma^{*-}})
\non\\   
   &=&-3 \Delta_{CP}(\bar B^0_s\to\Xi^{-}\overline{\Xi^{*-}})
     =-3 \Delta_{CP}(\bar B^0\to\Xi^{-}\overline{\Sigma^{*-}})
\non\\
   &=&-\Delta_{CP}(\bar B^0\to\Sigma^{-}\overline{\Delta^-}), 
   \non\\      
\Delta_{CP}(\bar B^0\to\Sigma^{+}\overline{\Sigma^{*+}})
    &=&\Delta_{CP}(\bar B^0\to\Xi^{0}\overline{\Xi^{*0}})
       =-\Delta_{CP}(\bar B^0_s\to p\overline{\Delta^+})
\non\\
   &=&-\Delta_{CP}(\bar B^0_s\to n\overline{\Delta^0})=0,
   \non\\   
\Delta_{CP}(\bar B^0\to p\overline{\Delta^+})
   &=&-\Delta_{CP}(\bar B^0_s\to\Sigma^{+}\overline{\Sigma^{*+}}),
\non\\
\Delta_{CP}(\bar B^0_s\to n\overline{\Sigma^{*0}})
    &=&-\Delta_{CP}(\bar B^0\to\Xi^{0}\overline{\Sigma^{*0}})
\non\\
\Delta_{CP}(\bar B^0_s\to p\overline{\Sigma^{*+}})
    &=&-\Delta_{CP}(\bar B^0\to \Sigma^{+}\overline{\Delta^+}).
\en
These relations are subjected to corrections from SU(3) breaking in the $|p_{cm}|^3$ factors and topological amplitudes.
The above relations are roughly satisfied by the results shown in Tables~\ref{tab:AcpBDDS=0} and \ref{tab:AcpBDDS=-1}
and the agreement can be improved when SU(3) breaking effects are taken into account.
One can make a quick but non-trivial check on the relative signs of these asymmetries 
and see that they are indeed in agreement with the above relations.~\footnote{Note that for the second relation, the $CP$ asymmetries of these modes all have vanishing central values, their relative signs cannot be read out from Tables~\ref{tab:AcpBDDS=0} and \ref{tab:AcpBDDS=-1}.}
Note that several vanishing $CP$ asymmetries as discussed previously are related to each other.
The fifth relation can be used to constrain the sizes of $CP$ asymmetries of $\Delta S=-1$ pure penguin modes model independently. For example, from the relation we can have 
\be
|\A(\bar B^0\to\Xi^{-}\overline{\Sigma^{*-}})|
&=&
\frac{1}{3}|\A(\bar B^0_s\to\Xi^{-}\overline{\Omega^-})|
\frac{\tau(B^0)}{\tau(B^0_s)}
\frac{{\cal B}(\bar B^0_s\to\Xi^{-}\overline{\Omega^-})}{{\cal B}(\bar B^0\to\Xi^{-}\overline{\Sigma^{*-}})}
\non\\
&\leq& \frac{1}{3}\frac{\tau(B^0)}{\tau(B^0_s)}
\frac{{\cal B}(\bar B^0_s\to\Xi^{-}\overline{\Omega^-})}{{\cal B}(\bar B^0\to\Xi^{-}\overline{\Sigma^{*-}})}
\simeq 4.1\%,
\en
which is satisfied by $\A(\bar B^0\to\Xi^{-}\overline{\Sigma^{*-}})$ shown in Table~\ref{tab:AcpBDDS=-1}. 
Note that the above inequality are generic and can be tested experimentally as the quantities on the both sides are measurable.

\begin{table}[t!]
\caption{\label{tab:AcpDBDS=0} Same as Table~\ref{tab:AcpBBDS=0}, but with $\Delta S=0$, $\overline B_q\to\DB$ modes.
}
\begin{ruledtabular}
\centering
{
\footnotesize
\begin{tabular}{lccclccc}
Mode
          & $\phi=0$
          & $\phi=\pm\pi/4$
          & $\phi=\pm\pi/2$
          & Mode
          & $\phi=0$
          & $\phi=\pm\pi/4$
          & $\phi=\pm\pi/2$
          \\
\hline $B^-\to \Delta^0\overline{p}$ 
          & $0\pm18.5$%
          & $\pm(26.7^{+19.1}_{-16.0})$%
          & $\pm(38.8^{+18.7}_{-14.0})$%
          & $\overline B{}^0_s\to \Delta^+\overline{\Sigma^{+}}$ 
          & $0\pm19.0$%
          & $\mp(36.0^{+18.8}_{-16.2})$%
          & $\mp(49.3^{+17.9}_{-14.2})$%
           \\
$B^-\to \Delta^-\overline{n}$
          & $0\pm35.7$%
          & $0\pm35.7$%
          & $0\pm35.7$%
          & $\overline B{}^0_s\to \Delta^0\overline{\Sigma^{0}}$ 
          & $0\pm30.9$%
          & $\mp(59.3^{+23.6}_{-25.6})$%
          & $\mp(79.5^{+16.2}_{-19.3})$%
           \\ 
$B^-\to \Sigma^{*0}\overline{\Sigma^{+}}$ 
          & $0\pm18.5$%
          & $\pm(26.7^{+19.1}_{-16.0})$%
          & $\pm(38.8^{+18.7}_{-14.0})$%
          &$\overline B{}^0_s\to \Delta^-\overline{\Sigma^{-}}$ 
          & $0\pm29.9$%
          & $0\pm29.9$%
          & $0\pm29.9$%
           \\            
$B^-\to \Sigma^{*-}\overline{\Sigma^{0}}$
          & $0\pm35.7$%
          & $0\pm35.7$%
          & $0\pm35.7$%
          & $\overline B{}^0_s\to \Sigma^{*0}\overline{\Xi^{0}}$ 
          & $0\pm11.0$%
          & $\mp(20.8^{+11.8}_{-9.5})$%
          & $\mp(28.9^{+12.0}_{-8.6})$%
           \\ 
$B^-\to \Xi^{*-}\overline{\Xi^{0}}$
          & $0\pm35.7$%
          & $0\pm35.7$%
          & $0\pm35.7$%
          & $\overline B{}^0_s\to \Sigma^{*-}\overline{\Xi^{-}}$ 
          & $0\pm29.9$%
          & $0\pm29.9$%
          & $0\pm29.9$%
           \\ 
$B^-\to \Sigma^{*-}\overline{\Lambda}$
          & $0\pm35.7$%
          & $0\pm35.7$%
          & $0\pm35.7$%
          & $\overline B{}^0_s\to \Delta^0\overline{\Lambda}$ 
          & $0\pm2.3$%
          & $\mp(4.4^{+2.6}_{-2.0})$%
          & $\mp(6.2^{+2.7}_{-1.9})$%
           \\ 
$\overline B{}^0\to \Delta^+\overline{p}$ 
          & $0\pm19.4$%
          & $\mp(36.0^{+19.3}_{-16.5})$%
          & $\mp(49.3^{+18.3}_{-14.4})$%
          & $\overline B{}^0\to \Sigma^{*+}\overline{\Sigma^{+}}$
          & $0$%
          & $0$%
          & $0$%
           \\
$\overline B{}^0\to \Delta^0\overline{n}$
          & $0\pm11.1$%
          & $\mp(20.8^{+12.0}_{-9.6})$%
          & $\mp(28.9^{+12.1}_{-8.7})$%
          & $\overline B{}^0\to \Sigma^{*0}\overline{\Sigma^{0}}$
          & $0\pm18.5$%
          & $\pm(26.7^{+19.1}_{-16.0})$%
          & $\pm(38.8^{+18.7}_{-14.0})$%
           \\
$\overline B{}^0\to \Xi^{*0}\overline{\Xi^{0}}$
          & $0$%
          & $0$%
          & $0$%
          & $\overline B{}^0\to \Sigma^{*-}\overline{\Sigma^{-}}$ 
          & $0\pm29.9$%
          & $0\pm29.9$%
          & $0\pm29.9$%
           \\      
$\overline B{}^0\to \Xi^{*-}\overline{\Xi^{-}}$
          & $0\pm29.9$%
          & $0\pm29.9$%
          & $0\pm29.9$%
          & $\overline B{}^0\to \Sigma^{*0}\overline{\Lambda}$ 
          & $0\pm19.4$%
          & $\mp(36.0^{+19.3}_{-16.5})$%
          & $\mp(49.3^{+18.3}_{-14.4})$%
          \\
\end{tabular}
}
\end{ruledtabular}
\end{table}

\subsubsection{$CP$ asymmetries of $\overline B_q\to \DB$ decays}

In Table~\ref{tab:AcpDBDS=0} we give results of direct $CP$ asymmetries of $\Delta S=0$, $\overline B_q\to\DB$ modes. 
The $CP$ asymmetries of two Group I modes,
$B^-\to \Delta^0\overline{p}$ 
and
$\overline B{}^0_s\to \Delta^0\overline{\Lambda}$ 
decays, 
are opposite in signs, while the former can reach $\pm39\%$
and is more sizable than the latter.
For Group II modes,      
$\A(\overline B{}^0\to \Delta^+\overline{p})$ 
and
$\A(\overline B{}^0\to \Sigma^{*0}\overline{\Lambda})$ 
are similar reaching $\mp49\%$,
while
$\A(\overline B{}^0_s\to \Delta^0\overline{\Sigma^{0}})$ 
can be as large as $\mp80\%$.
For Group III modes,       
$\A(\overline B{}^0_s\to \Sigma^{*0}\overline{\Xi^{0}})$ 
and
$\A(\overline B{}^0_s\to \Delta^+\overline{\Sigma^{+}})$ 
can reach $\mp29\%$ and $\mp49\%$, respectively, 
and have the same sign, 
while
$\A(B^-\to \Sigma^{*0}\overline{\Sigma^{+}})$ 
can be as large as $\pm39\%$, but with sign opposites to the other two's.

From Eqs. (\ref{eq: DBBm, DS=0}), (\ref{eq: DBB0, DS=0}) and (\ref{eq: DBBs, DS=0}),
we can easily identify the following relations
\be
\A(B^-\to\Delta^0\overline{p})
   &=&
     \A(B^-\to\Sigma^{*0}\overline{\Sigma^{+}}),
   \non\\
\A(B^-\to\Delta^-\overline{n})
   &=& 
   \A(B^-\to\Xi^{*-}\overline{\Xi^{0}})
  =\A(B^-\to\Sigma^{*-}\overline{\Sigma^{0}})
\non\\
   &=&\A(B^-\to\Sigma^{*-}\overline{\Lambda}),
\non\\
\A(\bar B^0\to\Sigma^{*-}\overline{\Sigma^{-}})
  & =&\A(\bar B^0\to\Xi^{*-}\overline{\Xi^{-}})
   =\A(\bar B^0_s\to\Delta^-\overline{\Sigma^{-}})
   \non\\
   &=&\A(\bar B^0_s\to\Sigma^{*-}\overline{\Xi^{-}}), 
\non\\ 
\A(\bar B^0\to\Sigma^{*+}\overline{\Sigma^{+}})
   &=&\A(\bar B^0\to\Xi^{*0}\overline{\Xi^{0}})=0.    
 \en  
We see from Table~\ref{tab:AcpDBDS=0} that these relations are satisfied. 
Note that these relations do not relay on the asymptotic limit, 
but are subjected to SU(3) breaking in the topological amplitudes.

Note that the central values of the $CP$ asymmetries of
$B^-\to \Delta^-\overline{n}$,
$B^-\to \Sigma^{*-}\overline{\Sigma^{0}}$,
$B^-\to \Xi^{*-}\overline{\Xi^{0}}$,
$B^-\to \Sigma^{*-}\overline{\Lambda}$,
$\overline B{}^0\to \Xi^{*-}\overline{\Xi^{-}}$,
$\overline B{}^0_s\to \Delta^-\overline{\Sigma^{-}}$,
$\overline B{}^0_s\to \Sigma^{*-}\overline{\Xi^{-}}$,
$\overline B{}^0\to \Sigma^{*-}\overline{\Sigma^{-}}$,
$\overline B{}^0\to \Xi^{*0}\overline{\Xi^{0}}$ and
$\overline B{}^0\to \Sigma^{*+}\overline{\Sigma^{+}}$ decays
are vanishing,
as they do not have any tree ($T_{i\DB}$) contribution. 
The 
$\overline B{}^0\to \Xi^{*-}\overline{\Xi^{-}}$,
$\overline B{}^0_s\to \Delta^-\overline{\Sigma^{-}}$,
$\overline B{}^0_s\to \Sigma^{*-}\overline{\Xi^{-}}$ and
$\overline B{}^0\to \Sigma^{*-}\overline{\Sigma^{-}}$ decays
are pure penguin modes, which only have $P_\DB$ and $P_{i EW\DB}$ contributions,
while 
$\overline B{}^0\to \Xi^{*0}\overline{\Xi^{0}}$ and
$\overline B{}^0\to \Sigma^{*+}\overline{\Sigma^{+}}$ decays
are pure exchange ($E_\DB$) modes,
the $CP$ asymmetries of the $\Delta S=0$ pure penguin modes need not be vanishing [see Eq.~(\ref{eq: PuPc})], while the $CP$ asymmetries of the pure exchange modes are always vanishing.

\begin{table}[t!]
\caption{\label{tab:AcpDBDS=-1} Same as Table~\ref{tab:AcpBBDS=0}, but with $\Delta S=-1$, $\overline B_q\to\DB$ modes.}
\begin{ruledtabular}
\centering
{
\footnotesize
\begin{tabular}{lccclccc}
Mode
          & $\phi=0$
          & $\phi=\pm\pi/4$
          & $\phi=\pm\pi/2$
          & Mode
          & $\phi=0$
          & $\phi=\pm\pi/4$
          & $\phi=\pm\pi/2$
          \\
\hline $B^-\to \Sigma^{*0}\overline{p}$ 
          & $0\pm19.4$%
          & $\mp(28.9^{+20.6}_{-17.2})$%
          & $\mp(45.0^{+21.5}_{-15.4})$%
          & $\overline B{}^0\to \Sigma^{*+}\overline{p}$ 
          & $0\pm14.8$%
          & $\pm(27.1^{+15.2}_{-12.3})$%
          & $\pm(35.3^{+14.2}_{-10.8})$%
           \\
$B^-\to \Sigma^{*-}\overline{n}$
          & $0\pm1.8$%
          & $0\pm1.8$%
          & $0\pm1.8$%
          & $\overline B{}^0\to \Sigma^{*0}\overline{n}$
          & $0\pm26.4$%
          & $\pm(47.8^{+22.0}_{-20.8})$%
          & $\pm(58.6^{+18.1}_{-16.1})$%
           \\ 
$B^-\to \Xi^{*0}\overline{\Sigma^{+}}$ 
          & $0\pm19.4$%
          & $\mp(28.9^{+20.6}_{-17.2})$%
          & $\mp(45.0^{+21.5}_{-15.4})$%
          &$\overline B{}^0\to \Xi^{*0}\overline{\Sigma^{0}}$
          & $0\pm19.2$%
          & $\mp(28.9^{+20.4}_{-16.9})$%
          & $\mp(45.0^{+21.3}_{-15.2})$%
           \\            
$B^-\to \Xi^{*-}\overline{\Sigma^{0}}$ 
          & $0\pm1.8$%
          & $0\pm1.8$%
          & $0\pm1.8$%
          & $\overline B{}^0\to \Xi^{*-}\overline{\Sigma^{-}}$ 
          & $0\pm1.5$%
          & $0\pm1.5$%
          & $0\pm1.5$%
           \\ 
$B^-\to \Omega^-\overline{\Xi^{0}}$ 
          & $0\pm1.8$%
          & $0\pm1.8$%
          & $0\pm1.8$%
          & $\overline B{}^0\to \Omega^-\overline{\Xi^{-}}$ 
          & $0\pm1.5$%
          & $0\pm1.5$%
          & $0\pm1.5$%
           \\ 
$B^-\to \Xi^{*-}\overline{\Lambda}$ 
          & $0\pm1.8$%
          & $0\pm1.8$%
          & $0\pm1.8$%
          & $\overline B{}^0\to \Xi^{*0}\overline{\Lambda}$ 
          & $0\pm14.8$%
          & $\pm(27.1^{+15.2}_{-12.3})$%
          & $\pm(35.3^{+14.2}_{-10.8})$%
           \\ 
$\overline B{}^0_s\to \Delta^+\overline{p}$
          & $0$%
          & $0$%
          & $0$%
          & $\overline B{}^0_s\to \Sigma^{*+}\overline{\Sigma^{+}}$ 
          & $0\pm15.1$%
          & $\pm(27.1^{+15.5}_{-12.5})$%
          & $\pm(35.3^{+14.5}_{-11.0})$%
           \\
$\overline B{}^0_s\to \Delta^0\overline{n}$
          & $0$%
          & $0$%
          & $0$%
          & $\overline B{}^0_s\to \Sigma^{*0}\overline{\Sigma^{0}}$ 
          & $0\pm7.9$%
          & $\pm(14.2^{+8.8}_{-6.7})$%
          & $\pm(19.2^{+9.0}_{-6.5})$%
           \\
$\overline B{}^0_s\to \Xi^{*0}\overline{\Xi^{0}}$ 
          & $0\pm26.7$%
          & $\pm(47.8^{+22.1}_{-21.0})$%
          & $\pm(58.6^{+18.2}_{-16.2})$%
          & $\overline B{}^0_s\to \Sigma^{*-}\overline{\Sigma^{-}}$ 
          & $0\pm1.5$%
          & $0\pm1.5$%
          & $0\pm1.5$%
           \\ 
$\overline B{}^0_s\to \Xi^{*-}\overline{\Xi^{-}}$ 
          & $0\pm1.5$%
          & $0\pm1.5$%
          & $0\pm1.5$%
          & $\overline B{}^0_s\to \Sigma^{*0}\overline{\Lambda}$
          & $0\pm41.6$%
          & $\pm(74.3^{+20.0}_{-29.8})$%
          & $\pm(84.7^{+12.8}_{-17.9})$%
           \\                                                                                                                                          
\end{tabular}
}
\end{ruledtabular}
\end{table}

In Table~\ref{tab:AcpDBDS=-1} we give results of direct $CP$ asymmetries of $\Delta S=-1$, $\overline B_q\to\DB$ modes. 
In the Group I modes,
$\A(\overline B{}^0\to \Xi^{*0}\overline{\Lambda)}$ 
and
$\A(\overline B{}^0\to \Sigma^{*+}\overline{p})$ 
are similar and sizable reaching $\pm35\%$,
while
$\A(\overline B{}^0\to \Omega^-\overline{\Xi^{-}})$ 
is vanishing and will be discussed later.
For Group II modes,
$\A(\overline B{}^0_s\to \Xi^{*0}\overline{\Xi^{0}})$, 
$\A(\overline B{}^0_s\to \Sigma^{*+}\overline{\Sigma^{+}})$, 
$\A(B^-\to \Xi^{*0}\overline{\Sigma^{+}})$ 
and
$\A(B^-\to \Sigma^{*0}\overline{p})$ 
are sizable reaching $\pm59\%$, $\pm35\%$, $\mp45\%$ and $\mp45\%$, respectively, 
and the signs of the first two asymmetries are opposite to the last two, 
while     
$\A(B^-\to \Omega^-\overline{\Xi^{0}})$, 
$\A(B^-\to \Xi^{*-}\overline{\Lambda})$ 
and
$\A(\overline B{}^0_s\to \Xi^{*-}\overline{\Xi^{-}})$ 
are vanishingly small and will be discussed later.   
For the Group III modes,
$\A(\overline B{}^0_s\to \Sigma^{*0}\overline{\Sigma^{0}})$ 
can be as large as $\pm 19\%$, but
$\A(B^-\to \Xi^{*-}\overline{\Sigma^{0}})$ 
is highly suppressed.

From Eqs. (\ref{eq: BDBm, DS=-1}), (\ref{eq: BDB0, DS=-1}) and (\ref{eq: BDBs, DS=-1}),
we can easily identify the following relations
\be
   \A(B^-\to\Sigma^{*0}\overline{p})
   &=&\A(B^-\to\Xi^{*0}\overline{\Sigma^{+}}),
   \non\\
\A(B^-\to\Sigma^{*-}\overline{n})
   &=&
\A(B^-\to\Xi^{*-}\overline{\Sigma^{0}})
   =  
\A(B^-\to\Omega^-\overline{\Xi^{0}})
\non\\
    &=&
\A(B^-\to\Xi^{*-}\overline{\Lambda}),   
\non\\
\A(\bar B^0_s\to\Sigma^{*-}\overline{\Sigma^{-}})
   &=&\A(\bar B^0_s\to\Xi^{*-}\overline{\Xi^{-}})
   =\A(\bar B^0\to\Xi^{*-}\overline{\Sigma^{-}})
\non\\   
   &=& \A(\bar B^0\to\Omega^-\overline{\Xi^{-}}), 
\non\\
\A(\bar B^0_s\to\Delta^+\overline{p})
   &=&\A(\bar B^0_s\to\Delta^0\overline{n})=0.
\en
We see from Table~\ref{tab:AcpDBDS=-1} that these relations are satisfied. 
Note that these relations do not relay on the asymptotic limit, but are subjected to SU(3) breaking in the topological amplitudes.  

The central values of $CP$ asymmetries of 
$B^-\to \Sigma^{*-}\overline{n}$, 
$B^-\to \Xi^{*-}\overline{\Sigma^{0}}$, 
$B^-\to \Omega^-\overline{\Xi^{0}}$, 
$B^-\to \Xi^{*-}\overline{\Lambda}$, 
$\overline B{}^0\to \Omega^-\overline{\Xi^{-}}$, 
$\overline B{}^0_s\to \Sigma^{*-}\overline{\Sigma^{-}}$,
$\overline B{}^0\to \Xi^{*-}\overline{\Sigma^{-}}$,
$\overline B{}^0_s\to \Xi^{*-}\overline{\Xi^{-}}$, 
$\overline B{}^0_s\to \Delta^+\overline{p}$ and
$\overline B{}^0_s\to \Delta^0\overline{n}$ decays
are vanishing,
as they do not have any tree ($T'_{i\DB}$) contribution. 
Since
$\overline B{}^0\to \Omega^-\overline{\Xi^{-}}$, 
$\overline B{}^0_s\to \Sigma^{*-}\overline{\Sigma^{-}}$,
$\overline B{}^0\to \Xi^{*-}\overline{\Sigma^{-}}$ and
$\overline B{}^0_s\to \Xi^{*-}\overline{\Xi^{-}}$ 
decays
are pure penguin modes, which only have $P'_\DB$ and $P'_{i EW\DB}$ contributions,
while 
$\overline B{}^0_s\to \Delta^+\overline{p}$ and
$\overline B{}^0_s\to \Delta^0\overline{n}$ decays
are pure exchange ($E'_\DB$) modes,
the $CP$ asymmetries of the $\Delta S=-1$ pure penguin modes are small [see Eq. (\ref{eq: A DeltaS penguin})], while the $CP$ asymmetries of the pure exchange modes are always vanishing.
These can be tests of the Standard model.
In particular, $\overline B{}^0\to \Omega^-\overline{\Xi^{-}}$ 
and $\overline B{}^0_s\to \Xi^{*-}\overline{\Xi^{-}}$ belonging to Group I and II modes, respectively,
their decay rates are unsuppressed (see Table~\ref{tab:DBDS=-1}). 
These modes should be searched for, 
but the latter mode requires $B_s$ tagging to search for its $CP$ asymmetry.

The $U$-spin relations for decuplet-antioctet modes are given by~\cite{Chua:2013zga}
\be
\Delta_{CP}(B^-\to\Delta^0\overline{p})
   &=&
     2\Delta_{CP}(B^-\to\Sigma^{*0}\overline{\Sigma^{+}})
   =-2 \Delta_{CP}(B^-\to\Sigma^{*0}\overline{p})
   \non\\
   &=&-\Delta_{CP}(B^-\to\Xi^{*0}\overline{\Sigma^{+}}),
   \non\\
\Delta_{CP}(B^-\to\Delta^-\overline{n})
   &=& 
   3 \Delta_{CP}(B^-\to\Xi^{*-}\overline{\Xi^{0}})
  =6 \Delta_{CP}(B^-\to\Sigma^{*-}\overline{\Sigma^{0}})
\non\\
   &=&2 \Delta_{CP}(B^-\to\Sigma^{*-}\overline{\Lambda})
   =  -3\Delta_{CP}(B^-\to\Sigma^{*-}\overline{n})
\non\\
   &=&
   -6 \Delta_{CP}(B^-\to\Xi^{*-}\overline{\Sigma^{0}})
   =   
   -\Delta_{CP}(B^-\to\Omega^-\overline{\Xi^{0}})
\non\\
    &=&
   -2 \Delta_{CP}(B^-\to\Xi^{*-}\overline{\Lambda}),   
\non\\
\Delta_{CP}(\bar B^0\to\Delta^+\overline{p})
   &=&
   -\Delta_{CP}(\bar B^0_s\to\Sigma^{*+}\overline{\Sigma^{+}}),
\non\\
3\Delta_{CP}(\bar B^0\to\Sigma^{*-}\overline{\Sigma^{-}})
  & =&3\Delta_{CP}(\bar B^0\to\Xi^{*-}\overline{\Xi^{-}})
   = \Delta_{CP}(\bar B^0_s\to\Delta^-\overline{\Sigma^{-}})
\non\\  
   &=&3\Delta_{CP}(\bar B^0_s\to\Sigma^{*-}\overline{\Xi^{-}})  
   =-3\Delta_{CP}(\bar B^0_s\to\Sigma^{*-}\overline{\Sigma^{-}})
\non\\   
   &=&-3\Delta_{CP}(\bar B^0_s\to\Xi^{*-}\overline{\Xi^{-}})  
   =-3 \Delta_{CP}(\bar B^0\to\Xi^{*-}\overline{\Sigma^{-}})
\non\\   
   &=& -\Delta_{CP}(\bar B^0\to\Omega^-\overline{\Xi^{-}}), 
\non\\   
\Delta_{CP}(\bar B^0\to\Sigma^{*+}\overline{\Sigma^{+}})
   &=&\Delta_{CP}(\bar B^0\to\Xi^{*0}\overline{\Xi^{0}})
   =-\Delta_{CP}(\bar B^0_s\to\Delta^+\overline{p})
\non\\   
   &=&-\Delta_{CP}(\bar B^0_s\to\Delta^0\overline{n})=0,
\non\\   
\Delta_{CP}(\bar B^0\to\Delta^0\overline{n})
   &=&
-\Delta_{CP}(\bar B^0_s\to\Xi^{*0}\overline{\Xi^{0}}),   
\non\\
\Delta_{CP}(\bar B^0_s\to \Delta^+\overline{\Sigma^{+}})
   &=&
-\Delta_{CP}(\bar B^0\to \Sigma^{*+}\overline{p}),
 \non\\
\Delta_{CP}(\bar B^0_s\to\Sigma^{*0}\overline{\Xi^{0}})
   &=&
-\Delta_{CP}(\bar B^0\to\Sigma^{*0}\overline{n}).
\en
These relations are subjected to corrections from SU(3) breaking in $|p_{cm}|^2$ and topological amplitudes. 
They are roughly satisfied by the results shown in Tables~\ref{tab:AcpDBDS=0} and \ref{tab:AcpDBDS=-1} and the agreement can be improved when SU(3) breaking effects are taken into account.
One can make a quick but non-trivial check on the relative signs of these asymmetries
and see that they are indeed agreed with the above relations.~\footnote{Note that for the second and forth relations, the $CP$ asymmetries of these modes all have vanishing central values, their relative signs cannot be read out from Tables~\ref{tab:AcpDBDS=0} and \ref{tab:AcpDBDS=-1}.}
The forth relation can be used to constrain the size of $CP$ asymmetries of $\Delta S=-1$ pure penguin modes model independently. For example, we can have
\be
|\A(\bar B^0\to\Omega^-\overline{\Xi^{-}})|
&=&
3|\A(\bar B^0_s\to\Sigma^{*-}\overline{\Xi^{-}})|
\frac{\tau(B^0)}{\tau(B^0_s)}
\frac{{\cal B}(\bar B^0_s\to\Sigma^{*-}\overline{\Xi^{-}})}{{\cal B}(\bar B^0\to\Omega^-\overline{\Xi^{-}})}
\non\\
&\leq&3 \frac{\tau(B^0)}{\tau(B^0_s)}
\frac{{\cal B}(\bar B^0_s\to\Sigma^{*-}\overline{\Xi^{-}})}{{\cal B}(\bar B^0\to\Omega^-\overline{\Xi^{-}})}
\simeq 6.1\%,
\en
which is satisfied by the result shown in Table~\ref{tab:AcpDBDS=-1}. Note that the two modes in the above inequality have final states that can cascadely decay to all charged particles. The inequality is generic and can be verified experimentally.

\begin{table}[t!]
\caption{\label{tab:AcpDDDS=0} Same as Table~\ref{tab:AcpBBDS=0}, but with $\Delta S=0$,
$\overline B_q\to\DD$ modes. 
}
\begin{ruledtabular}
\centering
{
\footnotesize
\begin{tabular}{lccclccc}
Mode
          & $\phi=0$
          & $\phi=\pm\pi/4$
          & $\phi=\pm\pi/2$
          & Mode
          & $\phi=0$
          & $\phi=\pm\pi/4$
          & $\phi=\pm\pi/2$
          \\
\hline $B^-\to \Delta^+ \overline{\Delta^{++}}$ 
          & $0\pm19.4$%
          & $\mp(36.0^{+19.3}_{-16.5})$%
          & $\mp(49.3^{+18.3}_{-14.4})$%
          & $\overline B{}^0_s\to \Delta^{+} \overline{\Sigma^{*+}}$ 
          & $0\pm19.0$%
          & $\mp(36.0^{+18.8}_{-16.2})$%
          & $\mp(49.3^{+17.9}_{-14.2})$%
           \\
$B^-\to \Delta^0 \overline{\Delta^{+}}$ 
          & $0\pm32.1$%
          & $\mp(59.3^{+24.4}_{-26.5})$%
          & $\mp(79.5^{+16.7}_{-19.9})$%
          & $\overline B{}^0_s\to \Delta^{0} \overline{\Sigma^{*0}}$ 
          & $0\pm30.9$%
          & $\mp(59.3^{+23.6}_{-25.6})$%
          & $\mp(79.5^{+16.2}_{-19.3})$%
           \\ 
$B^-\to \Delta^- \overline{\Delta^{0}}$
          & $0\pm35.7$%
          & $0\pm35.7$%
          & $0\pm35.7$%
          & $\overline B{}^0_s\to \Delta^{-} \overline{\Sigma^{*-}}$ 
          & $0\pm29.9$%
          & $0\pm29.9$%
          & $0\pm29.9$%
           \\            
$B^-\to \Sigma^{*0} \overline{\Sigma^{*+}}$ 
          & $0\pm32.1$%
          & $\mp(59.3^{+24.4}_{-26.5})$%
          & $\mp(79.5^{+16.7}_{-19.9})$%
          & $\overline B{}^0_s\to \Sigma^{*0} \overline{\Xi^{*0}}$ 
          & $0\pm30.9$%
          & $\mp(59.3^{+23.6}_{-25.6})$%
          & $\mp(79.5^{+16.2}_{-19.3})$%
           \\ 
$B^-\to \Sigma^{*-} \overline{\Sigma^{*0}}$
          & $0\pm35.7$%
          & $0\pm35.7$%
          & $0\pm35.7$%
          & $\overline B{}^0_s\to \Sigma^{*-} \overline{\Xi^{*-}}$ 
          & $0\pm29.9$%
          & $0\pm29.9$%
          & $0\pm29.9$%
           \\ 
$B^-\to \Xi^{*-} \overline{\Xi^{*0}}$
          & $0\pm35.7$%
          & $0\pm35.7$%
          & $0\pm35.7$%
          & $\overline B{}^0_s\to \Xi^{*-} \overline{\Omega^{-}}$ 
          & $0\pm29.9$%
          & $0\pm29.9$%
          & $0\pm29.9$%
           \\ 
$\overline B{}^0\to \Delta^{++} \overline{\Delta^{++}}$
          & $-100\sim100$%
          & $-100\sim100$%
          & $-100\sim100$%
          & $\overline B{}^0\to \Sigma^{*+} \overline{\Sigma^{*+}}$
          & $-100\sim100$%
          & $-100\sim100$%
          & $-100\sim100$%
           \\
$\overline B{}^0\to \Delta^{+} \overline{\Delta^{+}}$ 
          & $0\pm22.9$%
          & $\mp(36.0^{+22.4}_{-19.4})$%
          & $\mp(49.3^{+21.1}_{-16.8})$%
          & $\overline B{}^0\to \Sigma^{*0} \overline{\Sigma^{*0}}$ 
          & $0\pm37.1$%
          & $\mp(59.3^{+27.1}_{-31.2})$%
          & $\mp(79.5^{+17.9}_{-23.4})$%
           \\
$\overline B{}^0\to \Delta^{0} \overline{\Delta^{0}}$ 
          & $0\pm34.0$%
          & $\mp(59.3^{+25.4}_{-28.4})$%
          & $\mp(79.5^{+17.1}_{-21.4})$%
          & $\overline B{}^0\to \Sigma^{*-} \overline{\Sigma^{*-}}$ 
          & $0\pm30.6$%
          & $0\pm30.6$%
          & $0\pm30.6$%
           \\ 
$\overline B{}^0\to \Delta^{-} \overline{\Delta^{-}}$
          & $0\pm30.6$%
          & $0\pm30.6$%
          & $0\pm30.6$%
          & $\overline B{}^0\to \Xi^{*0} \overline{\Xi^{*0}}$
          & $-100\sim100$%
          & $-100\sim100$%
          & $-100\sim100$%
           \\ 
$\overline B{}^0\to \Omega^{-} \overline{\Omega^{-}}$
          & $0$%
          & $0$%
          & $0$%
          & $\overline B{}^0\to \Xi^{*-} \overline{\Xi^{*-}}$ 
          & $0\pm32.1$%
          & $0\pm32.1$%
          & $0\pm32.1$%
           \\
\end{tabular}
}
\end{ruledtabular}
\end{table}

\subsubsection{$CP$ asymmetries of $\overline B_q\to \DD$ decays}

In Table~\ref{tab:AcpDDDS=0}, we give results of direct $CP$ asymmetries of $\Delta S=0$, $\overline B_q\to\DD$ modes. 
For Group I modes,
$\A(\overline B{}^0\to \Delta^{0} \overline{\Delta^{0}})$ 
is sizable and can reach $\mp 80\%$,
but $\A(\overline B{}^0\to \Sigma^{*-} \overline{\Sigma^{*-}})$ 
is vanishing and will be discussed later.
For Group II modes,
$\A(B^-\to \Delta^0 \overline{\Delta^{+}})$, 
$\A(B^-\to \Sigma^{*0} \overline{\Sigma^{*+}})$, 
$\A(\overline B{}^0_s\to \Delta^{0} \overline{\Sigma^{*0}})$, 
$\A(\overline B{}^0_s\to \Sigma^{*0} \overline{\Xi^{*0}})$ 
and
$\A(\overline B{}^0\to \Sigma^{*0} \overline{\Sigma^{*0}})$ 
are similar and sizable, reaching $\mp80\%$,
$\A(B^-\to \Delta^+ \overline{\Delta^{++}})$ 
and
$\A(\overline B{}^0_s\to \Delta^{+} \overline{\Sigma^{*+}})$ 
are similar and sizable, reaching $\mp49\%$,
but
$\A(\overline B{}^0_s\to \Sigma^{*-} \overline{\Xi^{*-}})$ 
and
$\A(\overline B{}^0_s\to \Xi^{*-} \overline{\Omega^{-}})$ 
are vanishing and will be discussed later.
The Group III mode,
the
$\overline B{}^0\to \Delta^{+} \overline{\Delta^{+}}$ 
decay,
has sizable $CP$ asymmetry,
reaching $\mp49\%$.
Most of these $CP$ asymmetries basically share the sign of $\A(\overline B{}^0\to p\overline p)$.

From Eqs. (\ref{eq: DDBm, DS=0}), (\ref{eq: DDB0, DS=0}) and (\ref{eq: DDBs, DS=0}),
we can easily identify the following relations
\be
\A(B^-\to\Delta^-\overline{\Delta^{0}})
    &=&\A(B^-\to\Sigma^{*-}\overline{\Sigma^{*0}})
      =\A(B^-\to\Xi^{*-}\overline{\Xi^{*0}}),
\non\\
\A(B^-\to\Delta^0\overline{\Delta^{+}})
      &=&\A(B^-\to\Sigma^{*0}\overline{\Sigma^{*+}}),
\non\\
\A(\overline{B^0_s}\to\Delta^{0}\overline{\Sigma^{*0}})
     &=&\A(\overline{B^0_s}\to \Sigma^{*0}\overline{\Xi^{*0}}),
\non\\
\A(\overline{B^0_s}\to\Delta^{-}\overline{\Sigma^{*-}})
     &=&\A(\overline{B^0_s}\to\Xi^{*-}\overline{\Omega^-})
      =\A(\overline{B^0_s}\to\Sigma^{*-}\overline{\Xi^{*-}}). 
\en
We see from Table~\ref{tab:AcpDDDS=0} that these relations are satisfied. 
These relations do not relay on the asymptotic limit and are not corrected from phase space ratio, but are subjected to SU(3) breaking in the topological amplitudes.  
Note that the $CP$ asymmetries of several Group I modes, $B^-\to\Delta^0\overline{\Delta^{+}}$, $B^-\to\Sigma^{*0}\overline{\Sigma^{*+}}$,
$\overline{B^0_s}\to\Delta^{0}\overline{\Sigma^{*0}}$ and $\overline{B^0_s}\to \Sigma^{*0}\overline{\Xi^{*0}}$ decays are related.

The central values of the $CP$ asymmetries of
$B^-\to \Delta^- \overline{\Delta^{0}}$,
$B^-\to \Sigma^{*-} \overline{\Sigma^{*0}}$,
$B^-\to \Xi^{*-} \overline{\Xi^{*0}}$,
$\overline B{}^0\to \Delta^{-} \overline{\Delta^{-}}$,
$\overline B{}^0\to \Sigma^{*-} \overline{\Sigma^{*-}}$, 
$\overline B{}^0\to \Xi^{*-} \overline{\Xi^{*-}}$,
$\overline B{}^0_s\to \Delta^{-} \overline{\Sigma^{*-}}$,
$\overline B{}^0_s\to \Sigma^{*-} \overline{\Xi^{*-}}$, 
$\overline B{}^0_s\to \Xi^{*-} \overline{\Omega^{-}}$, 
and
$\overline B{}^0\to \Omega^{-} \overline{\Omega^{-}}$ decays
are vanishing,
as they do not have any tree ($T_{\DD}$) contribution. 
The 
$\overline B{}^0\to \Delta^{++} \overline{\Delta^{++}}$,
$\overline B{}^0\to \Sigma^{*+} \overline{\Sigma^{*+}}$ and
$\overline B{}^0\to \Xi^{*0} \overline{\Xi^{*0}}$ decays
are subleading modes,
which only have $E_\DD$ and $P_\DD$ contributions.
Their $CP$ asymmetries can be any value.
The 
$\overline B{}^0\to \Delta^{-} \overline{\Delta^{-}}$,
$\overline B{}^0\to \Sigma^{*-} \overline{\Sigma^{*-}}$, 
$\overline B{}^0\to \Xi^{*-} \overline{\Xi^{*-}}$,
$\overline B{}^0_s\to \Delta^{-} \overline{\Sigma^{*-}}$,
$\overline B{}^0_s\to \Sigma^{*-} \overline{\Xi^{*-}}$ and 
$\overline B{}^0_s\to \Xi^{*-} \overline{\Omega^{-}}$ 
decays
are $\Delta S=0$ pure penguin modes, 
which only have $P_\DD$, $P_{EW\DD}$ and $PA_\DD$ contributions, 
while the
$\overline B{}^0\to \Omega^{-} \overline{\Omega^{-}}$ decay
is a pure penguin annihilation ($PA_\DD$) mode.
The $CP$ asymmetries of the $\Delta S=0$ penguin modes need not be vanishing [see Eq.~(\ref{eq: PuPc})], while the $CP$ asymmetry of the pure penguin annihilation mode is always vanishing.

\begin{table}[t!]
\caption{\label{tab:AcpDDDS=-1} Same as Table~\ref{tab:AcpBBDS=0}, but with $\Delta S=-1$, $\overline B_q\to\DD$ modes.}
\begin{ruledtabular}
\centering
{
\footnotesize
\begin{tabular}{lccclccc}
Mode
          & $\phi=0$
          & $\phi=\pm\pi/4$
          & $\phi=\pm\pi/2$
          & Mode
          & $\phi=0$
          & $\phi=\pm\pi/4$
          & $\phi=\pm\pi/2$
          \\
\hline $B^-\to \Sigma^{*+} \overline{\Delta^{++}}$ 
          & $0\pm15.1$%
          & $\pm(27.1^{+15.5}_{-12.5})$%
          & $\pm(35.3^{+14.5}_{-11.0})$%
          & $\overline B{}^0\to \Sigma^{*+} \overline{\Delta^{+}}$ 
          & $0\pm14.8$%
          & $\pm(27.1^{+15.2}_{-12.3})$%
          & $\pm(35.3^{+14.2}_{-10.8})$%
           \\          
$B^-\to \Sigma^{*0} \overline{\Delta^{+}}$ 
          & $0\pm8.0$%
          & $\pm(14.2^{+8.9}_{-6.8})$%
          & $\pm(19.2^{+9.1}_{-6.6})$%
          & $\overline B{}^0\to \Sigma^{*0} \overline{\Delta^{0}}$ 
          & $0\pm7.7$%
          & $\pm(14.2^{+8.6}_{-6.6})$%
          & $\pm(19.2^{+8.8}_{-6.4})$%
           \\ 
$B^-\to \Sigma^{*-} \overline{\Delta^{0}}$ 
          & $0\pm1.8$%
          & $0\pm1.8$%
          & $0\pm1.8$%
          &$\overline B{}^0\to \Sigma^{*-} \overline{\Delta^{-}}$ 
          & $0\pm1.5$%
          & $0\pm1.5$%
          & $0\pm1.5$%
           \\            
$B^-\to \Xi^{*0} \overline{\Sigma^{*+}}$ 
          & $0\pm8.0$%
          & $\pm(14.2^{+8.9}_{-6.8})$%
          & $\pm(19.2^{+9.1}_{-6.6})$%
          & $\overline B{}^0\to \Xi^{*0} \overline{\Sigma^{*0}}$ 
          & $0\pm7.7$%
          & $\pm(14.2^{+8.6}_{-6.6})$%
          & $\pm(19.2^{+8.8}_{-6.4})$%
           \\ 
$B^-\to \Xi^{*-} \overline{\Sigma^{*0}}$ 
          & $0\pm1.8$%
          & $0\pm1.8$%
          & $0\pm1.8$%
          & $\overline B{}^0\to \Xi^{*-} \overline{\Sigma^{*-}}$ 
          & $0\pm1.5$%
          & $0\pm1.5$%
          & $0\pm1.5$%
           \\ 
$B^-\to \Omega^{-} \overline{\Xi^{*0}}$ 
          & $0\pm1.8$%
          & $0\pm1.8$%
          & $0\pm1.8$%
          & $\overline B{}^0\to \Omega^{-} \overline{\Xi^{*-}}$ 
          & $0\pm1.5$%
          & $0\pm1.5$%
          & $0\pm1.5$%
           \\ 
$\overline B{}^0_s\to \Delta^{++} \overline{\Delta^{++}}$
          & $-100\sim100$%
          & $-100\sim100$%
          & $-100\sim100$%
          & $\overline B{}^0_s\to \Sigma^{*+} \overline{\Sigma^{*+}}$ 
          & $0\pm18.3$%
          & $\pm(27.1^{+19.0}_{-14.4})$%
          & $\pm(35.3^{+17.8}_{-12.3})$%
           \\
$\overline B{}^0_s\to \Delta^{+} \overline{\Delta^{+}}$
          & $-100\sim100$%
          & $-100\sim100$%
          & $-100\sim100$%
          & $\overline B{}^0_s\to \Sigma^{*0} \overline{\Sigma^{*0}}$ 
          & $0\pm9.6$%
          & $\pm(14.2^{+11.0}_{-7.7})$%
          & $\pm(19.2^{+11.2}_{-7.2})$%
           \\
$\overline B{}^0_s\to \Delta^{0} \overline{\Delta^{0}}$
          & $-100\sim100$%
          & $-100\sim100$%
          & $-100\sim100$%
          & $\overline B{}^0_s\to \Sigma^{*-} \overline{\Sigma^{*-}}$ 
          & $0\pm1.6$%
          & $0\pm1.6$%
          & $0\pm1.6$%
           \\ 
$\overline B{}^0_s\to \Delta^{-} \overline{\Delta^{-}}$
          & $0$%
          & $0$%
          & $0$%
          & $\overline B{}^0_s\to \Xi^{*0} \overline{\Xi^{*0}}$ 
          & $0\pm8.6$%
          & $\pm(14.2^{+9.8}_{-7.2})$%
          & $\pm(19.2^{+10.0}_{-6.8})$%
           \\ 
$\overline B{}^0_s\to \Omega^{-} \overline{\Omega^{-}}$ 
          & $0\pm1.5$%
          & $0\pm1.5$%
          & $0\pm1.5$%
          & $\overline B{}^0_s\to \Xi^{*-} \overline{\Xi^{*-}}$ 
          & $0\pm1.6$%
          & $0\pm1.6$%
          & $0\pm1.6$%
           \\ 
\end{tabular}
}
\end{ruledtabular}
\end{table}

In Table~\ref{tab:AcpDDDS=-1} we give results of direct $CP$ asymmetries of $\Delta S=-1$, $\overline B_q\to\DD$ modes. 
For Group I modes,
$\A(B^-\to \Sigma^{*+} \overline{\Delta^{++}})$ 
and
$\A(\overline B{}^0_s\to \Sigma^{*+} \overline{\Sigma^{*+}})$ 
are similar and can be as large as $\pm35\%$,
$\A(B^-\to \Xi^{*0} \overline{\Sigma^{*+}})$ 
and
$\A(\overline B{}^0_s\to \Xi^{*0} \overline{\Xi^{*0}})$ 
are similar and can be as large as $\pm19\%$,
but
$\A(\overline B{}^0_s\to \Omega^{-} \overline{\Omega^{-}})$, 
$\A(B^-\to \Omega^{-} \overline{\Xi^{*0}})$, 
$\A(B^-\to \Sigma^{*-} \overline{\Delta^{0}})$ 
and
$\A(\overline B{}^0_s\to \Sigma^{*-} \overline{\Sigma^{*-}})$ 
are vanishingly small and will be discussed later.
For Group II modes,
$\A(\overline B{}^0\to \Sigma^{*+} \overline{\Delta^{+}})$ 
can reach $\pm35\%$,      
$\A(\overline B{}^0\to \Sigma^{*0} \overline{\Delta^{0}})$ 
and
$\A(\overline B{}^0\to \Xi^{*0} \overline{\Sigma^{*0}})$ 
are similar and can be $\pm19\%$,
but
$\A(\overline B{}^0\to \Xi^{*-} \overline{\Sigma^{*-}})$ 
and
$\A(\overline B{}^0\to \Omega^{-} \overline{\Xi^{*-}})$ 
are vanishing. 
For Group III modes,
$\A(B^-\to \Sigma^{*0} \overline{\Delta^{+}})$ 
and         
$\A(\overline B{}^0_s\to \Sigma^{*0} \overline{\Sigma^{*0}})$ 
are similar and can be $\pm19\%$,
but
$\A(\overline B{}^0_s\to \Xi^{*-} \overline{\Xi^{*-}})$ 
and
$\A(B^-\to \Xi^{*-} \overline{\Sigma^{*0}})$ 
are vanishingly small and will be discussed later.
 
From Eqs. (\ref{eq: DDBm, DS=-1}), (\ref{eq: DDB0, DS=-1}) and (\ref{eq: DDBs, DS=-1}),
we can easily identify the following relations
\be
\A(B^-\to\Sigma^{*-}\overline{\Delta^0})
     &=&   
\A(B^-\to\Omega^-\overline{\Xi^{*0}})
      =\A(B^-\to\Xi^{*-}\overline{\Sigma^{*0}}),
\non\\
\A(B^-\to\Sigma^{*0}\overline{\Delta^+})
     &=&\A(B^-\to\Xi^{*0}\overline{\Sigma^{*+}}),
\non\\
\A(\bar B^0\to\Sigma^{*0}\overline{\Delta^0})
   &=&\A(\bar B^0\to\Xi^{*0}\overline{\Sigma^{*0}}),
\non\\
\A(\bar B^0\to\Sigma^{*-}\overline{\Delta^-})
   &=& \A(\bar B^0\to\Xi^{*-}\overline{\Sigma^{*-}})
   =\A(\bar B^0\to\Omega^-\overline{\Xi^{*-}}).
\en
We see from Table~\ref{tab:AcpDDDS=-1} that these relations are satisfied. 
Note that these relations do not relay on the asymptotic limit, but are subjected to SU(3) breaking in the topological amplitudes.

The central values of the $CP$ asymmetries of 
$B^-\to \Sigma^{*-} \overline{\Delta^{0}}$, 
$B^-\to \Xi^{*-} \overline{\Sigma^{*0}}$, 
$B^-\to \Omega^{-} \overline{\Xi^{*0}}$, 
$\overline B{}^0\to \Sigma^{*-} \overline{\Delta^{-}}$,
$\overline B{}^0\to \Xi^{*-} \overline{\Sigma^{*-}}$, 
$\overline B{}^0\to \Omega^{-} \overline{\Xi^{*-}}$, 
$\overline B{}^0_s\to \Sigma^{*-} \overline{\Sigma^{*-}}$, 
$\overline B{}^0_s\to \Omega^{-} \overline{\Omega^{-}}$, 
$\overline B{}^0_s\to \Xi^{*-} \overline{\Xi^{*-}}$ 
and 
$\overline B{}^0_s\to \Delta^{-} \overline{\Delta^{-}}$ decays
are vanishing,
as they do not have any tree ($T'_{\DD}$) contribution. 
The subleading modes,
$\overline B{}^0_s\to \Delta^{++} \overline{\Delta^{++}}$
$\overline B{}^0_s\to \Delta^{+} \overline{\Delta^{+}}$ and
$\overline B{}^0_s\to \Delta^{0} \overline{\Delta^{0}}$ decays
have $E'_\DD$ and $P'_\DD$ contributions.
Their $CP$ asymmetries can be any value.
Since the 
$\overline B{}^0\to \Sigma^{*-} \overline{\Delta^{-}}$,
$\overline B{}^0\to \Xi^{*-} \overline{\Sigma^{*-}}$, 
$\overline B{}^0\to \Omega^{-} \overline{\Xi^{*-}}$, 
$\overline B{}^0_s\to \Sigma^{*-} \overline{\Sigma^{*-}}$, 
$\overline B{}^0_s\to \Omega^{-} \overline{\Omega^{-}}$ and 
$\overline B{}^0_s\to \Xi^{*-} \overline{\Xi^{*-}}$ 
decays
are $\Delta S=-1$ pure penguin modes, which only have $P'_\DD$, $P'_{EW\DD}$ and $PA'_\DD$ contributions,
and the
$\overline B{}^0_s\to \Delta^{-} \overline{\Delta^{-}}$ decay
is a pure penguin annihilation ($PA'_\DD$) mode,
their $CP$ asymmetries are small [see Eq.~(\ref{eq: A DeltaS penguin})] or always vanishing.
These can be tests of the Standard Model.
Note that some of these modes have relatively good detectability.
These include
two Group I modes, 
$\overline B{}^0_s\to \Omega^{-} \overline{\Omega^{-}}$ 
and 
$\overline B{}^0_s\to \Sigma^{*-} \overline{\Sigma^{*-}}$ 
decays,
two Group II modes,
$\overline B{}^0\to \Xi^{*-} \overline{\Sigma^{*-}}$ 
and
$\overline B{}^0\to \Omega^{-} \overline{\Xi^{*-}}$ 
decays,
and
a Group III modes, 
the $\overline B{}^0_s\to \Xi^{*-} \overline{\Xi^{*-}}$ 
decay, where all have rates of orders $10^{-7}$ (see Table~\ref{tab:DDDS=-1}).
It will be interesting to search for these modes and use their $CP$ asymmetries as the null tests of the Standard Model.

The $U$-spin relations for decuplet-antidecuplet modes are given by~\cite{Chua:2013zga}~\footnote{A typo in the fifth relation is corrected.}
\be
2\Delta_{CP}(B^-\to\Delta^-\overline{\Delta^{0}})
    &=&3\Delta_{CP}(B^-\to\Sigma^{*-}\overline{\Sigma^{*0}})
       =\Delta_{CP}(B^-\to\Xi^{*-}\overline{\Xi^{*0}})
\non\\
    &=&-6 \Delta_{CP}(B^-\to\Sigma^{*-}\overline{\Delta^0})
       = -2 \Delta_{CP}(B^-\to\Omega^-\overline{\Xi^{*0}})
\non\\
    &=&-3 \Delta_{CP}(B^-\to\Xi^{*-}\overline{\Sigma^{*0}}), 
\non\\
\Delta_{CP}(B^-\to\Delta^0\overline{\Delta^{+}})
     &=&2\Delta_{CP}(B^-\to\Sigma^{*0}\overline{\Sigma^{*+}})
        =-2 \Delta_{CP}(B^-\to\Sigma^{*0}\overline{\Delta^+})
\non\\
      &=&-\Delta_{CP}(B^-\to\Xi^{*0}\overline{\Sigma^{*+}}), 
\non\\
\Delta_{CP}(B^-\to\Delta^+\overline{\Delta^{++}})
      &=&-\Delta_{CP}(B^-\to \Sigma^{*+}\overline{\Delta^{++}}), 
\non\\
\Delta_{CP}(\overline{B^0_s}\to\Delta^{0}\overline{\Sigma^{*0}})
     &=&\Delta_{CP}(\overline{B^0_s}\to \Sigma^{*0}\overline{\Xi^{*0}})
       =- \Delta_{CP}(\bar B^0\to\Sigma^{*0}\overline{\Delta^0})
\non\\
     &=&-\Delta_{CP}(\bar B^0\to\Xi^{*0}\overline{\Sigma^{*0}}), 
\non\\
4 \Delta_{CP}(\overline{B^0_s}\to\Delta^{-}\overline{\Sigma^{*-}})
    &=&4 \Delta_{CP}(\overline{B^0_s}\to\Xi^{*-}\overline{\Omega^-})
       =3 \Delta_{CP}(\overline{B^0_s}\to\Sigma^{*-}\overline{\Xi^{*-}})
\non\\
    &=& -4 \Delta_{CP}(\bar B^0\to\Sigma^{*-}\overline{\Delta^-})
      =-3\Delta_{CP} (\bar B^0\to\Xi^{*-}\overline{\Sigma^{*-}})
\non\\
     &=& -4 \Delta_{CP}(\bar B^0\to\Omega^-\overline{\Xi^{*-}}), 
\non\\
\Delta_{CP}(\overline{B^0_s}\to\Delta^+\overline{\Sigma^{*+}})
      &=&-\Delta_{CP}(\bar B^0\to \Sigma^{*+}\overline{\Delta^+}), 
\non\\
\Delta_{CP}(\overline{B^0}\to \Xi^{*0}\overline{\Xi^{*0}})
     &=&-\Delta_{CP}(\bar B^0_s\to\Delta^0\overline{\Delta^0}), 
\non\\
\Delta_{CP}(\overline{B^0}\to\Xi^{*-}\overline{\Xi^{*-}})
    &=& -\Delta_{CP}(\bar B^0_s\to\Sigma^{*-}\overline{\Sigma^{*-}}), 
\non\\
\Delta_{CP}(\overline{B^0}\to\Sigma^{*0}\overline{\Sigma^{*0}})
    &=&-\Delta_{CP}(\bar B^0_s\to\Sigma^{*0}\overline{\Sigma^{*0}}), 
\non\\
\Delta_{CP}(\overline{B^0}\to\Omega^-\overline{\Omega^-})
   &=&-\Delta_{CP}(\bar B^0_s\to\Delta^-\overline{\Delta^-})=0, 
\non\\
\Delta_{CP}(\overline{B^0}\to\Delta^{++}\overline{\Delta^{++}})
   &=& -\Delta_{CP}(\bar B^0_s\to \Delta^{++}\overline{\Delta^{++}}), 
\non\\
\Delta_{CP}(\overline{B^0}\to\Sigma^{*+}\overline{\Sigma^{*+}})
  &=&-\Delta_{CP}(\bar B^0_s\to\Delta^+\overline{\Delta^+}), 
\non\\
\Delta_{CP}(\overline{B^0}\to\Delta^{-}\overline{\Delta^{-}})
  &=&-\Delta_{CP}(\bar B^0_s\to\Omega^{-}\overline{\Omega^{-}}), 
\non\\   
\Delta_{CP}(\overline{B^0}\to\Sigma^{*-}\overline{\Sigma^{*-}})
  &=&-\Delta_{CP}(\bar B^0_s\to\Xi^{*-}\overline{\Xi^{*-}}), 
\non\\
\Delta_{CP}(\overline{B^0}\to\Delta^+\overline{\Delta^{+}})
  &=&-\Delta_{CP}(\bar B^0_s\to\Sigma^{*+}\overline{\Sigma^{*+}}), 
\non\\
\Delta_{CP}(\overline{B^0}\to\Delta^0\overline{\Delta^{0}})
  &=&-\Delta_{CP}(\bar B^0_s\to\Xi^{*0}\overline{\Xi^{*0}}). 
  \label{eq: Uspin DD}
\en
These relations are roughly satisfied by the results shown in Tables~\ref{tab:AcpDDDS=0} and \ref{tab:AcpDDDS=-1}
and the agreement can be improved when SU(3) breaking effects are taken into account.
For example, one can make a quick but non-trivial check on the relative signs of these asymmetries
and see that they are indeed agreed with the above relations. ~\footnote{Note that for the first relation, the $CP$ asymmetries of these modes all have vanishing central values, their relative signs cannot be read out from Tables~\ref{tab:AcpDDDS=0} and \ref{tab:AcpDDDS=-1}. 
The relative signs of $\Delta_{CP}(\overline{B^0}\to \Xi^{*0}\overline{\Xi^{*0}}$ and $\Delta_{CP}(\bar B^0_s\to\Delta^0\overline{\Delta^0})$ 
$\Delta_{CP}(\overline{B^0}\to\Delta^{++}\overline{\Delta^{++}})$
and $\Delta_{CP}(\bar B^0_s\to \Delta^{++}\overline{\Delta^{++}})$,
$\Delta_{CP}(\overline{B^0}\to\Sigma^{*+}\overline{\Sigma^{*+}})$
and $\Delta_{CP}(\bar B^0_s\to\Delta^+\overline{\Delta^+})$ 
cannot be read out from Tables~\ref{tab:AcpDDDS=0} and \ref{tab:AcpDDDS=-1}. The signs of modes in the fifth, eighth, thirteenth and fourteenth relations cannot be read out from the tables.}
Note that several vanishing $CP$ asymmetries as discussed previously are related to each other.
For example, from the last relation of the above equation, we have
\be
{\cal A}(\bar B^0\to\Delta^0\overline{\Delta^0})
&=&-{\cal A}(\bar B^0_s\to\Xi^{*0}\overline{\Xi^{*0}})\frac{\tau(B^0)}{\tau(B^0_s)}
   \frac{{\cal B}(\bar B^0_s\to\Xi^{*0}\overline{\Xi^{*0}})}{{\cal B}(\bar B^0\to\Delta^0\overline{\Delta^{0}})}
\non\\
&\simeq&-(3.7) {\cal A}(\bar B^0_s\to\Xi^{*0}\overline{\Xi^{*0}}).
\label{eq: Usipin DD1}
\en 
The results in Tables~\ref{tab:AcpDDDS=0} and \ref{tab:AcpDDDS=-1} agree with it. 
Note that these two modes are both Group I modes.  We will return to this later.
The fifth, eighth, thirteenth and fourteenth relations can be used to constrains the sizes of $CP$ asymmetries of $\Delta S=-1$ pure penguin modes model independently. For example, using the fifth relation we can have
\be
|{\cal A}(\bar B^0\to\Omega^-\overline{\Xi^{*-}})|
&=&\frac{3}{4}|{\cal A}(\bar B^0_s\to\Sigma^{*-}\overline{\Xi^{*-}})|\frac{\tau(B^0)}{\tau(B^0_s)}
   \frac{{\cal B}(\bar B^0_s\to\Sigma^{*-}\overline{\Xi^{*-}})}{{\cal B}(\bar B^0\to\Omega^-\overline{\Xi^{*-}})}
\non\\
&\leq&\frac{3}{4}\frac{\tau(B^0)}{\tau(B^0_s)}
   \frac{{\cal B}(\bar B^0_s\to\Sigma^{*-}\overline{\Xi^{*-}})}{{\cal B}(\bar B^0\to\Omega^-\overline{\Xi^{*-}})}
   \simeq 5.5\%,
\en
which is satisfied by the result shown in Table \ref{tab:AcpDDDS=-1}. Note that the above two modes are both Group II modes and one does not need $B_s$ tagging to test the inequality experimentally.

\begin{figure}[h!]
\centering
 \subfigure[]{
  \includegraphics[width=0.315\textwidth]{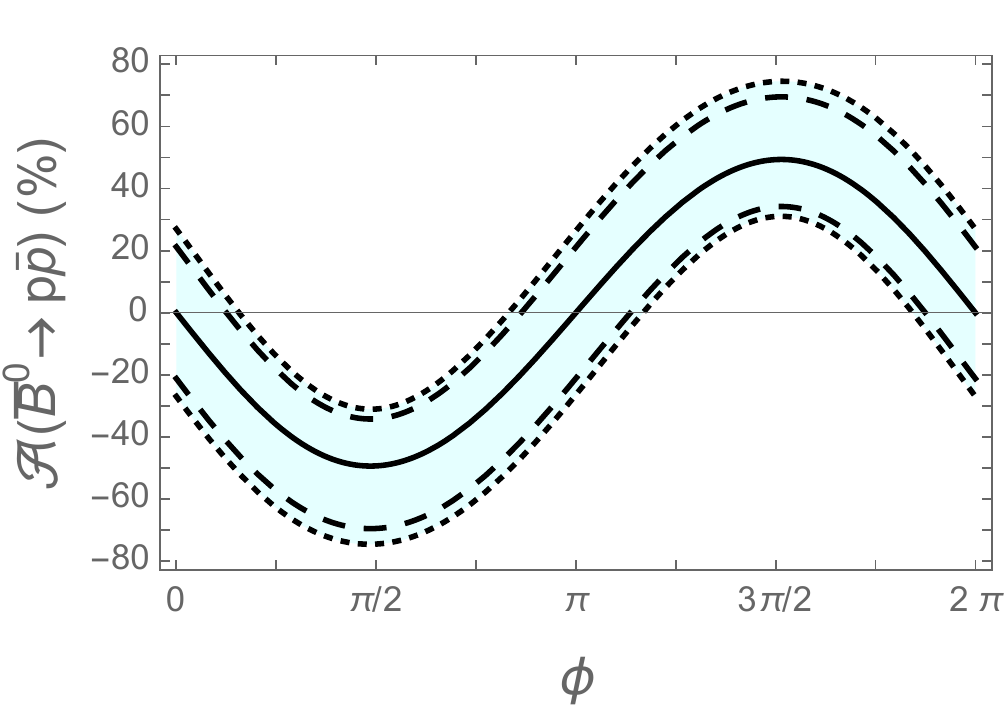}
}\hspace{0.7cm}
\subfigure[]{
  \includegraphics[width=0.315\textwidth]{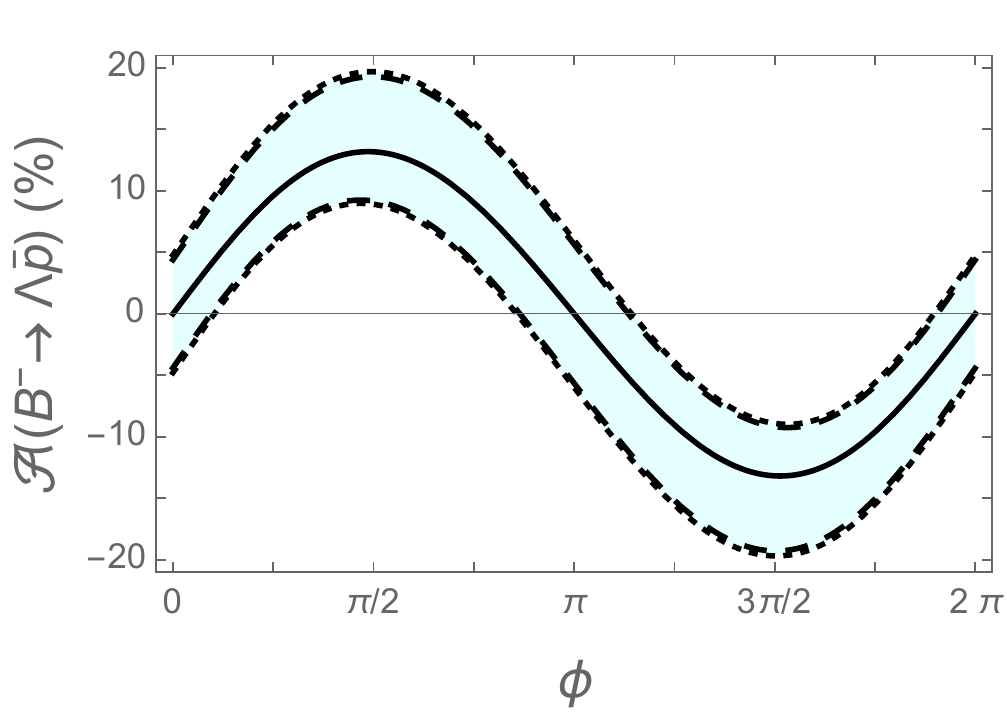}
}\\\subfigure[]{
  \includegraphics[width=0.315\textwidth]{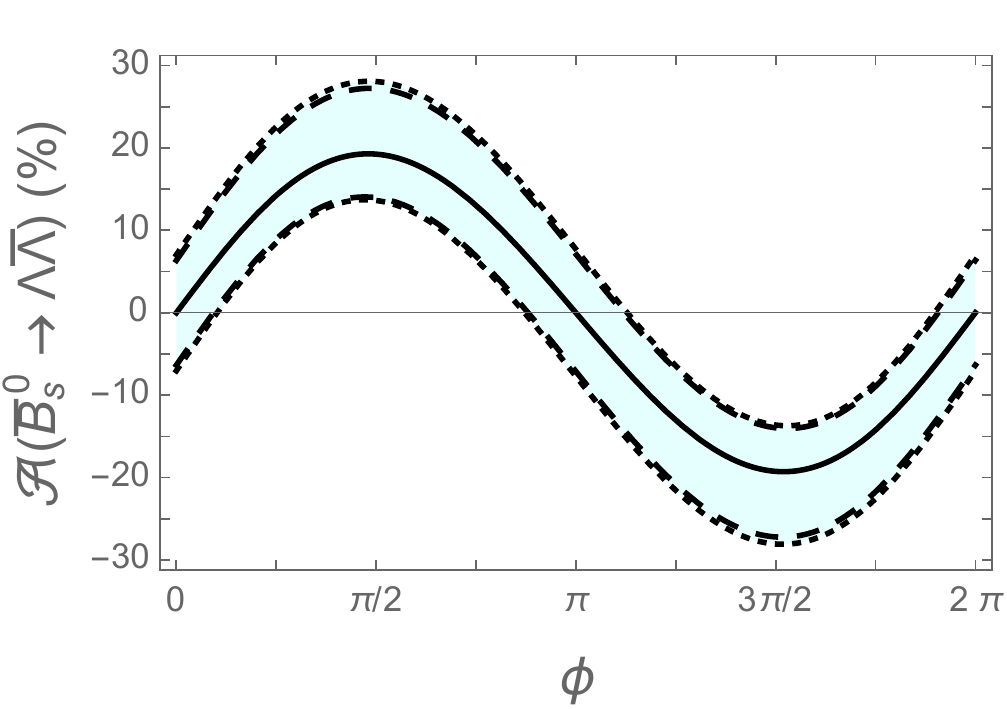}
}\hspace{0.7cm}
\subfigure[]{
  \includegraphics[width=0.315\textwidth]{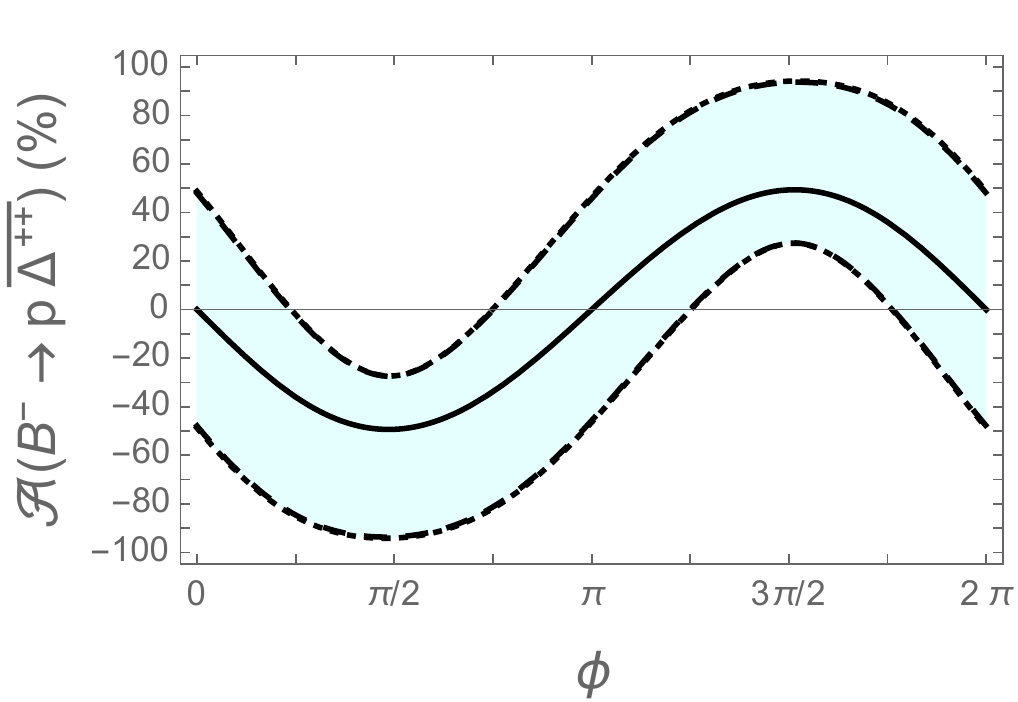}
}
\\\subfigure[]{
  \includegraphics[width=0.315\textwidth]{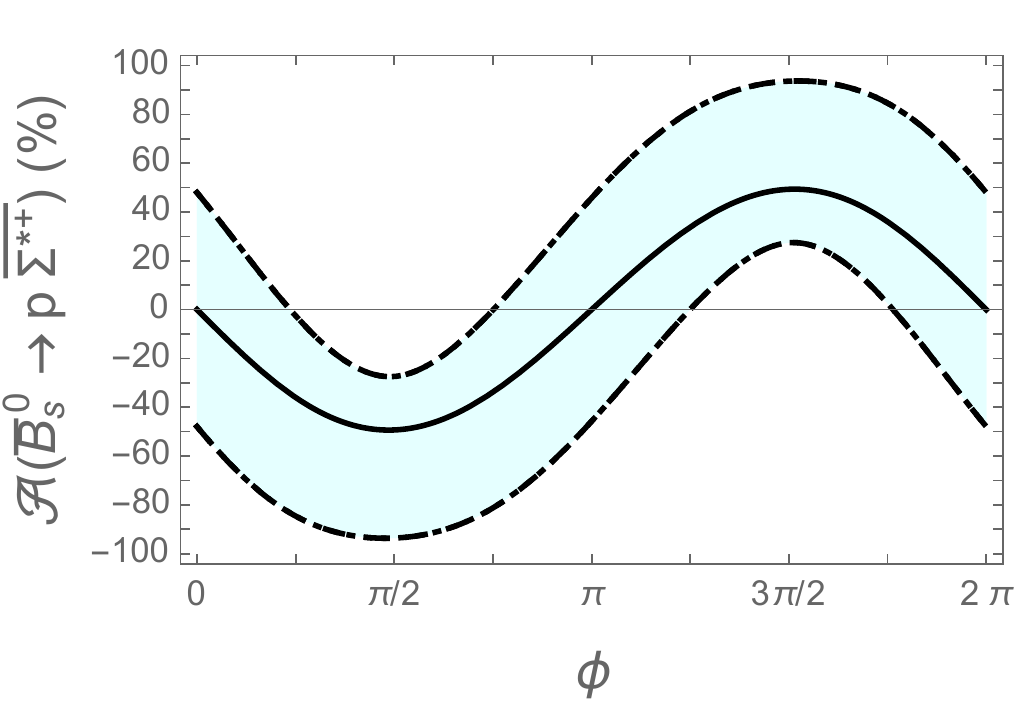}
}\hspace{0.7cm}
\subfigure[]{
  \includegraphics[width=0.315\textwidth]{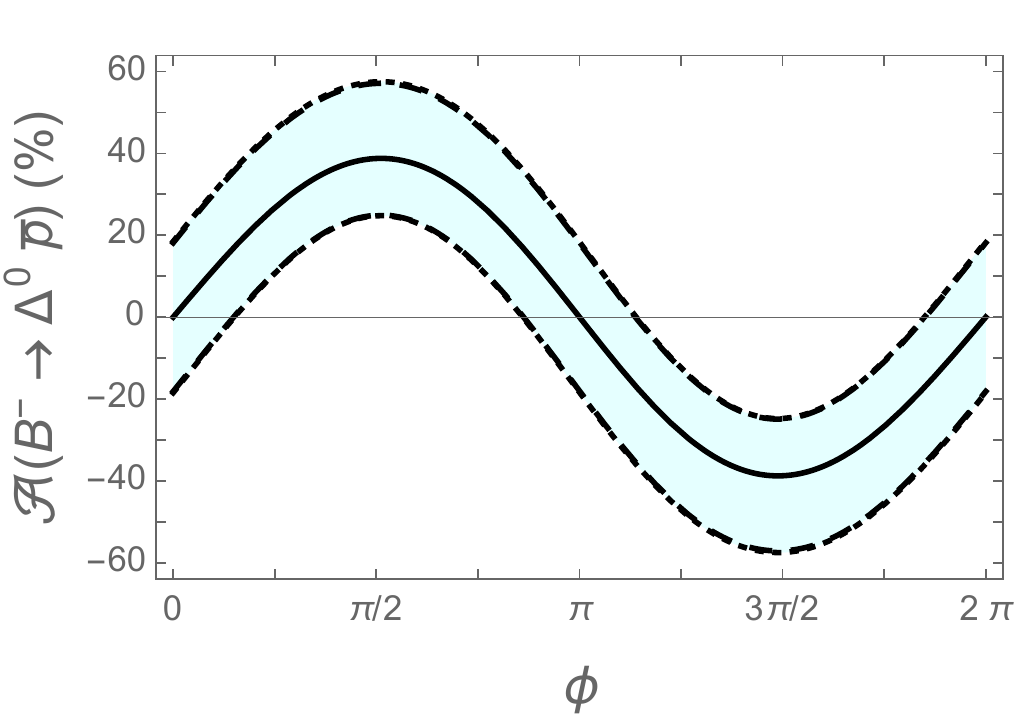}
}
\\\subfigure[]{
  \includegraphics[width=0.315\textwidth]{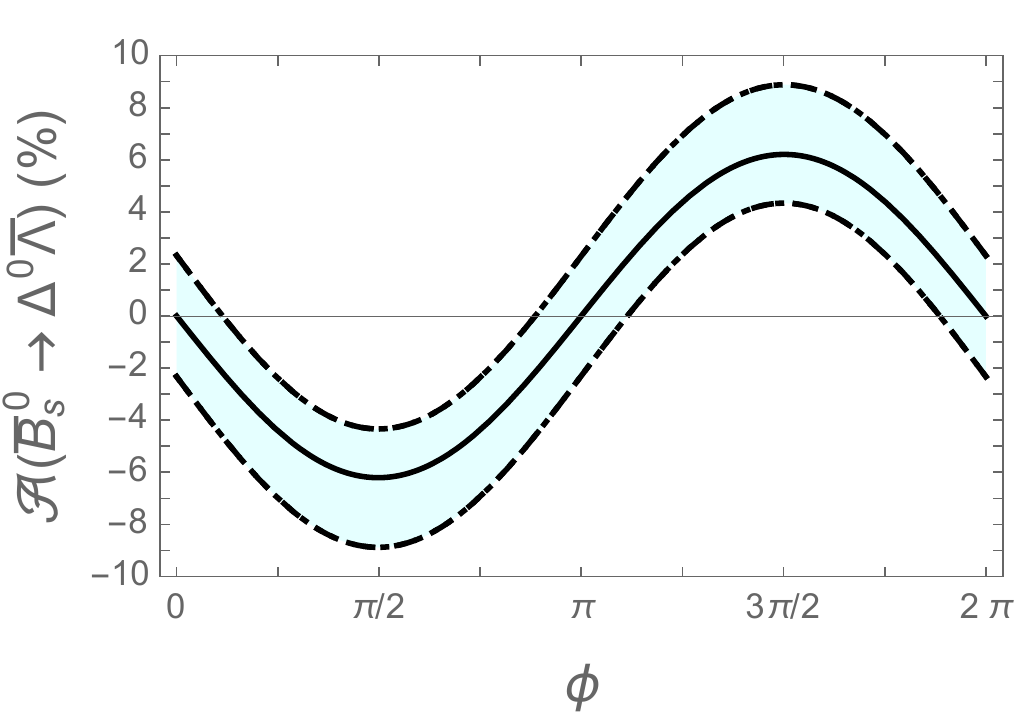}
}
\caption{Direct $CP$ asymmetries of some interesting modes are plotted with respect to the penguin-tree relative strong phase $\phi$. 
The solid lines are from tree-penguin interferences using the asymptotic relation.
The bands between the dashed (dotted) lines are with contributions from corrections to the asymptotic relation without (with) contributions from sub-leading terms.
} \label{fig:ACPI}
\end{figure}

\begin{figure}[h!]
\centering
 \subfigure[]{
  \includegraphics[width=0.315\textwidth]{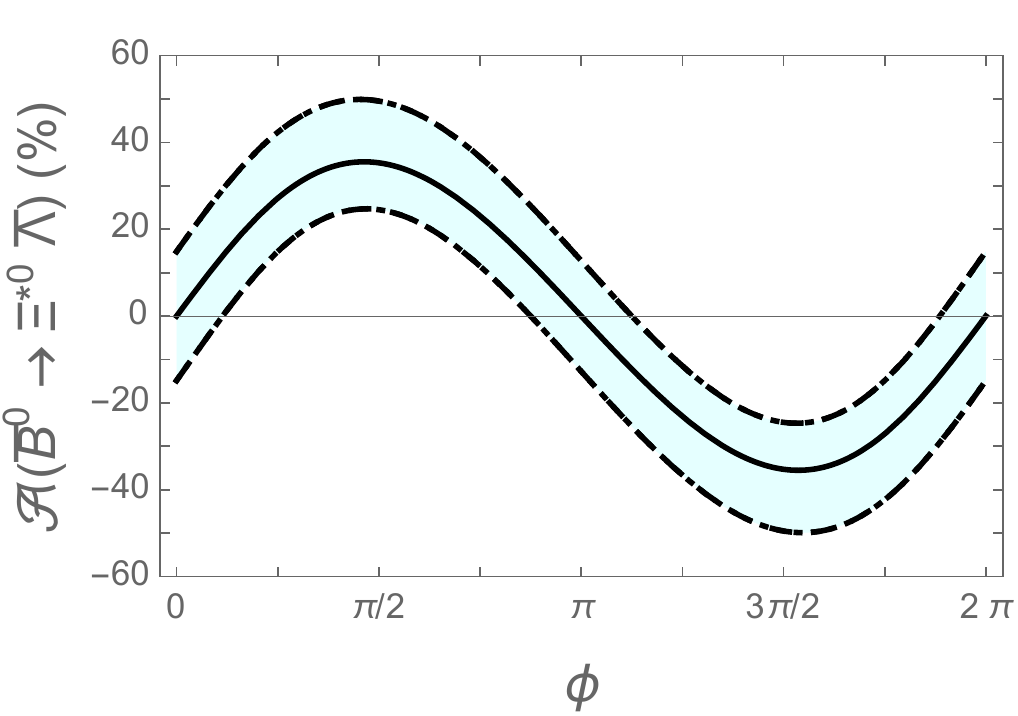}
}\hspace{0.7cm}
\subfigure[]{
  \includegraphics[width=0.315\textwidth]{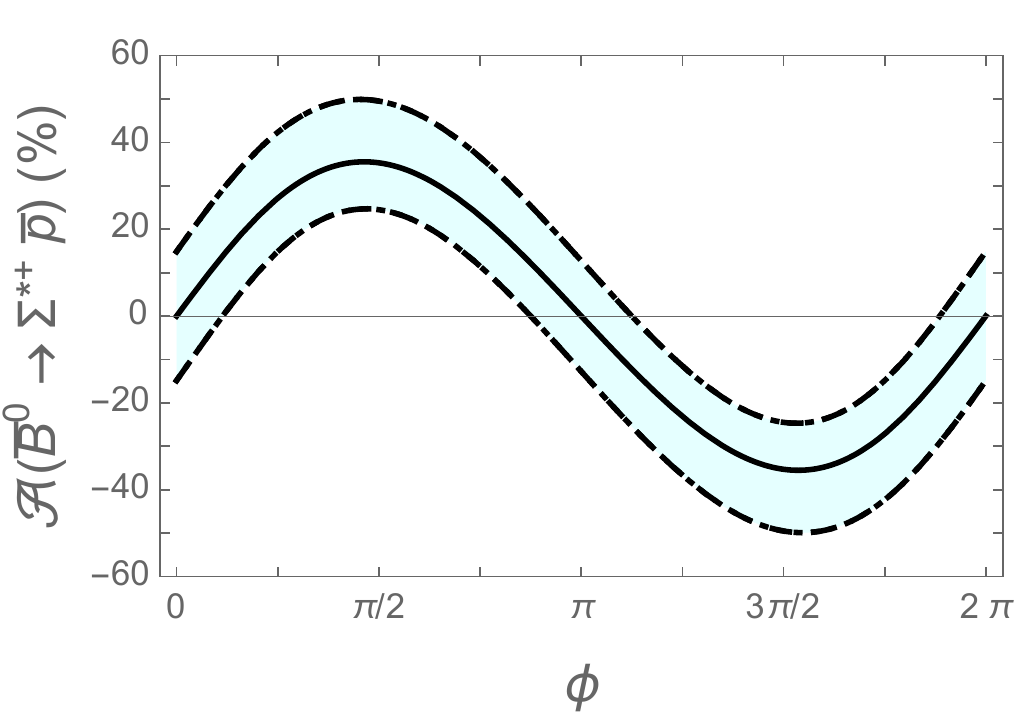}
}\\\subfigure[]{
  \includegraphics[width=0.315\textwidth]{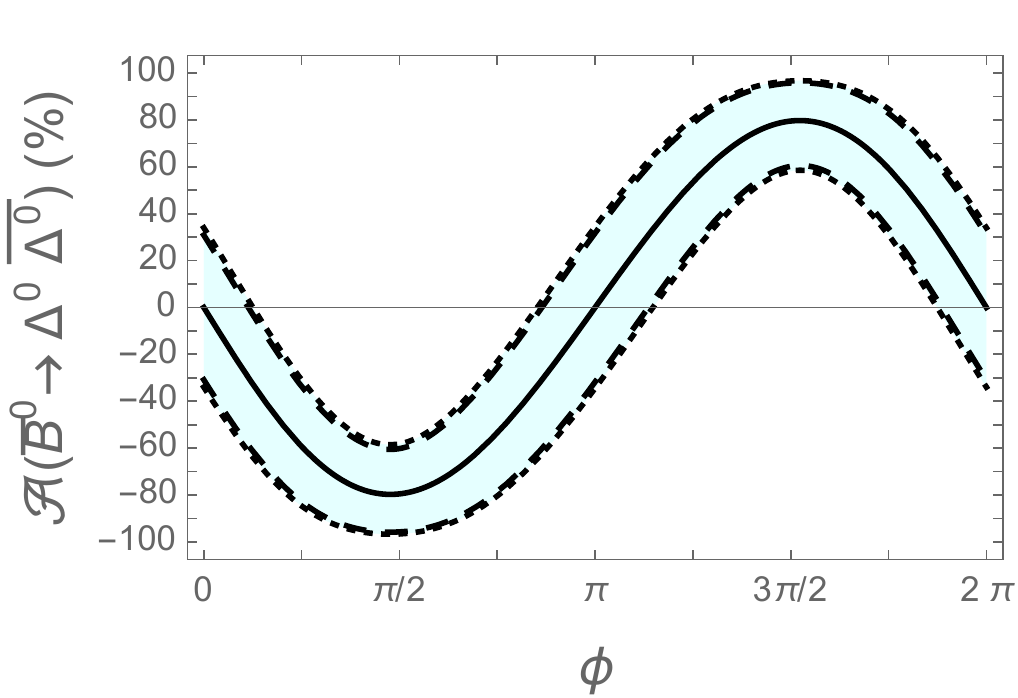}
}\hspace{0.7cm}
\subfigure[]{
  \includegraphics[width=0.315\textwidth]{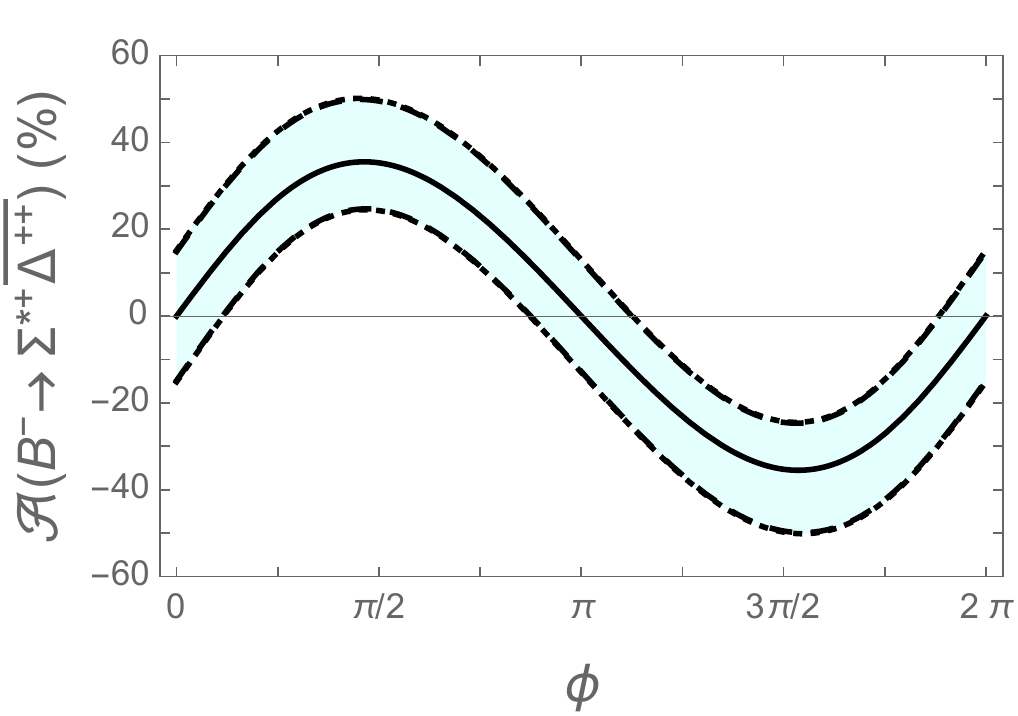}
}
\\\subfigure[]{
  \includegraphics[width=0.315\textwidth]{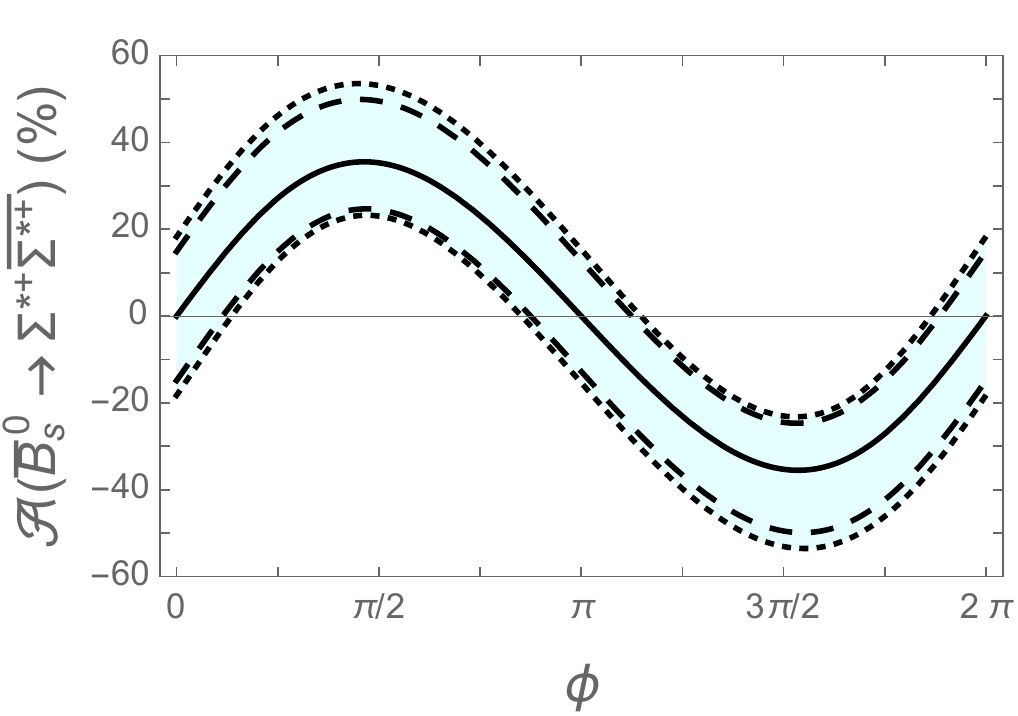}
}\hspace{0.7cm}
\subfigure[]{
  \includegraphics[width=0.315\textwidth]{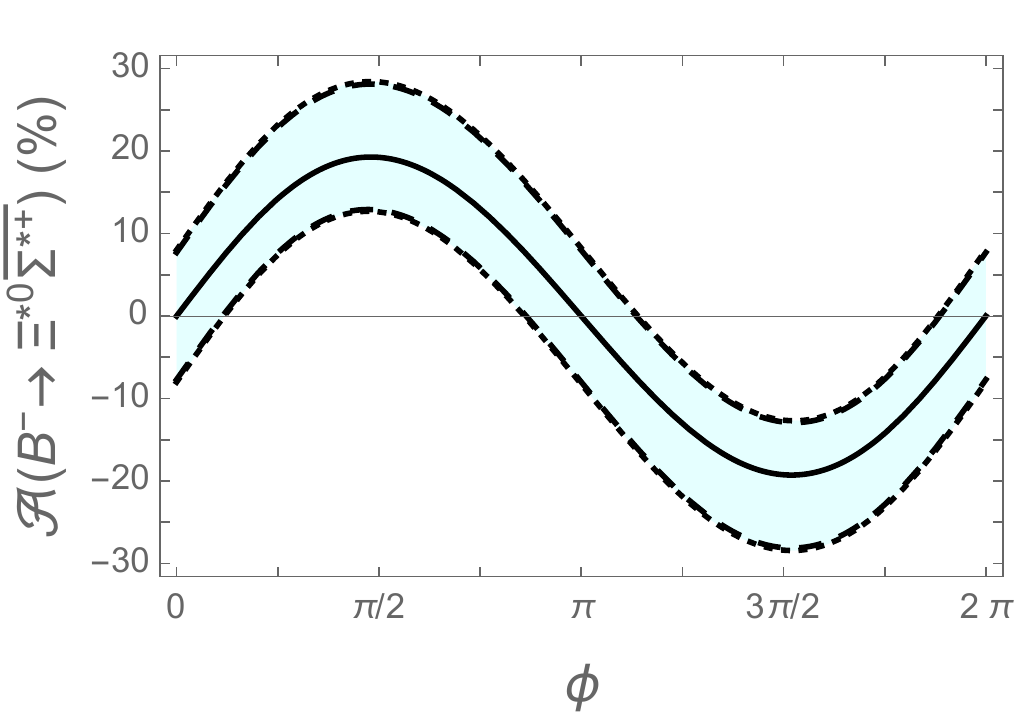}
}
\\\subfigure[]{
  \includegraphics[width=0.315\textwidth]{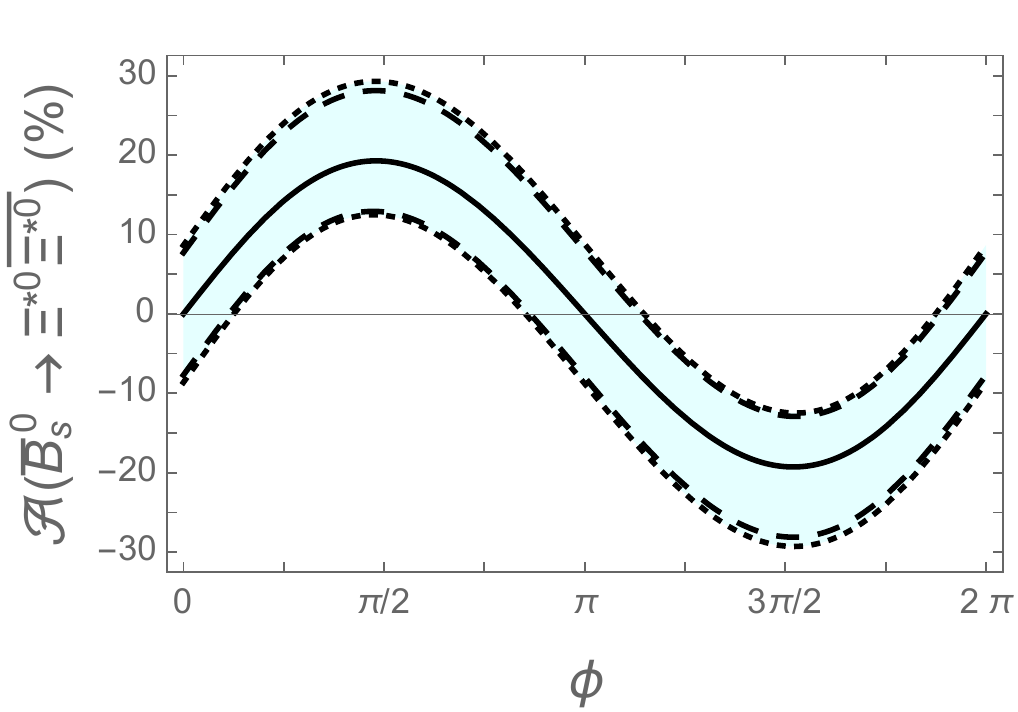}
}
\caption{Direct $CP$ asymmetries of some interesting modes are plotted with respect to the penguin-tree relative strong phase $\phi$.  
The solid lines are from tree-penguin interferences using the asymptotic relation.
The bands between the dashed (dotted) lines are with contributions from corrections to the asymptotic relation without (with) contributions from sub-leading terms.
} \label{fig:ACPII}
\end{figure}

In Tables~\ref{tab:AcpBBDS=0} to \ref{tab:AcpDDDS=-1}, we show results on $CP$ asymmetries for some specify values of the penguin-tree relative strong phase $\phi$ (0, $\pm\pi/4$ and $\pm\pi/2$).
It will be useful to plot the $CP$ asymmetries of some interesting modes in the full range of $\phi$.
In Fig.~\ref{fig:ACPI}, we plot the $CP$ asymmetries of 
$\overline B{}^0\to p\overline{p}$, 
$B^-\to \Lambda\overline{p}$, 
$\overline B{}^0_s\to \Lambda\overline{\Lambda}$, 
$B^-\to p\overline{\Delta^{++}}$, 
$\overline B{}^0_s\to p\overline{\Sigma^{*+}}$, 
$B^-\to \Delta^0\overline{p}$ 
and
$\overline B{}^0_s\to \Delta^0\overline{\Lambda}$ 
decays.
In Fig.~\ref{fig:ACPII}, we plot the $CP$ asymmetries of 
$\overline B{}^0\to \Xi^{*0}\overline{\Lambda}$, 
$\overline B{}^0\to \Sigma^{*+}\overline{p}$, 
$\overline B{}^0\to \Delta^{0} \overline{\Delta^{0}}$, 
$B^-\to \Sigma^{*+} \overline{\Delta^{++}}$, 
$\overline B{}^0_s\to \Sigma^{*+} \overline{\Sigma^{*+}}$, 
$B^-\to \Xi^{*0} \overline{\Sigma^{*+}}$ 
and
$\overline B{}^0_s\to \Xi^{*0} \overline{\Xi^{*0}}$ 
decays.
These are direct $CP$ asymmetries of several Group I modes, which have unsuppressed rates and can cascadely decay to all charged final states.
The solid lines are from tree-penguin interferences using the asymptotic relation.
The bands between the dashed (dotted) lines in the figures are with contributions from corrections to the asymptotic relation without (with) contributions from sub-leading terms.
For the figures we see that corrections to the asymptotic relation dominate the uncertainties.

Note that the remaining Group I modes, including 
$\overline B{}^0_s\to \Xi^{-}\overline{\Xi^{-}}$, 
$B^-\to \Xi^{-}\overline{\Lambda}$, 
$\overline B{}^0\to \Xi^{-}\overline{\Sigma^{*-}}$, 
$\overline B{}^0\to \Omega^-\overline{\Xi^{-}}$, 
$\overline B{}^0\to \Sigma^{*-} \overline{\Sigma^{*-}}$, 
$\overline B{}^0_s\to \Omega^{-} \overline{\Omega^{-}}$, 
$B^-\to \Omega^{-} \overline{\Xi^{*0}}$, 
$B^-\to \Sigma^{*-} \overline{\Delta^{0}}$ 
and
$\overline B{}^0_s\to \Sigma^{*-} \overline{\Sigma^{*-}}$ 
decays 
do not depends on $\phi$, hence, their $CP$ asymmetries are not plotted.
In particular, as noted before, 
$\A(\overline B{}^0_s\to \Xi^{-}\overline{\Xi^{-}})$, 
$\A(\overline B{}^0\to \Xi^{-}\overline{\Sigma^{*-}})$, 
$\A(\overline B{}^0\to \Omega^-\overline{\Xi^{-}})$, 
$\A(\overline B{}^0_s\to \Omega^{-} \overline{\Omega^{-}})$, 
and
$\A(\overline B{}^0_s\to \Sigma^{*-} \overline{\Sigma^{*-}})$ 
are small and can be used to test the Standard Model,
especially these modes have good detectability in rates. 

We now return to Figs.~\ref{fig:ACPI} and \ref{fig:ACPII}.
Most of these asymmetries are sizable. 
We can classify these modes, according to the dependence of $\phi$, into two groups.
The 
$\A(\overline B{}^0\to p\overline{p})$, 
$\A(B^-\to p\overline{\Delta^{++}})$, 
$\A(\overline B{}^0_s\to p\overline{\Sigma^{*+}})$, 
$\A(\overline B{}^0_s\to \Delta^0\overline{\Lambda})$ 
and
$\A(\overline B{}^0\to \Delta^{0} \overline{\Delta^{0}})$ 
have similar behavior, while
$\A(B^-\to \Lambda\overline{p})$, 
$\A(\overline B{}^0_s\to \Lambda\overline{\Lambda})$, 
$\A(B^-\to \Delta^0\overline{p})$, 
$\A(\overline B{}^0\to \Xi^{*0}\overline{\Lambda)}$, 
$\A(\overline B{}^0\to \Sigma^{*+}\overline{p})$, 
$\A(B^-\to \Sigma^{*+} \overline{\Delta^{++}})$, 
$\A(\overline B{}^0_s\to \Sigma^{*+} \overline{\Sigma^{*+}})$, 
$\A(B^-\to \Xi^{*0} \overline{\Sigma^{*+}})$ 
and
$\A(\overline B{}^0_s\to \Xi^{*0} \overline{\Xi^{*0}})$ 
have similar behavior, but different from the first group.
The asymmetries in these two groups basically have opposite signs for $\phi$ away from 0, $\pi$, $2\pi$.
Note that through $U$-spin relation,
$\A(\overline{B^0}\to\Delta^0\overline{\Delta^{0}})$ and
$\A(\bar B^0_s\to\Xi^{*0}\overline{\Xi^{*0}})$ are related by Eq.~(\ref{eq: Uspin DD}).
We see from Fig.~\ref{fig:ACPII} (c) and (g) that they indeed respect the relation.

It will be interesting to measure the $CP$ asymmetries of these Group I modes and compare to the predictions plotted in Figs.~\ref{fig:ACPI} and \ref{fig:ACPII}. Measuring these asymmetries can provide useful information on the decay amplitudes.

\section{Conclusion}

With the experimental evidences of $\overline B {}^0\to p \overline{p}$ and $B^-\to\Lambda\bar p$ decays, 
it is now possible to extract both tree and penguin amplitudes of charmless two-body baryonic decays
for the first time.
The extracted penguin-tree ratio agrees with the expectation.
Predictions on all $\overline B_q\to \BB$, $\BD$, $\DB$ and $\DD$ decay rates are given.
It is non-trivial that the results do not violate any existing experimental upper limit.
From the results, it is understandable that why $\overline B {}^0\to p \overline{p}$ and $B^-\to\Lambda\bar p$ modes are the first two modes with experimental evidences.
Relations on rates are verified.
There are 23 modes that have relatively sizable rates and can cascadely decay to all charged final states,
including
$\overline B{}^0\to p\bar p$,
$B^-\to \Lambda\bar p$, $\Xi^{-}\overline{\Lambda}$,
$\overline B{}^0_s\to \Lambda\overline{\Lambda}$, 
$\Xi^-\overline{\Xi^-}$;
$B^-\to p\overline{\Delta^{++}}$,
$\overline B{}^0_s\to p\overline{\Sigma^{*+}}$,
$\overline B{}^0\to \Xi^-\overline{\Sigma^{*-}}$;
$B^-\to \Delta^0\overline p$,
$\overline B{}^0_s\to\Delta^0\overline\Lambda$,
$\overline B{}^0\to \Sigma^{*-}\overline p$, 
$\Omega^-\overline{\Xi^-}$, $\Xi^{*0}\overline\Lambda$;
$\overline{B}{}^0\to \Delta^0\overline{\Delta^0}$, 
$\overline{B}{}^0\to \Sigma^{*-}\overline{\Sigma^{*-}}$,
$B^-\to\Sigma^{*+}\overline {\Delta^{++}}$, 
$\Xi^{*0}\overline{\Sigma^{*+}}$, 
$\Omega^-\overline{\Xi^{*0}}$, 
$\Sigma^{*-} \overline{\Delta^{0}}$, 
$\overline B{}^0_s\to\Omega^-\overline{\Omega^-}$,
$\Xi^{*0}\overline{\Xi^{*0}}$, 
$\Sigma^{*+} \overline{\Sigma^{*+}}$
and
$\Sigma^{*-} \overline{\Sigma^{*-}}$ 
decays.
With $\pi^0$ and $\gamma$ another 38 modes can be searched for, while with $\pi^0\pi^0$, $\pi^0\gamma$ and $\gamma\gamma$ 38 more modes can be searched for in the future.
In particular, we note that the predicted $B^-\to p\overline{\Delta^{++}}$ rate is close to the experimental bound,
which has not been updated in the last ten years~\cite{Wei:2007fg}.
The bounds on $B^-\to \Delta^0\overline{p}$ and $\overline B{}^0\to \Sigma^{*+}\overline{p}$ rates have not been updated in the last ten years~\cite{Wei:2007fg, Wang:2007as} and the bound on $\overline B{}^0\to \Delta^{0} \overline{\Delta^{0}}$ rate has not been updated in about three decades~\cite{Bortoletto:1989mu},
while their rates are predicted to be of the order of $10^{-8}$.
Also note that the $\overline B{}^0_s\to \Omega^{-} \overline{\Omega^{-}}$ rate is predicted to be the highest rate.
The analysis of this work can be improved systematically when more modes are measured.

Direct $CP$ asymmetries of all $\overline B_q\to {\cal B} \overline {\cal B}$, 
${\cal B} \overline {\cal D}$, 
${\cal D} \overline {\cal B}$ and 
${\cal D} \overline {\cal D}$ modes are explored.
Relations on $CP$ asymmetries are verified.
Results of $CP$ asymmetries for modes with relatively good detectability in rates are highlighted.
In particular, the direct $CP$ asymmetry of $\overline B {}^0\to p \overline{p}$ decay can be as large as $\pm 50\%$.
Some of the $CP$ asymmetries are small or vanishing. 
For $\overline B\to \BB$, $\Delta S=-1$ decays,
$\overline B{}^0\to  \Xi^{-}\overline{\Sigma^{-}}$,
$\overline B{}^0_s\to \Sigma^{-}\overline{\Sigma^{-}}$
and
$\overline B{}^0_s\to \Xi^{-}\overline{\Xi^{-}}$ 
decays are pure penguin modes.
For $\overline B\to \BD$, $\Delta S=0$ decays,
$\overline B{}^0\to \Sigma^{+}\overline{\Sigma^{*+}}$ and
$\overline B{}^0\to \Xi^{0}\overline{\Xi^{*0}}$ decays
are pure exchange modes. 
For $\overline B\to \BD$, $\Delta S=-1$ decays,
$\overline B{}^0\to \Sigma^{-}\overline{\Delta^-}$,
$\overline B{}^0\to \Xi^{-}\overline{\Sigma^{*-}}$, 
$\overline B{}^0_s\to \Sigma^{-}\overline{\Sigma^{*-}}$ and
$\overline B{}^0_s\to \Xi^{-}\overline{\Xi^{*-}}$ 
decays are pure penguin modes
and
$\overline B{}^0_s\to p\overline{\Delta^+}$ and
$\overline B{}^0_s\to n\overline{\Delta^0}$
decays are pure exchange modes.
For $\overline B\to \DB$, $\Delta S=0$ decays,
$\overline B{}^0\to \Xi^{*0}\overline{\Xi^{0}}$ and
$\overline B{}^0\to \Sigma^{*+}\overline{\Sigma^{+}}$ decays
are pure exchange modes.
For $\overline B\to \DB$, $\Delta S=-1$ decays, 
$\overline B{}^0\to \Omega^-\overline{\Xi^{-}}$, 
$\overline B{}^0_s\to \Sigma^{*-}\overline{\Sigma^{-}}$,
$\overline B{}^0\to \Xi^{*-}\overline{\Sigma^{-}}$ and
$\overline B{}^0_s\to \Xi^{*-}\overline{\Xi^{-}}$ 
decays
are pure penguin modes, 
while 
$\overline B{}^0_s\to \Delta^+\overline{p}$ and
$\overline B{}^0_s\to \Delta^0\overline{n}$ decays
are pure exchange modes.
For $\overline B\to \DD$, $\Delta S=0$ decays, 
$\overline B{}^0\to \Omega^{-} \overline{\Omega^{-}}$ decay
is a pure penguin annihilation mode.
For $\overline B\to \DD$, $\Delta S=-1$ decays, 
$\overline B{}^0\to \Sigma^{*-} \overline{\Delta^{-}}$,
$\overline B{}^0\to \Xi^{*-} \overline{\Sigma^{*-}}$, 
$\overline B{}^0\to \Omega^{-} \overline{\Xi^{*-}}$, 
$\overline B{}^0_s\to \Sigma^{*-} \overline{\Sigma^{*-}}$, 
$\overline B{}^0_s\to \Omega^{-} \overline{\Omega^{-}}$ and 
$\overline B{}^0_s\to \Xi^{*-} \overline{\Xi^{*-}}$ 
decays
are pure penguin modes, 
and the
$\overline B{}^0_s\to \Delta^{-} \overline{\Delta^{-}}$ decay
is a pure penguin annihilation mode.
The $CP$ asymmetries of the above modes are small, following from the hierarchy of the CKM factors, or vanishing.
They can be added to the list of the tests of the Standard Model. 
Note that some of these modes have relatively good detectability in rates.
These include 5 Group I modes,
$\overline B{}^0_s\to \Xi^{-}\overline{\Xi^{-}}$, 
$\overline B{}^0\to \Xi^{-}\overline{\Sigma^{*-}}$, 
$\overline B{}^0\to \Omega^-\overline{\Xi^{-}}$, 
$\overline B{}^0_s\to \Sigma^{*-} \overline{\Sigma^{*-}}$ 
and
$\overline B{}^0_s\to \Omega^{-} \overline{\Omega^{-}}$ 
decays,
4 Group II modes,
$\overline B{}^0_s\to \Xi^{-}\overline{\Xi^{*-}}$, 
$\overline B{}^0_s\to \Xi^{*-}\overline{\Xi^{-}}$, 
$\overline B{}^0\to \Xi^{*-} \overline{\Sigma^{*-}}$ 
and
$\overline B{}^0\to \Omega^{-} \overline{\Xi^{*-}}$ 
decays,
and a Group III mode,
the $\overline B{}^0_s\to \Xi^{*-} \overline{\Xi^{*-}}$ 
decay,
but some require $B_s$ tagging to search for its $CP$ asymmetry.
It will be interesting to search for these modes and use their $CP$ asymmetries to search for New Physics.
Furthermore, since these modes are rare decay modes and all of them are pure penguin modes, 
they are expected to be sensitive to New Physics contributions.

\section{Acknowledgments}

The author thanks Paoti Chang and Eduardo Rodrigues for discussions and useful comments.
This research was supported in part by the Ministry of Science and Technology of R.O.C. under Grant
Nos. 103-2112-M-033-002-MY3.

\appendix

\section{Topological amplitudes of two-body charmless baryonic $B$ decays}\label{appendix:amplitudes}

We collect all the $\overline B\to\DD$, $\DB$, $\BD$, $\BB$ decay amplitudes obtained in Ref.~\cite{Chua:2013zga}. Note that Eq.~(\ref{eq: DDB0, DS=-1}) is the corrected version.

\subsection{$\overline B$ to octet-anti-octet baryonic decays}

The full $\bar B\to\BB$ decay amplitudes for $\Delta S=0$ processes are given by
\be
A(B^-\to n\overline{p})
   &=&-T_{1\BB}-5 P_{1\BB}
           +\frac{2}{3}(P_{1EW\BB}
           -P_{3EW\BB}+P_{4EW\BB})
           -5 A_{1\BB},
   \non\\
A(B^-\to\Sigma^{0}\overline{\Sigma^{+}})
   &=&\sq2T_{3\BB}
         +\frac{1}{\sq2}(5P_{1\BB}-P_{2\BB})
         +\frac{1}{3\sq2}(P_{1EW\BB}+P_{2EW\BB}+2P_{3EW\BB}
   \non\\
   &&-2P_{4EW\BB})
         +\frac{1}{\sq2}(5 A_{1\BB}-A_{2\BB}),
   \non\\   
A(B^-\to\Sigma^{-}\overline{\Sigma^{0}})
   &=&-\frac{1}{\sq2}(5P_{1\BB}-P_{2\BB})
           -\frac{1}{3\sq2}(P_{1EW\BB}+P_{2EW\BB}-4P_{3EW\BB}-2P_{4EW\BB})
   \non\\
   &&-\frac{1}{\sq2}(5 A_{1\BB}- A_{2\BB}),
   \non\\
A(B^-\to\Sigma^{-}\overline{\Lambda})
   &=&-\frac{1}{\sq6}(5P_{1\BB}+P_{2\BB})
           -\frac{1}{3\sq6}(P_{1EW\BB}-P_{2EW\BB}-4P_{3EW\BB}-2P_{4EW\BB})
   \non\\
   &&-\frac{1}{\sq6}(5 A_{1\BB}+ A_{2\BB}),
   \non\\  
A(B^-\to\Xi^{-}\overline{\Xi^{0}})
   &=&-P_{2\BB}+\frac{1}{3}P_{2EW\BB}- A_{2\BB},
   \non\\
A(B^-\to\Lambda\overline{\Sigma^+})
   &=&-\sq{\frac{2}{3}}(T_{1\BB}-T_{3\BB})
          -\frac{1}{\sq6}(5P_{1\BB}+P_{2\BB})
           +\frac{1}{3\sq6}(5P_{1EW\BB}+P_{2EW\BB}
   \non\\
   &&   -4P_{3EW\BB}+2P_{4EW\BB})
           -\frac{1}{\sq6}(5 A_{1\BB}+A_{2\BB}),
\label{eq: BBBm, DS=0}           
\en
\be
A(\bar B^0\to p\overline{p})
   &=&-T_{2\BB}+2T_{4\BB}
          +P_{2\BB}
           +\frac{2}{3}P_{2EW\BB}
          -5 E_{1\BB}+ E_{2\BB}
          -9PA_\BB,
   \non\\
A(\bar B^0\to n\overline{n})
   &=&-(T_{1\BB}+T_{2\BB})
          -(5P_{1\BB}-P_{2\BB})
           +\frac{2}{3}(P_{1EW\BB}+P_{2EW\BB}
   \non\\
   &&-P_{3EW\BB}-2P_{4EW\BB})
         + E_{2\BB}
         -9 PA_\BB,
   \non\\
A(\bar B^0\to\Sigma^{+}\overline{\Sigma^{+}})
   &=&-5 E_{1\BB}+ E_{2\BB}
          -9 PA_\BB,
   \non\\
A(\bar B^0\to\Sigma^{0}\overline{\Sigma^{0}})
   &=&-T_{3\BB}
          -\frac{1}{2}(5P_{1\BB}-P_{2\BB})
           -\frac{1}{6}(P_{1EW\BB}+P_{2EW\BB}+2P_{3EW\BB}-2P_{4EW\BB})
   \non\\
   &&-\frac{1}{2}(5 E_{1\BB}- E_{2\BB})
          -9 PA_\BB,
   \non\\
A(\bar B^0\to\Sigma^{0}\overline{\Lambda})
   &=&\frac{1}{\sq3}(T_{3\BB}+2T_{4\BB})
          +\frac{1}{2\sq3}(5P_{1\BB}+P_{2\BB})
           +\frac{1}{6\sq3}(P_{1EW\BB}-P_{2EW\BB}
   \non\\
   &&+2P_{3EW\BB}+10P_{4EW\BB})
        -\frac{1}{2\sqrt3}(5 E_{1\BB}+ E_{2\BB}),
   \non\\
A(\bar B^0\to\Sigma^{-}\overline{\Sigma^{-}})
   &=&-(5P_{1\BB}-P_{2\BB})
           -\frac{1}{3}(P_{1EW\BB}+P_{2EW\BB}-4P_{3EW\BB}-2P_{4EW\BB})
   \non\\
   &&-9 PA_\BB,
   \non\\   
A(\bar B^0\to\Xi^{0}\overline{\Xi^{0}})
   &=& E_{2\BB}
          -9 PA_\BB,
   \non\\
A(\bar B^0\to\Xi^{-}\overline{\Xi^{-}})
   &=&P_{2\BB}
           -\frac{1}{3}P_{2EW\BB}
           -9PA_\BB,
   \non\\
A(\bar B^0\to\Lambda\overline{\Sigma^{0}})
   &=&\frac{1}{\sq3}(T_{1\BB}-T_{3\BB})
          +\frac{1}{2\sq3}(5P_{1\BB}+P_{2\BB})
           -\frac{1}{6\sq3}(5P_{1EW\BB}+P_{2EW\BB}
   \non\\
   &&-2P_{3EW\BB}+2P_{4EW\BB})
       -\frac{1}{2\sq3}(5 E_{1\BB}+E_{2\BB}), 
   \non\\
A(\bar B^0\to\Lambda\overline{\Lambda})
   &=&-\frac{1}{3}(T_{1\BB}+2T_{2\BB}-T_{3\BB}-2T_{4\BB})
          -\frac{5}{6}(P_{1\BB}-P_{2\BB})          
   \non\\
   &&+\frac{1}{18}(5P_{1EW\BB}+7P_{2EW\BB}-2P_{3EW\BB}-10P_{4EW\BB})
        -\frac{5}{6}(E_{1\BB}-E_{2\BB})
   \non\\ 
   &&-9PA_\BB,  
\label{eq: BBB0, DS=0}   
\en
and
\be
A(\bar B^0_s\to p\overline{\Sigma^{+}})
   &=&T_{2\BB}-2T_{4\BB}
          -P_{2\BB}
          -\frac{2}{3}P_{2EW\BB},
   \non\\
A(\bar B^0_s\to n\overline{\Sigma^{0}})
   &=&-\frac{1}{\sq2}T_{2\BB}
          +\frac{1}{\sq2}P_{2\BB}          
          +\frac{\sq2}{3}(P_{2EW\BB}-3P_{4EW\BB}),
   \non\\
A(\bar B^0_s\to n\overline{\Lambda})
   &=&\frac{1}{\sq6}(2T_{1\BB}+T_{2\BB})
          +\frac{1}{\sq6}(10P_{1\BB}-P_{2\BB})          
   \non\\
   &&-\frac{1}{3}\sqrt{\frac{2}{3}}(2P_{1EW\BB}+P_{2EW\BB}-2P_{3EW\BB}-P_{4EW\BB}),
   \non\\   
A(\bar B^0_s\to\Sigma^{0}\overline{\Xi^{0}})
   &=&\sq2(T_{3\BB}+T_{4\BB})
          +\frac{5}{\sq2}P_{1\BB}
          +\frac{1}{3\sq2}(P_{1EW\BB}+2P_{3EW\BB}+4P_{4EW\BB}),
   \non\\
A(\bar B^0_s\to\Sigma^{-}\overline{\Xi^{-}})
   &=&-5P_{1\BB}          
           +\frac{1}{3}(-P_{1EW\BB}+4P_{3EW\BB}+2P_{4EW\BB}),
   \non\\   
A(\bar B^0_s\to\Lambda\overline{\Xi^0})
   &=&-\sq{\frac{2}{3}}(T_{1\BB}+T_{2\BB}-T_{3\BB}-T_{4\BB})
          -\frac{1}{\sq6}(5P_{1\BB}-2P_{2\BB})          
   \non\\
   &&+\frac{1}{3\sq6}(5P_{1EW\BB}+4P_{2EW\BB}-2P_{3EW\BB}-4P_{4EW\BB}),
\label{eq: BBBs, DS=0}
\en
while those for $\Delta S=1$ transitions are given by
\be
A(B^-\to\Sigma^{0}\overline{p})
   &=&-\frac{1}{\sq2}(T'_{1\BB}-2T'_{3\BB})
           -\frac{1}{\sq2} P'_{2\BB}
           +\frac{1}{3\sq2}(3P'_{1EW\BB}+P'_{2EW\BB})
           -\frac{1}{\sq2}A'_{2\BB},
   \non\\
A(B^-\to\Sigma^{-}\overline{n})
   &=&-P'_{2\BB}
          +\frac{1}{3}P'_{2EW\BB}
          - A'_{2\BB},
   \non\\
A(B^-\to\Xi^{0}\overline{\Sigma^{+}})
   &=&-T'_{1\BB}
           -5 P'_{1\BB}
           +\frac{2}{3}(P'_{1EW\BB}-P'_{3EW\BB}+P'_{4EW\BB})
           -5 A'_{1\BB},
   \non\\
A(B^-\to\Xi^{-}\overline{\Sigma^{0}})
   &=&-\frac{5}{\sq2}P'_{1\BB}
           -\frac{1}{3\sq2}(P'_{1EW\BB}-4P'_{3EW\BB}-2P'_{4EW\BB})
           -\frac{5}{\sq2}A'_{1\BB},
   \non\\  
A(B^-\to\Xi^{-}\overline{\Lambda})
   &=&-\frac{1}{\sqrt6}(5P'_{1\BB}-2P'_{2\BB})
           -\frac{1}{3\sq6}(P'_{1EW\BB}+2P'_{2EW\BB}-4P'_{3EW\BB}-2P'_{4EW\BB})
   \non\\
   &&-\frac{1}{\sq6}(5 A'_{1\BB}-2 A'_{2\BB}),
   \non\\      
A(B^-\to\Lambda\overline{p})
   &=&\frac{1}{\sq6}(T'_{1\BB}+2T'_{3\BB})
           +\frac{1}{\sqrt6}(10P'_{1\BB}-P'_{2\BB})
           -\frac{1}{3\sq6}(P'_{1EW\BB}-P'_{2EW\BB}
   \non\\
   &&-4P'_{3EW\BB}+4P'_{4EW\BB})
         +\frac{1}{\sq6}(10 A'_{1\BB}- A'_{2\BB}), 
\label{eq: BBBm, DS=-1}                
\en
\be
A(\bar B^0\to \Sigma^{+}\overline{p})
   &=& T'_{2\BB}-2T'_{4\BB}
           -P'_{2\BB}
           -\frac{2}{3}P'_{2EW\BB},
   \non\\
A(\bar B^0\to\Sigma^{0}\overline{n})
   &=&-\frac{1}{\sq2}(T'_{1\BB}+T'_{2\BB}-2T'_{3\BB}-2T'_{4\BB})
          +\frac{1}{\sq2}P'_{2\BB}
    \non\\
    &&+\frac{1}{3\sq2}(3P'_{1EW\BB}+2P'_{2EW\BB}),
   \non\\
A(\bar B^0\to\Xi^{0}\overline{\Sigma^{0}})
   &=&\frac{1}{\sq2}T'_{1\BB}
           +\frac{5}{\sq2}P'_{1\BB}
           -\frac{\sq2}{3}(P'_{1EW\BB}-P'_{3EW\BB}+P'_{4EW\BB}),
   \non\\
A(\bar B^0\to\Xi^{0}\overline{\Lambda})
   &=&-\frac{1}{\sq6}(T'_{1\BB}+2T'_{2\BB})
           -\frac{1}{\sq6}(5P'_{1\BB}-2P'_{2\BB})
    \non\\
    &&+\frac{1}{3}\sq{\frac{2}{3}}(P'_{1EW\BB}+2P'_{2EW\BB}-P'_{3EW\BB}-5P'_{4EW\BB}),
   \non\\ 
A(\bar B^0\to\Xi^{-}\overline{\Sigma^{-}})
   &=&-5P'_{1\BB}
           -\frac{1}{3}(P'_{1EW\BB}-4P'_{3EW\BB}-2P'_{4EW\BB}),
   \non\\   
A(\bar B^0\to\Lambda\overline{n})
   &=&\frac{1}{\sq6}(T'_{1\BB}+T'_{2\BB}+2T'_{3\BB}+2T'_{4\BB})
           +\frac{1}{\sq6}(10P'_{1\BB}-P'_{2\BB})
    \non\\
    &&-\frac{1}{3\sq6}(P'_{1EW\BB}+2P'_{2EW\BB}-4P'_{3EW\BB}-8P'_{4EW\BB}),
\label{eq: BBB0, DS=-1} 
\en
and
\be
A(\bar B^0_s\to p\overline{p})
   &=&-5E'_{1\BB}+ E'_{2\BB}-9 PA'_\BB,
   \non\\
A(\bar B^0_s\to n\overline{n})
   &=& E'_{2\BB}-9PA'_\BB,
   \non\\
A(\bar B^0_s\to\Sigma^{+}\overline{\Sigma^{+}})
   &=&-T'_{2\BB}+2T'_{4\BB}
           +P'_{2\BB}
           +\frac{2}{3}P'_{2EW\BB}
           -5 E'_{1\BB}+ E'_{2\BB}
           -9PA'_\BB,
   \non\\
A(\bar B^0_s\to\Sigma^{0}\overline{\Sigma^{0}})
   &=&-\frac{1}{2}(T'_{2\BB}-2T'_{4\BB})
           +P'_{2\BB}
           +\frac{1}{6}P'_{2EW\BB}
           -\frac{1}{2}(5 E'_{1\BB}- E'_{2\BB})
           -9 PA'_\BB,
   \non\\
A(\bar B^0_s\to\Sigma^{0}\overline{\Lambda})
   &=&\frac{1}{2\sq3}(2T'_{1\BB}+T'_{2\BB}-4T'_{3\BB}-2T'_{4\BB})
           -\frac{1}{2\sq3}(2P'_{1EW\BB}+P'_{2EW\BB})
    \non\\
    &&-\frac{1}{2\sq3}(5 E'_{1\BB}+ E'_{2\BB}),
     \non\\     
A(\bar B^0_s\to\Sigma^{-}\overline{\Sigma^{-}})
   &=&P'_{2\BB}
           -\frac{1}{3}P'_{2EW\BB}
           -9 PA'_\BB,
   \non\\   
A(\bar B^0_s\to\Xi^{0}\overline{\Xi^{0}})
   &=&-T'_{1\BB}-T'_{2\BB}
           -(5P'_{1\BB}-P'_{2\BB})
           +\frac{2}{3}(P'_{1EW\BB}+P'_{2EW\BB}-P'_{3EW\BB}
    \non\\
    &&-2P'_{4EW\BB})
           + E'_{2\BB}
           -9 PA'_\BB,
   \non\\
A(\bar B^0_s\to\Xi^{-}\overline{\Xi^{-}})
   &=&-(5P'_{1\BB}-P'_{2\BB})
          -\frac{1}{3}(P'_{1EW\BB}+P'_{2EW\BB}-4P'_{3EW\BB}-2P'_{4EW\BB})
   \non\\
    &&-9 PA'_\BB,
    \non\\     
A(\bar B^0_s\to\Lambda\overline{\Sigma^{0}})
   &=&\frac{1}{2\sq3}(T'_{2\BB}+2T'_{4\BB})
           +\frac{1}{2\sq3}(-P'_{2EW\BB}+4P'_{4EW\BB})
    \non\\       
    &&-\frac{1}{2\sq3}(5 E'_{1\BB}+E'_{2\BB}),
   \non\\
A(\bar B^0_s\to\Lambda\overline{\Lambda})
   &=&-\frac{1}{6}(2T'_{1\BB}+T'_{2\BB}+4T'_{3\BB}+2T'_{4\BB})
           -\frac{1}{3}(10P'_{1\BB}-P'_{2\BB})
   \non\\       
   &&+\frac{1}{18}(2P'_{1EW\BB}+P'_{2EW\BB}-8P'_{3EW\BB}-4P'_{4EW\BB})
   \non\\      
   &&-\frac{5}{6}(E'_{1\BB}-E'_{2\BB})
         -9PA'_\BB.
\label{eq: BBBs, DS=-1}
\en

\subsection{$\overline B$ to octet-anti-decuplet baryonic decays}

The full $\bar B\to\BD$ decay amplitudes for $\Delta S=0$ processes are given by
\be
A(B^-\to p\overline{\Delta^{++}})
   &=&-\sq6 (T_{1\BD}-2T_{2\BD})+\sq6 P_\BD+2\sq{\frac{2}{3}}P_{1EW\BD}+\sq6 A_\BD,
   \non\\
A(B^-\to n\overline{\Delta^+})
   &=&-\sq2T_{1\BD}+\sq2 P_\BD+\frac{2\sq2}{3}(P_{1EW\BD}-3P_{2EW\BD})+\sq2A_\BD,
   \non\\
A(B^-\to\Sigma^0\overline{\Sigma^{*+}})
   &=&-2T_{2\BD}-P_\BD+\frac{1}{3}(P_{1EW\BD}-6P_{2EW\BD})- A_\BD,
   \non\\ 
A(B^-\to\Sigma^-\overline{\Sigma^{*0}})
   &=&-P_\BD+\frac{1}{3}P_{1EW\BD}- A_\BD,
   \non\\      
A(B^-\to\Xi^{-}\overline{\Xi^{*0}})
   &=&-\sq2 P_\BD+\frac{\sq2}{3}P_{1EW\BD}-\sq2A_\BD,
   \non\\
A(B^-\to\Lambda\overline{\Sigma^{*+}})
   &=&\frac{2}{\sq3}(T_{1\BD}-T_{2\BD})-\sq3 P_\BD-\frac{1}{\sq3}(P_{1EW\BD}-2P_{2EW\BD})-\sq3 A_\BD,
   \non\\
\label{eq: BDBm, DS=0} 
\en
\be
A(\bar B^0\to p\overline{\Delta^+})
   &=&-\sq2(T_{1\BD}-2T_{2\BD})+\sq2 P_\BD+\frac{2\sq2}{3}P_{1EW\BD}-\sq2 E_\BD,
   \non\\
A(\bar B^0\to n\overline{\Delta^0})
   &=&-\sq2T_{1\BD}+\sq2 P_\BD+\frac{2\sq2}{3}(P_{1EW\BD}-3P_{2EW\BD})-\sq2 E_\BD,
   \non\\
A(\bar B^0\to\Sigma^{+}\overline{\Sigma^{*+}})
   &=&\sq2 E_\BD,
   \non\\
A(\bar B^0\to\Sigma^{0}\overline{\Sigma^{*0}})
   &=&-\sq2T_{2\BD}-\frac{1}{\sq2}P_\BD+\frac{1}{3\sq2}(P_{1EW\BD}-6P_ {2EW\BD})-\frac{1}{\sq2} E_\BD,
   \non\\
A(\bar B^0\to\Sigma^{-}\overline{\Sigma^{*-}})
   &=&-\sq2P_\BD+\frac{\sq2}{3}P_{1EW\BD},
   \non\\
A(\bar B^0\to\Xi^{0}\overline{\Xi^{*0}})
   &=&\sq2E_{\BD},
   \non\\
A(\bar B^0\to\Xi^{-}\overline{\Xi^{*-}})
   &=&-\sq2 P_\BD+\frac{\sq2}{3}P_{1EW\BD},
   \non\\
A(\bar B^0\to\Lambda\overline{\Sigma^{*0}})
   &=&\sq{\frac{2}{3}}(T_{1\BD}-T_{2\BD})-\sq{\frac{3}{2}} P_\BD
   -\frac{1}{\sq6}(P_{1EW\BD}-2P_{2EW\BD})+\sq{\frac{3}{2}}E_\BD,
   \non\\   
\label{eq: BDB0, DS=0} 
\en
and
\be
A(\bar B^0_s\to p\overline{\Sigma^{*+}})
   &=&-\sq2 (T_{1\BD}-2T_{2\BD})+\sq2 P_\BD+\frac{2\sq2}{3}P_{1EW\BD},
   \non\\
A(\bar B^0_s\to n\overline{\Sigma^{*0}})
   &=&-T_{1\BD}+P_\BD+\frac{2}{3}(P_{1EW\BD}-3P_{2EW\BD}),
   \non\\
A(\bar B^0_s\to\Sigma^{0}\overline{\Xi^{*0}})
   &=&-2T_{2\BD}-P_\BD+\frac{1}{3}(P_{1EW\BD}-6P_{2EW\BD}),
   \non\\
A(\bar B^0_s\to\Sigma^{-}\overline{\Xi^{*-}})
   &=&-\sq2P_\BD+\frac{\sq2}{3}P_{1EW\BD},
   \non\\   
A(\bar B^0_s\to\Xi^{-}\overline{\Omega^-})
   &=&-\sq6 P_\BD+\sq{\frac{2}{3}}P_{1EW\BD},
   \non\\
A(\bar B^0_s\to\Lambda\overline{\Xi^{*0}})
   &=&\frac{2}{\sq3}(T_{1\BD}-T_{2\BD})
         -\sq3 P_\BD
         -\frac{1}{\sq3}(P_{1EW\BD}-2P_{2EW\BD}),
\label{eq: BDBs, DS=0} 
\en
while those for $\Delta S=1$ transitions are given by
\be
A(B^-\to \Sigma^+\overline{\Delta^{++}})
   &=&\sq6 (T'_{1\BD}-2T'_{2\BD})
          -\sq6P'_\BD
          -2\sq{\frac{2}{3}}P'_{1EW\BD}
          -\sq6 A'_\BD,
   \non\\
A(B^-\to\Sigma^0\overline{\Delta^+})
   &=&-T'_{1\BD}+2T'_{2\BD}
   +2P'_\BD
   +\frac{1}{3}P'_{1EW\BD}
   +2 A'_\BD,
   \non\\
A(B^-\to\Sigma^-\overline{\Delta^0})
   &=&\sq2 P'_\BD
         -\frac{\sq2}{3}P'_{1EW\BD}
         +\sq2 A'_\BD,
   \non\\
A(B^-\to\Xi^{0}\overline{\Sigma^{*+}})
   &=&\sq2T'_{1\BD}
         -\sq2 P'_\BD
         -\frac{2\sq2}{3}(P'_{1EW\BD}-3P'_{2EW\BD})
         -\sq2 A'_\BD,
   \non\\   
A(B^-\to\Xi^{-}\overline{\Sigma^{*0}})
   &=&P'_\BD-\frac{1}{3}P'_{1EW\BD}+ A'_\BD,
   \non\\
A(B^-\to\Lambda\overline{\Delta^{+}})
   &=&\frac{1}{\sq3}(T'_{1\BD}+2T'_{2\BD})
         -\frac{1}{\sq3}(P'_{1EW\BD}-4P'_{2EW\BD}),
\label{eq: BDBm, DS=-1}
\en
\be
A(\bar B^0\to \Sigma^{+}\overline{\Delta^+})
   &=&\sq2 (T'_{1\BD}-2 T'_{2\BD})
          -\sq2 P'_\BD
          -\frac{2\sq2}{3}P'_{1EW\BD},
   \non\\
A(\bar B^0\to\Sigma^{0}\overline{\Delta^0})
   &=&-T'_{1\BD}+2T'_{2\BD}
         +2P'_\BD
         +\frac{1}{3}P'_{1EW\BD},
   \non\\
A(\bar B^0\to\Sigma^{-}\overline{\Delta^-})
   &=&\sq6 P'_\BD
         -\sq{\frac{2}{3}}P'_{1EW\BD},
   \non\\
A(\bar B^0\to\Xi^{0}\overline{\Sigma^{*0}})
   &=&T'_{1\BD}
         -P'_\BD
         -\frac{2}{3}(P'_{1EW\BD}-3P'_{2EW\BD}),
   \non\\
A(\bar B^0\to\Xi^{-}\overline{\Sigma^{*-}})
   &=&\sq2P'_\BD
         -\frac{\sq2}{3}P'_{1EW\BD},
   \non\\   
A(\bar B^0\to\Lambda\overline{\Delta^0})
   &=&\frac{1}{\sq3}(T'_{1\BD}+2T'_{2\BD})
         -\frac{1}{\sq3}(P'_{1EW\BD}-4P'_{2EW\BD}),
\label{eq: BDB0, DS=-1}
\en
and
\be
A(\bar B^0_s\to p\overline{\Delta^+})
   &=&-\sq2E'_\BD,
   \non\\
A(\bar B^0_s\to n\overline{\Delta^0})
   &=&-\sq2E'_\BD,
   \non\\
A(\bar B^0_s\to\Sigma^{+}\overline{\Sigma^{*+}})
   &=&\sq2(T'_{1\BD}-2T'_{2\BD})
         -\sq2P'_\BD
         -\frac{2\sq2}{3}P'_{1EW\BD}
        +\sq2E'_\BD,
   \non\\
A(\bar B^0_s\to\Sigma^{0}\overline{\Sigma^{*0}})
   &=&-\frac{1}{\sq2}(T'_{1\BD}-2T'_{2\BD})
         +\sq2 P'_\BD
         +\frac{1}{3\sq2}P'_{1EW\BD}
         -\frac{1}{\sq2} E'_\BD,
   \non\\
A(\bar B^0_s\to\Sigma^{-}\overline{\Sigma^{*-}})
   &=&\sq2P'_\BD
         -\frac{\sq2}{3}P'_{1EW\BD},
   \non\\   
A(\bar B^0_s\to\Xi^{0}\overline{\Xi^{*0}})
   &=&\sq2T'_{1\BD}
         -\sq2P'_\BD
         -\frac{2\sq2}{3}(P'_{1EW\BD}-3P'_{2EW\BD})
         +\sq2 E'_\BD,
   \non\\
A(\bar B^0_s\to\Xi^{-}\overline{\Xi^{*-}})
   &=&\sq2P'_\BD
         -\frac{\sq2}{3}P'_{1EW\BD},
   \non\\
A(\bar B^0_s\to\Lambda\overline{\Sigma^{*0}})
   &=&\frac{1}{\sq6}(T'_{1\BD}+2T'_{2\BD})
          -\frac{1}{\sq6}(P'_{1EW\BD}-4P'_{2EW\BD})
          +\sq{\frac{3}{2}} E'_\BD.   
\label{eq: BDBs, DS=-1}
\en

\subsection{$\overline B$ to decuplet-anti-octet baryonic decays}

The full $\bar B\to\DB$ decay amplitudes for $\Delta S=0$ processes are given by
\be
A(B^-\to\Delta^0\overline{p})
   &=&\sq2T_{1\DB}
          -\sq2 P_\DB
          +\frac{\sq2}{3}(3P_{1EW\DB}+P_{2EW\DB})
          -\sq2 A_\DB,
   \non\\
A(B^-\to\Delta^-\overline{n})
   &=&-\sq6 P_\DB
         +\sq{\frac{2}{3}}P_{2EW\DB}
         -\sq6 A_\DB,
   \non\\
A(B^-\to\Sigma^{*0}\overline{\Sigma^{+}})
   &=&-T_{1\DB}
          +P_\DB
          -\frac{1}{3}(3P_{1EW\DB}+P_{2EW\DB})
          +A_\DB,
   \non\\   
A(B^-\to\Sigma^{*-}\overline{\Sigma^{0}})
   &=&-P_\DB
          +\frac{1}{3}P_{2EW\DB}
          - A_\DB,
   \non\\
A(B^-\to\Xi^{*-}\overline{\Xi^{0}})
   &=&\sq2 P_\DB
         -\frac{\sq2}{3}P_{2EW\DB}
         +\sq2A_\DB,
   \non\\
A(B^-\to\Sigma^{*-}\overline{\Lambda})
   &=&\sq3 P_\DB
          -\frac{1}{\sq3}P_{2EW\DB}
          +\sq3 A_\DB,
\label{eq: DBBm, DS=0}
\en
\be
A(\bar B^0\to\Delta^+\overline{p})
   &=&\sq2T_{2\DB}
          +\sq2 P_\DB
          +\frac{2\sq2}{3}P_{2EW\DB}
          -\sq2E_\DB,
   \non\\
A(\bar B^0\to\Delta^0\overline{n})
   &=&\sq2(T_{1\DB}+T_{2\DB})
         +\sq2 P_\DB
         +\frac{\sq2}{3}(3P_{1EW\DB}+2P_{2EW\DB})
         -\sq2E_\DB,
   \non\\
A(\bar B^0\to\Sigma^{*+}\overline{\Sigma^{+}})
   &=&\sq2 E_\DB,
   \non\\
A(\bar B^0\to\Sigma^{*0}\overline{\Sigma^{0}})
   &=&\frac{1}{\sq2}T_{1\DB}
         -\frac{1}{\sq2}P_\DB
         +\frac{1}{3\sq2}(3P_{1EW\DB}+P_{2EW\DB})
         -\frac{1}{\sq2} E_\DB,
   \non\\
A(\bar B^0\to\Sigma^{*-}\overline{\Sigma^{-}})
   &=&-\sq2P_\DB
          +\frac{\sq2}{3}P_{2EW\DB},
   \non\\   
A(\bar B^0\to\Xi^{*0}\overline{\Xi^{0}})
   &=&\sq2E_{\DB},
   \non\\
A(\bar B^0\to\Xi^{*-}\overline{\Xi^{-}})
   &=&-\sq2 P_\DB
          +\frac{\sq2}{3}P_{2EW\DB},
   \non\\
A(\bar B^0\to\Sigma^{*0}\overline{\Lambda})
   &=&-\frac{1}{\sq6}(T_{1\DB}+2T_{2\DB})
          -\sq{\frac{3}{2}}P_\DB
          -\frac{1}{\sq6}(P_{1EW\DB}+P_{2EW\DB})
          +\sq{\frac{3}{2}} E_\DB,   
   \non\\
\label{eq: DBB0, DS=0}
\en
and
\be
A(\bar B^0_s\to \Delta^+\overline{\Sigma^{+}})
   &=&-\sq2 T_{2\DB}
          -\sq2 P_\DB
          -\frac{2\sq2}{3}P_{2EW\DB},
   \non\\
A(\bar B^0_s\to\Delta^0\overline{\Sigma^{0}})
   &=&T_{2\DB}
         +2 P_\DB
         +\frac{1}{3}P_{2EW\DB},
   \non\\
A(\bar B^0_s\to\Delta^-\overline{\Sigma^{-}})
   &=&\sq6 P_\DB
          -\sq{\frac{2}{3}}P_{2EW\DB},
   \non\\
A(\bar B^0_s\to\Sigma^{*0}\overline{\Xi^{0}})
   &=&-(T_{1\DB}+T_{2\DB})
          -P_\DB
          -\frac{1}{3}(3P_{1EW\DB}+2P_{2EW\DB}),
   \non\\
A(\bar B^0_s\to\Sigma^{*-}\overline{\Xi^{-}})
   &=&\sq2P_\DB
          -\frac{\sq2}{3}P_{2EW\DB},
   \non\\   
A(\bar B^0_s\to\Delta^0\overline{\Lambda})
   &=&-\frac{1}{\sq3}(2T_{1\DB}+T_{2\DB})
          -\frac{1}{\sq3}(2P_{1EW\DB}+P_{2EW\DB}),
\label{eq: DBBs, DS=0}
\en
while those for $\Delta S=1$ transitions are given by
\be
A(B^-\to\Sigma^{*0}\overline{p})
   &=&T'_{1\DB}- P'_\DB
          +\frac{1}{3}(3P'_{1EW\DB}+P'_{2EW\DB})
          -A'_\DB,
   \non\\
A(B^-\to\Sigma^{*-}\overline{n})
   &=&-\sq2 P'_\DB
          +\frac{\sq2}{3}P'_{2EW\DB}
          -\sq2A'_\DB,
   \non\\
A(B^-\to\Xi^{*0}\overline{\Sigma^{+}})
   &=&-\sq2T'_{1\DB}
         +\sq2 P'_\DB
         -\frac{\sq2}{3}(3P'_{1EW\DB}+P'_{2EW\DB})
         +\sq2 A'_\DB,
   \non\\
A(B^-\to\Xi^{*-}\overline{\Sigma^{0}})
   &=&-P'_\DB
          +\frac{1}{3}P'_{2EW\DB}
          -A'_\DB,
   \non\\   
A(B^-\to\Omega^-\overline{\Xi^{0}})
   &=&\sq6 P'_\DB
          -\sq{\frac{2}{3}}P'_{2EW\DB}
          +\sq6 A'_\DB,
   \non\\
A(B^-\to\Xi^{*-}\overline{\Lambda})
   &=&\sq3 P'_\DB
          -\frac{1}{\sq3}P'_{2EW\DB}
          +\sq3A'_\DB,    
\label{eq: DBBm, DS=-1}
\en
\be
A(\bar B^0\to \Sigma^{*+}\overline{p})
   &=&\sq2 T'_{2\DB}
         +\sq2 P'_\DB
         +\frac{2\sq2}{3}P'_{2EW\DB},
   \non\\
A(\bar B^0\to\Sigma^{*0}\overline{n})
   &=&T'_{1\DB}+T'_{2\DB}
          +P'_\DB
          +\frac{1}{3}(3P'_{1EW\DB}+2P'_{2EW\DB}),
   \non\\
A(\bar B^0\to\Xi^{*0}\overline{\Sigma^{0}})
   &=&T'_{1\DB}
          -P'_\DB
          +\frac{1}{3}(3P'_{1EW\DB}+P'_{2EW\DB}),
   \non\\
A(\bar B^0\to\Xi^{*-}\overline{\Sigma^{-}})
   &=&-\sq2P'_\DB
          +\frac{\sq2}{3}P'_{2EW\DB},
   \non\\   
A(\bar B^0\to\Omega^-\overline{\Xi^{-}})
   &=&-\sq6 P'_\DB
          +\sq{\frac{2}{3}}P_{2EW\DB},
      \non\\
A(\bar B^0\to\Xi^{*0}\overline{\Lambda})
   &=&-\frac{1}{\sq3}(T'_{1\DB}+2T'_{2\DB})
          -\sq3 P'_\DB
          -\frac{1}{\sq3}(P'_{1EW\DB}+P'_{2EW\DB}),
\label{eq: DBB0, DS=-1}
\en
and
\be
A(\bar B^0_s\to\Delta^+\overline{p})
   &=&-\sq2 E'_\DB,
   \non\\
A(\bar B^0_s\to\Delta^0\overline{n})
   &=&-\sq2E'_\DB,
   \non\\
A(\bar B^0_s\to\Sigma^{*+}\overline{\Sigma^{+}})
   &=&-\sq2T'_{2\DB}
          -\sq2P'_\DB
          -\frac{2\sq2}{3}P'_{2EW\DB}
          +\sq2 E'_\DB,
   \non\\
A(\bar B^0_s\to\Sigma^{*0}\overline{\Sigma^{0}})
   &=& \frac{1}{\sq2}T'_{2\DB}
         +\sq2P'_\DB
         +\frac{1}{3\sq2}P'_{2EW\DB}
         -\frac{1}{\sq2} E'_\DB,
   \non\\
A(\bar B^0_s\to\Sigma^{*-}\overline{\Sigma^{-}})
   &=&\sq2P'_\DB
          -\frac{\sq2}{3}P'_{2EW\DB},
   \non\\   
A(\bar B^0_s\to\Xi^{*0}\overline{\Xi^{0}})
   &=&-\sq2(T'_{1\DB}+T'_{2\DB})
          -\sq2P'_\DB
          -\frac{\sq2}{3}(3P'_{1EW\DB}+2P'_{2EW\DB})
          +\sq2 E'_\DB,
   \non\\
A(\bar B^0_s\to\Xi^{*-}\overline{\Xi^{-}})
   &=&\sq2 P'_\DB
         -\frac{\sq2}{3}P'_{2EW\DB},
   \non\\
A(\bar B^0_s\to\Sigma^{*0}\overline{\Lambda})
   &=&-\frac{1}{\sq6}(2T'_{1\DB}+T'_{2\DB})
          -\frac{1}{\sq6}(2P'_{1EW\DB}+P'_{2EW\DB})
          +\sq{\frac{3}{2}} E'_\DB.
\label{eq: DBBs, DS=-1}
\en

\subsection{$\overline B$ to decuplet-anti-decuplet baryonic decays}

The full $\bar B\to\DD$ decay amplitudes for $\Delta S=0$ processes are given by
\be
A(B^-\to \Delta^+\overline{\Delta^{++}})
   &=&2\sq3 T_\DD+2\sq3 P_\DD+\frac{4}{\sq3}P_{EW\DD}+2\sq3A_\DD,
   \non\\
A(B^-\to\Delta^0\overline{\Delta^+})
   &=&2T_\DD+4 P_\DD+\frac{2}{3}P_{EW\DD}+4A_\DD,
   \non\\
A(B^-\to\Delta^-\overline{\Delta^0})
   &=&2\sq3 P_\DD-\frac{2}{\sq3}P_{EW\DD}+2\sq3A_\DD,
   \non\\
A(B^-\to\Sigma^{*0}\overline{\Sigma^{*+}})
   &=&\sq2T_\DD+2\sq2 P_\DD+\frac{\sq2}{3}P_{EW\DD}+2\sq2A_\DD,
   \non\\   
A(B^-\to\Sigma^{*-}\overline{\Sigma^{*0}})
   &=&2\sq2 P_\DD-\frac{2\sq2}{3}P_{EW\DD}+2\sq2A_\DD,
   \non\\
A(B^-\to\Xi^{*-}\overline{\Xi^{*0}})
   &=&2 P_\DD-\frac{2}{3}P_{EW\DD}+2A_\DD,
\label{eq: DDBm, DS=0}
\en
\be
A(\bar B^0\to \Delta^{++}\overline{\Delta^{++}})
   &=&6E_\DD+18PA_\DD,
   \non\\
A(\bar B^0\to\Delta^+\overline{\Delta^+})
   &=&2T_\DD+2 P_\DD+\frac{4}{3}P_{EW\DD}+4E_\DD+18PA_\DD,
   \non\\
A(\bar B^0\to\Delta^0\overline{\Delta^0})
   &=&2T_\DD+4 P_\DD+\frac{2}{3}P_{EW\DD}+2E_\DD+18PA_\DD,
   \non\\
A(\bar B^0\to\Delta^-\overline{\Delta^-})
   &=&6P_\DD-2P_{EW\DD}+18PA_\DD,
   \non\\
A(\bar B^0\to\Sigma^{*+}\overline{\Sigma^{*+}})
   &=&\frac{2}{3}6E_\DD+18PA_\DD,
   \non\\
A(\bar B^0\to\Sigma^{*0}\overline{\Sigma^{*0}})
   &=&T_\DD+2P_\DD+\frac{1}{3}P_{EW\DD}+2E_\DD+18PA_\DD,
   \non\\
A(\bar B^0\to\Sigma^{*-}\overline{\Sigma^{*-}})
   &=&4P_\DD-\frac{4}{3}P_{EW\DD}+18PA_\DD,
   \non\\   
A(\bar B^0\to\Xi^{*0}\overline{\Xi^{*0}})
   &=&\frac{1}{3}E_{\DD}+18PA_\DD,
   \non\\
A(\bar B^0\to\Xi^{*-}\overline{\Xi^{*-}})
   &=&2 P_\DD-\frac{2}{3}P_{EW\DD}+18PA_\DD,
   \non\\
A(\bar B^0\to\Omega^{-}\overline{\Omega^{-}})
   &=&18PA_\DD,      
\label{eq: DDB0, DS=0}
\en
and
\be
A(\bar B^0_s\to \Delta^+\overline{\Sigma^{*+}})
   &=&2 T_\DD+2 P_\DD+\frac{4}{3}P_{EW\DD},
   \non\\
A(\bar B^0_s\to\Delta^0\overline{\Sigma^{*0}})
   &=&\sq2T_\DD+2\sq2 P_\DD+\frac{\sq2}{3}P_{EW\DD},
   \non\\
A(\bar B^0_s\to\Delta^-\overline{\Sigma^{*-}})
   &=&2\sq3 P_\DD-\frac{2}{\sq3}P_{EW\DD},
   \non\\
A(\bar B^0_s\to\Sigma^{*0}\overline{\Xi^{*0}})
   &=&\sq2T_\DD+2\sq2 P_\DD+\frac{\sq2}{3}P_{EW\DD},
   \non\\
A(\bar B^0_s\to\Sigma^{*-}\overline{\Xi^{*-}})
   &=&4P_\DD-\frac{4}{3}P_{EW\DD},
   \non\\   
A(\bar B^0_s\to\Xi^{*-}\overline{\Omega^-})
   &=&2\sq3 P_\DD-\frac{2}{\sq3}P_{EW\DD},
\label{eq: DDBs, DS=0}
\en
while those for $\Delta S=1$ transitions are given by
\be
A(B^-\to \Sigma^{*+}\overline{\Delta^{++}})
   &=&2\sq3 T'_\DD+2\sq3 P'_\DD+\frac{4}{\sq3}P'_{EW\DD}+2{\sq3}A'_\DD,
   \non\\
A(B^-\to\Sigma^{*0}\overline{\Delta^+})
   &=&\sq2T'_\DD+2\sq2 P'_\DD+\frac{\sq2}{3}P'_{EW\DD}+2{\sq2}A'_\DD,
   \non\\
A(B^-\to\Sigma^{*-}\overline{\Delta^0})
   &=&2 P'_\DD-\frac{2}{3}P'_{EW\DD}+2A'_\DD,
   \non\\
A(B^-\to\Xi^{*0}\overline{\Sigma^{*+}})
   &=&2T'_\DD+4 P'_\DD+\frac{2}{3}P'_{EW\DD}+4A'_\DD,
   \non\\
A(B^-\to\Xi^{*-}\overline{\Sigma^{*0}})
   &=&2\sq2 P'_\DD-\frac{2\sq2}{3}P'_{EW\DD}+2{\sq2}A'_\DD,
   \non\\   
A(B^-\to\Omega^-\overline{\Xi^{*0}})
   &=&2\sq3 P'_\DD-\frac{2}{\sq3}P'_{EW\DD}+2{\sq3}A'_\DD,
\label{eq: DDBm, DS=-1}
\en
\be
A(\bar B^0\to \Sigma^{*+}\overline{\Delta^+})
   &=&2 T'_\DD+2 P'_\DD+\frac{4}{3}P'_{EW\DD},
   \non\\
A(\bar B^0\to\Sigma^{*0}\overline{\Delta^0})
   &=&\sq2T'_\DD+2\sq2 P'_\DD+\frac{\sq2}{3}P'_{EW\DD},
   \non\\
A(\bar B^0\to\Sigma^{*-}\overline{\Delta^-})
   &=&2\sq3 P'_\DD-\frac{2}{\sq3}P'_{EW\DD},
   \non\\
A(\bar B^0\to\Xi^{*0}\overline{\Sigma^{*0}})
   &=&\sq2T'_\DD+2\sq2 P'_\DD+\frac{\sq2}{3}P'_{EW\DD},
   \non\\
A(\bar B^0\to\Xi^{*-}\overline{\Sigma^{*-}})
   &=&4P'_\DD-\frac{4}{3}P'_{EW\DD},
   \non\\   
A(\bar B^0\to\Omega^-\overline{\Xi^{*-}})
   &=&2\sq3 P'_\DD-\frac{2}{\sq3}P'_{EW\DD},
\label{eq: DDB0, DS=-1}
\en
and
\be
A(\bar B^0_s\to \Delta^{++}\overline{\Delta^{++}})
   &=&6E'_\DD+18PA'_\DD,
   \non\\
A(\bar B^0_s\to\Delta^+\overline{\Delta^+})
   &=&4E'_\DD+18PA'_\DD,
   \non\\
A(\bar B^0_s\to\Delta^0\overline{\Delta^0})
   &=&2E'_\DD+18PA'_\DD,
   \non\\
A(\bar B^0_s\to\Delta^-\overline{\Delta^-})
   &=&18PA'_\DD,
   \non\\
A(\bar B^0_s\to\Sigma^{*+}\overline{\Sigma^{*+}})
   &=&2T'_\DD+2P'_\DD+\frac{4}{3}P'_{EW\DD}+4E'_\DD+18PA'_\DD,
   \non\\
A(\bar B^0_s\to\Sigma^{*0}\overline{\Sigma^{*0}})
   &=&T'_\DD+2P'_\DD+\frac{1}{3}P'_{EW\DD}+2E'_\DD+18PA'_\DD,
   \non\\
A(\bar B^0_s\to\Sigma^{*-}\overline{\Sigma^{*-}})
   &=&2P'_\DD-\frac{2}{3}P'_{EW\DD}+18PA'_\DD,
   \non\\   
A(\bar B^0_s\to\Xi^{*0}\overline{\Xi^{*0}})
   &=&2T'_\DD+4P'_\DD+\frac{2}{3}P'_{EW\DD}+2E'_{\DD}+18PA'_\DD,
   \non\\
A(\bar B^0_s\to\Xi^{*-}\overline{\Xi^{*-}})
   &=&4 P'_\DD-\frac{4}{3}P'_{EW\DD}+18PA'_\DD,
   \non\\
A(\bar B^0_s\to\Omega^{-}\overline{\Omega^{-}})
   &=&6 P_\DD-2P_{EW\DD}+18PA'_\DD.
\label{eq: DDBs, DS=-1}
\en

\section{Formulas for decay amplitudes and rates}\label{appendix:formulas}

In general the decay amplitudes of $\overline B$ to final states with octet
baryon (${\cal B}$) and decuplet baryons (${\cal D}$) can be
expressed as~\cite{Jarfi:1990ej}
\begin{eqnarray}
A(\overline B\to {\cal B}_1 \overline {\cal B}_2)&=&\bar
u_1(A_{{\cal B}\overline {\cal B}}+\gamma_5 B_{{\cal B}\overline
{\cal B}}) v_2,
\nonumber\\
A(\overline B\to {\cal D}_1 \overline {\cal B}_2)&=&i
\frac{q^\mu}{m_B} \bar u^\mu_1(A_{{\cal D}\overline {\cal
B}}+\gamma_5 B_{{\cal D}\overline {\cal B}}) v_2,
\nonumber\\
A(\overline B\to {\cal B}_1 \overline {\cal D}_2)&=&i
\frac{q^\mu}{m_B}\bar u_1(A_{{\cal B}\overline {\cal D}}+\gamma_5
B_{{\cal B}\overline {\cal D}}) v^\mu_2,
\nonumber\\
A(\overline B\to {\cal D}_1 \overline {\cal D}_2)&=&\bar
u^\mu_1(A_{{\cal D}\overline {\cal D}}+\gamma_5 B_{{\cal
D}\overline {\cal D}}) v_{2\mu}+\frac{q^\mu q^\nu}{m^2_B}\bar
u^\mu_1(C_{{\cal D}\overline {\cal D}}+\gamma_5 D_{{\cal
D}\overline {\cal D}}) v_{2\nu},
\label{eq: A1}
\end{eqnarray}
where $q=p_1-p_2$ and $u^\mu,\,v^\mu$ are the Rarita-Schwinger
vector spinors for a spin-$\frac{3}{2}$ particle, where~\cite{Moroi:1995fs}
\be
u_\mu(\pm\frac{3}{2})&=&\epsilon_\mu(\pm1) u(\pm\frac{1}{2})
\non\\
u_\mu(\pm\frac{1}{2})&=&(\epsilon_\mu(\pm1)
u(\mp\frac{1}{2})+\sqrt{2}\,\epsilon_\mu(0)
u(\pm\frac{1}{2}))/\sqrt3,
\en 
with $\epsilon_\mu(\lambda)$ the usual polarization vector and $u(s)$ the spinor.
Note that
\be
q\cdot\epsilon(\lambda)_{1,2}=\mp\,\delta_{\lambda,0}\,m_B
p_c/m_{1,2},
\en 
where $p_c$ is the baryon momentum in the $B$ rest
frame and
$\epsilon^*_1(0)\cdot\epsilon_2(0)=(m_B^2-m^2_1-m^2_2)/2m_1 m_2$
is the largest product among $\epsilon^*_1(\lambda_1)\cdot
\epsilon_2(\lambda_2)$. 
We can now express the amplitudes in Eq. (\ref{eq: A1}) as
\begin{eqnarray}
A(\overline B\to {\cal D}_1 \overline {\cal B}_2)&=&-i
\sqrt{\frac{2}{3}}\frac{p_{cm}}{m_1} \bar u_1(A_{{\cal D}\overline
{\cal B}}+\gamma_5 B_{{\cal D}\overline {\cal B}}) v_2,
\nonumber\\
A(\overline B\to {\cal B}_1 \overline {\cal D}_2)&=&i
\sqrt{\frac{2}{3}}\frac{p_{cm}}{m_2}\bar u_1(A_{{\cal B}\overline
{\cal D}}+\gamma_5 B_{{\cal B}\overline {\cal D}}) v_2,
\nonumber\\
A(\overline B\to {\cal D}_1 \overline {\cal
D}_2)&\simeq&\frac{m_B^2}{3m_1m_2}\bar u_1(A^\prime_{{\cal
D}\overline {\cal D}}+\gamma_5 B^\prime_{{\cal D}\overline {\cal
D}})v_2, \label{eq:largemB}
\end{eqnarray}
where $A^\prime_{{\cal D}\overline {\cal D}}=A_{{\cal D}\overline
{\cal D}}-2(p_{cm}/m_B)^2 C_{{\cal D}\overline {\cal D}}$ and
$B^\prime_{{\cal D}\overline {\cal D}}=B_{{\cal D}\overline {\cal
D}}-2(p_{cm}/m_B)^2 D_{{\cal D}\overline {\cal D}}$ and decuplets are only or dominantly in the $\pm\frac{1}{2}$-helicity states. 
All four $\overline B\to {\mathbf B}_1\overline {\mathbf B}_2$
(${\mathbf B}\overline {\mathbf B}={\cal B} \overline {\cal B}$,
${\cal D} \overline {\cal B}$, ${\cal B} \overline {\cal D}$,
${\cal D} \overline {\cal D}$) decays can be effectively expressed as
\begin{equation}
A(\overline B\to {\mathbf B}_1\overline {\mathbf B}_2)
 =\bar u_1({\bf A}+\gamma_5 {\bf B})v_2,
 \label{eq:asymptoticform}
\end{equation}
and the rates are given by
\be
\Gamma(\overline B\to {\mathbf B}_1\overline {\mathbf B}_2)
=\frac{p_{cm}}{8 \pi m_B^2} [(2 m_B^2 - 2 (m_{{\mathbf B}_1}+ m_{{\mathbf B}_1})^2) {\bf A}^2+
(2 m_B^2 - 2 (m_{{\mathbf B}_1}- m_{{\mathbf B}_1})^2) {\bf B}^2],
\en
where
\be
{\bf A}_{\BB}=A_{\BB},\,
{\bf A}_{\BD}=i
\sqrt{\frac{2}{3}}\frac{p_{cm}}{m_{\overline{\cal D}}}A_{\DB},\,
{\bf A}_{\DB}=-i
\sqrt{\frac{2}{3}}\frac{p_{cm}}{m_{\cal D}}A_{\BD},\,
{\bf A}_{\DD}=\frac{m_B^2}{3m_{\cal D}m_{\overline{\cal D}}}A'_{\DD},
\en
and
\be
{\bf B}_{\BB}=B_{\BB},\,
{\bf B}_{\BD}=i
\sqrt{\frac{2}{3}}\frac{p_{cm}}{m_{\overline{\cal D}}}B_{\DB},\,
{\bf B}_{\DB}=-i
\sqrt{\frac{2}{3}}\frac{p_{cm}}{m_{\cal D}}B_{\BD},\,
{\bf B}_{\DD}=\frac{m_B^2}{3m_{\cal D}m_{\overline{\cal D}}}B'_{\DD}.
\en
Note that in the asymptotic limit, $m_B\gg m_{\mathcal B}$, one has 
$p_{cm}\to {m_B}/{2}$.




\begin{thebibliography}{99}

\bibitem{Aaij:2016xfa} 
  R.~Aaij {\it et al.} [LHCb Collaboration],
  ``Evidence for the two-body charmless baryonic decay $B^+ \to p \bar\Lambda$,''
  arXiv:1611.07805 [hep-ex].


\bibitem{Cheng:2001tr} 
  H.~Y.~Cheng and K.~C.~Yang,
  ``Charmless exclusive baryonic B decays,''
  Phys.\ Rev.\ D {\bf 66}, 014020 (2002)
  doi:10.1103/PhysRevD.66.014020
  [hep-ph/0112245].
  
\bibitem{Chua:2013zga} 
  C.~K.~Chua,
  ``Charmless Two-body Baryonic $B_{u,d,s}$ Decays Revisited,''
  Phys.\ Rev.\ D {\bf 89}, no. 5, 056003 (2014)
  [arXiv:1312.2335 [hep-ph]].
  

\bibitem{Aaij:2013fta} 
  R. Aaij {\it et al.}  [LHCb Collaboration],
  ``First evidence for the two-body charmless baryonic decay $B^0 \to p \bar{p}$,''
  JHEP {\bf 1310}, 005 (2013)
  [arXiv:1308.0961 [hep-ex]].

\bibitem{Hsiao:2014zza} 
  Y.~K.~Hsiao and C.~Q.~Geng,
  ``Violation of partial conservation of the axial-vector current and two-body baryonic $B$ and D$_s$ decays,''
  Phys.\ Rev.\ D {\bf 91}, no. 7, 077501 (2015)
  doi:10.1103/PhysRevD.91.077501
  [arXiv:1407.7639 [hep-ph]].


\bibitem{Deshpande:1987nc}
N.~G.~Deshpande, J.~Trampetic and A.~Soni,
``Remarks On B Decays Into Baryonic Modes And Possible Implications For V(Ub),''
Mod.\ Phys.\ Lett.\  {\bf 3A}, 749 (1988).


\bibitem{Jarfi:1990ej}
M.~Jarfi, O.~Lazrak, A.~Le Yaouanc, L.~Oliver, O.~Pene and
J.~C.~Raynal,
``Decays Of B Mesons Into Baryon - Anti-Baryon,''
Phys.\ Rev.\ D {\bf 43}, 1599 (1991).


\bibitem{Cheng:2001ub}
H.~Y.~Cheng and K.~C.~Yang,
``Charmful baryonic B decays anti-B0 $\to$ Lambda/c anti-p and anti-B $\to$  Lambda/c anti-p pi (rho),''
Phys.\ Rev.\ D {\bf 65}, 054028 (2002) [Erratum-ibid.\ D {\bf 65},
099901 (2002)] [arXiv:hep-ph/0110263].


\bibitem{Chernyak:ag}
V.~L.~Chernyak and I.~R.~Zhitnitsky,
``B Meson Exclusive Decays Into Baryons,''
Nucl.\ Phys.\ B {\bf 345}, 137 (1990).


\bibitem{Ball:1990fw}
P.~Ball and H.~G.~Dosch,
``Branching Ratios Of Exclusive Decays Of Bottom Mesons Into Baryon - Anti-Baryon Pairs,''
Z.\ Phys.\ C {\bf 51}, 445 (1991).


\bibitem{Chang:2001jt}
C.~H.~Chang and W.~S.~Hou,
``B meson decays to baryons in the diquark model,''
Eur.\ Phys.\ J.\ C {\bf 23}, 691 (2002) [arXiv:hep-ph/0112219].


\bibitem{Gronau:1987xq}
M.~Gronau and J.~L.~Rosner,
``Charmless B Decays Involving Baryons,''
Phys.\ Rev.\ D {\bf 37}, 688 (1988).


\bibitem{He:re}
X.~G.~He, B.~H.~McKellar and D.~d.~Wu,
``SU(6) Prediction Of $\Lambda_c$ Branching Ratio in B Meson Decays,''
Phys.\ Rev.\ D {\bf 41}, 2141 (1990).


\bibitem{Sheikholeslami:fa}
S.~M.~Sheikholeslami and M.~P.~Khanna,
``B Meson Weak Decays Into Baryon Anti-Baryon Pairs in SU(3),''
Phys.\ Rev.\ D {\bf 44}, 770 (1991).


\bibitem{Luo:2003pv}
Z.~Luo and J.~L.~Rosner,
  ``Final state phases in $B\to$ baryon anti-baryon decays,''
  Phys.\ Rev.\ D {\bf 67}, 094017 (2003)
  [hep-ph/0302110].


\bibitem{review}
H.~-Y.~Cheng and J.~G.~Smith,
  ``Charmless Hadronic B-Meson Decays,''
  Ann.\ Rev.\ Nucl.\ Part.\ Sci.\  {\bf 59}, 215 (2009)
  [arXiv:0901.4396 [hep-ph]].


\bibitem{review1}
 H.~-Y.~Cheng,
  ``Exclusive baryonic B decays Circa 2005,''
  Int.\ J.\ Mod.\ Phys.\ A {\bf 21}, 4209 (2006)
  [hep-ph/0603003].


\bibitem{Cheng:2014qxa} 
  H.~Y.~Cheng and C.~K.~Chua,
  ``On the smallness of Tree-dominated Charmless Two-body Baryonic $B$ Decay Rates,''
  Phys.\ Rev.\ D {\bf 91}, no. 3, 036003 (2015)
  [arXiv:1412.8272 [hep-ph]].


  
\bibitem{Zeppenfeld:1980ex}
D.~Zeppenfeld,
``SU(3) Relations For B Meson Decays,''
Z.\ Phys.\ C {\bf 8}, 77 (1981).


\bibitem{Chau:tk}
L.~L.~Chau and H.~Y.~Cheng,
``Analysis Of Exclusive Two-Body Decays Of Charm Mesons Using The Quark Diagram Scheme,''
Phys.\ Rev.\ D {\bf 36}, 137 (1987).


\bibitem{Chau:1990ay}
L.~L.~Chau, H.~Y.~Cheng, W.~K.~Sze, H.~Yao and B.~Tseng,
``Charmless Nonleptonic Rare Decays Of B Mesons,''
Phys.\ Rev.\ D {\bf 43}, 2176 (1991) [Erratum-ibid.\ D {\bf 58},
019902 (1998)].


\bibitem{Gronau:1994rj}
M.~Gronau, O.~F.~Hernandez, D.~London and J.~L.~Rosner,
``Decays of B mesons to two light pseudoscalars,''
Phys.\ Rev.\ D {\bf 50}, 4529 (1994) [arXiv:hep-ph/9404283].


\bibitem{Gronau:1995hn}
M.~Gronau, O.~F.~Hernandez, D.~London and J.~L.~Rosner,
``Electroweak penguins and two-body B decays,''
Phys.\ Rev.\ D {\bf 52}, 6374 (1995) [arXiv:hep-ph/9504327].

\bibitem{Cheng:2014rfa} 
  H.~Y.~Cheng, C.~W.~Chiang and A.~L.~Kuo,
  ``Updating $B\to PP,VP$ decays in the framework of flavor symmetry,''
  Phys.\ Rev.\ D {\bf 91}, no. 1, 014011 (2015)
  doi:10.1103/PhysRevD.91.014011
  [arXiv:1409.5026 [hep-ph]].

\bibitem{Savage:ub}
M.~J.~Savage and M.~B.~Wise,
``SU(3) Predictions For Nonleptonic B Meson Decays,''
Phys.\ Rev.\ D {\bf 39}, 3346 (1989) [Erratum-ibid.\ D {\bf 40},
3127 (1989)].

\bibitem{Chua:2003it} 
  C.~-K.~Chua,
  ``Charmless two body baryonic B decays,''
  Phys.\ Rev.\ D {\bf 68}, 074001 (2003)
  [hep-ph/0306092].
  


\bibitem{Brodsky:1980sx}
S.J.~Brodsky, G.P.~Lepage and S.A.~Zaidi,
``Weak And Electromagnetic Form-Factors Of Baryons At Large Momentum
Transfer,''
Phys.\ Rev.\ D {\bf 23}, 1152 (1981).


\bibitem{Beneke:2001ev} 
  M.~Beneke, G.~Buchalla, M.~Neubert and C.~T.~Sachrajda,
``QCD factorization in B $\to \pi K, \pi \pi$ decays and extraction of Wolfenstein parameters,''
  Nucl.\ Phys.\ B {\bf 606}, 245 (2001)
  [hep-ph/0104110].
 
 
\bibitem{Tsai:2007pp} 
  Y.~-T.~Tsai {\it et al.}  [BELLE Collaboration],
  ``Search for $B^0\to  p \bar p$, $\Lambda \bar\Lambda$ and $B^ +\to p \bar\Lambda$ at Belle,''
  Phys.\ Rev.\ D {\bf 75}, 111101 (2007)
  [hep-ex/0703048].



\bibitem{PDG} 
  C.~Patrignani {\it et al.} [Particle Data Group Collaboration],
  ``Review of Particle Physics,''
  Chin.\ Phys.\ C {\bf 40}, no. 10, 100001 (2016).
  doi:10.1088/1674-1137/40/10/100001

\bibitem{Wei:2007fg} 
  J.~T.~Wei {\it et al.}  [BELLE Collaboration],
  ``Study of $B^+\to p\bar p K^+$ and $B^+ \to p \bar p \pi^+$,''
  Phys.\ Lett.\ B {\bf 659}, 80 (2008)
  [arXiv:0706.4167 [hep-ex]].


\bibitem{FSI}
H.~Y.~Cheng, C.~K.~Chua and A.~Soni,
  ``Final state interactions in hadronic B decays,''
  Phys.\ Rev.\ D {\bf 71}, 014030 (2005)
  doi:10.1103/PhysRevD.71.014030
  [hep-ph/0409317].


\bibitem{FSI1}
C.~K.~Chua,
  ``Rescattering effects in charmless $\bar B_{u,d,s}\to P P$ decays,''
  Phys.\ Rev.\ D {\bf 78}, 076002 (2008)
  doi:10.1103/PhysRevD.78.076002
  [arXiv:0712.4187 [hep-ph]].


\bibitem{Wang:2007as} 
  M.~-Z.~Wang {\it et al.}  [Belle Collaboration],
  ``Study of $B^+\to p \bar\Lambda \gamma$, $p$ $\bar \Lambda$ $\pi^0$ and $B^0\to  p \bar\Lambda \pi^-$,''
  Phys.\ Rev.\ D {\bf 76}, 052004 (2007)
  [arXiv:0704.2672 [hep-ex]].


\bibitem{Beneke:2003zv} 
  M.~Beneke and M.~Neubert,
  ``QCD factorization for $B \to PP$ and $B \to PV$ decays,''
  Nucl.\ Phys.\ B {\bf 675}, 333 (2003)
  doi:10.1016/j.nuclphysb.2003.09.026
  [hep-ph/0308039].
  
\bibitem{Uspin}
N.~G.~Deshpande and X.~-G.~He,
  ``$CP$ asymmetry relations between $\bar B^0 \to \pi \pi$ and $\bar B^0 \to \pi K$ rates,''
  Phys.\ Rev.\ Lett.\  {\bf 75}, 1703 (1995)
  [hep-ph/9412393].
  
\bibitem{Uspin1}  
  M.~Gronau,
  ``$U$ spin symmetry in charmless $B$ decays,''
  Phys.\ Lett.\ B {\bf 492}, 297 (2000)
  [hep-ph/0008292].

\bibitem{Bortoletto:1989mu} 
  D.~Bortoletto {\it et al.} [CLEO Collaboration],
  ``A Search for $b \to u$ Transitions in Exclusive Hadronic $B$ Meson Decays,''
  Phys.\ Rev.\ Lett.\  {\bf 62}, 2436 (1989).
  doi:10.1103/PhysRevLett.62.2436
  
\bibitem{Moroi:1995fs} 
  T.~Moroi,
  ``Effects of the gravitino on the inflationary universe,''
  hep-ph/9503210.

\end{thebibliography}
\end{document}